%% file: ms.tex
\documentclass[twocolumn]{aastex61}

\usepackage{multirow}

\newcommand{\blue}[1]{\textcolor{blue}{#1}}

\def\enzo{\texttt{Enzo}}
\def\yt{\texttt{yt}}
\def\omega{\texttt{OMEGA}}
\def\omegap{\texttt{OMEGA+}}
\def\gammac{\texttt{GAMMA}}

\submitjournal{ApJ}

\shorttitle{Validating Semi-Analytic Models}
\shortauthors{C\^ot\'e et al.}

\begin{document}

\title{Validating Semi-Analytic Models of High-Redshift Galaxy Formation\\using Radiation Hydrodynamical Simulations}

\correspondingauthor{Benoit C\^ot\'e}
\email{bcote@uvic.ca, benoit.cote@csfk.mta.hu}

\author[0000-0002-9986-8816]{Benoit C\^ot\'e}
\affil{National Superconducting Cyclotron Laboratory, Michigan State University, MI, 48823, USA}
\affiliation{Konkoly Observatory, Research Centre for Astronomy and Earth Sciences, Hungarian Academy of Sciences, Konkoly Thege Miklos ut 15-17, H-1121 Budapest, Hungary}
\affiliation{Joint Institute for Nuclear Astrophysics - Center for the Evolution of the Elements, USA}
\affiliation{NuGrid Collaboration, \url{http://nugridstars.org}}

\author[0000-0002-4109-9313]{Devin W. Silvia}
\affiliation{Department of Physics and Astronomy, Michigan State University, MI, 48823, USA}
\affiliation{Department of Computational Mathematics, Science and Engineering, Michigan State University, MI, 48823, USA}
\affiliation{National Science Foundation, Astronomy and Astrophysics Postdoctoral Fellow, USA}

\author[0000-0002-2786-0348]{Brian W. O'Shea}
\affiliation{National Superconducting Cyclotron Laboratory, Michigan State University, MI, 48823, USA}
\affiliation{Department of Physics and Astronomy, Michigan State University, MI, 48823, USA}
\affiliation{Department of Computational Mathematics, Science and Engineering, Michigan State University, MI, 48823, USA}
\affiliation{Joint Institute for Nuclear Astrophysics - Center for the Evolution of the Elements, USA}

\author[0000-0002-6804-630X]{Britton Smith}
\affiliation{San Diego Supercomputer Center, University of California, San Diego, CA 92093, USA}

\author[0000-0003-1173-8847]{John H. Wise}
\affiliation{Center for Relativistic Astrophysics, Georgia Institute of Technology, 837 State Street, Atlanta, GA 30332, USA}



\begin{abstract}

We use a cosmological hydrodynamic simulation calculated with \enzo\ and the semi-analytic galaxy formation model (SAM) \gammac\ to address the chemical evolution of
dwarf galaxies in the early universe. The long-term goal of the project is to better understand the origin of metal-poor stars and the formation of dwarf galaxies and the
Milky Way halo by cross-validating these theoretical approaches. We combine \gammac\ with the merger tree of the most massive galaxy found in the hydrodynamic simulation
and compare the star formation rate, the metallicity distribution function (MDF), and the age-metallicity relationship predicted by the two approaches.
We found that the SAM can reproduce the global trends of the hydrodynamic simulation.  However, there are degeneracies between the model parameters
and more constraints (e.g., star formation efficiency, gas flows) need to be extracted from the simulation to isolate the correct semi-analytic solution.
Stochastic processes such as bursty star formation histories and 
star formation triggered by supernova explosions cannot be reproduced by the
current version of \gammac.
Non-uniform mixing in the galaxy's interstellar medium, coming primarily from self-enrichment by local supernovae, causes a broadening in the MDF that can be emulated in the SAM by convolving its predicted MDF
with a Gaussian function having a standard deviation of $\sim$0.2~dex.
We found that the most massive galaxy in the simulation retains nearby 100\,\% of its baryonic
mass within its virial radius, which is in agreement with what is needed in \gammac\ to reproduce
the global trends of the simulation.

\end{abstract}




\section{Introduction} \label{sec:intro}
\input{introduction}

\section{Defining the Experiment} \label{sec:def_exp}
\input{def_experiment}

\section{Hydrodynamic Simulation} \label{sec:hydrosims}
\input{hydro_sims}

\section{Chemical Evolution Code} \label{sec:chemcode}
\input{gamma}

\section{Impact of Input Parameters} \label{sec:comparison}
\input{input_parameter_impact}

\section{Non-Uniform Mixing of Metals} \label{sec:numixing}
\input{non_uniform_mixing}

\section{Reproducing the Hydrodynamic Simulation} \label{sec:degeneracy}
\input{reproducing_hydro_sims}

\section{Discussion} \label{sec:discussion}
\input{discussion}

\section{Conclusion} \label{sec:conclusion}
\input{conclusion}

\acknowledgments

We are thankful to the anonymous referee for providing constructive feedback
and to Alex Ji, Falk Herwig, Brad Gibson, Hendrik Schatz, and Frank Timmes for their feedback on our paper prior submission.
This research is supported by the National Science Foundation (NSF;
USA) under grant No. PHY-1430152 (JINA Center for the Evolution of the
Elements) and by the ERC Consolidator Grant (Hungary) funding scheme
(project RADIOSTAR, G.A. n.\,724560).  B.W.O. was supported by the
National Aeronautics and Space Administration (NASA) through grant
NNX15AP39G and Hubble Theory Grant HST-AR-13261.01-A, and by the NSF
through grant AST-1514700.  DS was supported by the NSF Astronomy and
Astrophysics Postdoctoral Fellowship program.  BDS is supported by
NSF AST-1615848.  JHW is supported by NSF grants AST-1333360 and
AST-1614333, NASA grant NNX17AG23G, and Hubble theory grants
HST-AR-13895 and HST-AR-14326.

\enzo\ and \yt\ are developed by a large number of
independent researchers from numerous institutions around the
world. Their commitment to open science has helped make this work
possible.

%

\vspace{5mm}


\software{\texttt{GAMMA} (this work),
                \texttt{OMEGA} \citep{2017ApJ...835..128C},
                \texttt{NuPyCEE} \citep{2016ascl.soft10015R},
                \texttt{Enzo} \citep{2014ApJS..211...19B},
                \texttt{Rockstar} \citep{2013ApJ...762..109B},
                \texttt{ytree} (Smith et al., in preparation),
                \texttt{consistent-trees} \citep{2013ApJ...763...18B},
                \texttt{yt} \citep{2011ApJS..192....9T},
                \texttt{grafic} \citep{2001ApJS..137....1B},
                \texttt{Moray} \citep{Wise11_Moray},
                \texttt{Grackle} \citep{2017MNRAS.466.2217S},
                \texttt{NumPy} \citep{2011arXiv1102.1523V},
                \texttt{matplotlib} (\url{https://matplotlib.org}),
                \texttt{pydot} (\url{https://pypi.python.org/pypi/pydot}).}



\bibliographystyle{yahapj}
\bibliography{apj-jour,ms}



\end{document}

%% file: introduction.tex
The hierarchical nature and physical complexity of galaxy formation makes it challenging to create physically realistic and computationally tractable models that incorporate all stages of a galaxy's history, from the very first stars to the present day.  A further layer of difficulty arises from the fact that observations of the early universe via campaigns such as the Hubble Ultra-Deep Field and the Hubble Frontier Fields \citep[e.g.,][]{1996AJ....112.1335W,2006AJ....132.1729B,2013ApJS..209....6I,2014ApJ...786...60A,2015ApJ...799...12I} tend to measure different and less detailed quantities than observations of the Milky Way and its neighbors galaxies \citep{2015ARA&A..53..631F,2016ARA&A..54..529B}.  As a result, using observations to inform the models is a non-trivial endeavour.  In this paper, we present the first steps toward developing reliable chemical evolution models for the early universe, using high-redshift cosmological hydrodynamic simulations of galaxy formation to calibrate them.

Theoretical studies of galaxy formation typically take one of two approaches to build upon analytic efforts, each of which has its own strengths and limitations. The first approach uses cosmological simulations that employ as much physics as is computationally tractable and typically include dark matter dynamics, (magneto)hydrodynamics, plasma heating microphysics, cooling, chemistry, sub-grid prescriptions for star formation and stellar feedback, and, to a greater extent in the modern era than in the past, radiation transport, cosmic rays, and additional plasma processes.  This method has been successful in reproducing many of the observable properties of low- and high-redshift ($z$) galaxies \citep[e.g.,][]{2014MNRAS.444.1518V,2014MNRAS.445..581H,2014ApJ...795..144C,2014MNRAS.442.2560W,2015MNRAS.446..521S,2016ApJ...824...79A}.

However, due to the costs and challenging nature of the computations, an individual simulation must make compromises in the implemented physics, the dynamic range in resolution, and/or in the number of distinct galaxies simulated.  This typically means that an individual calculation either models galaxies in the high-redshift universe at very high spatial resolution with complex physics, a small number of low-redshift galaxies at reasonably good spatial resolution with less sophisticated physics, or many low-redshift galaxies at moderate spatial resolution with similarly reduced physics.  Such compromises are necessary given the finite computational resources, but they make predictions regarding the full range of stellar populations within the Local Group impossible from individual simulations. Calculations that can reach $z=0$ have mass and spatial resolutions that are too poor to adequately resolve the earliest generations of galaxy formation.  Moreover, the high computational cost of each simulation prevents exploring variations in parameters and physical processes in a thorough way.

The second approach to the theoretical study of galaxy formation is to use semi-analytical models (SAMs).  These models use merger trees based on either the extended Press-Schechter formalism or on dark matter cosmological simulations to capture the hierarchical nature of structure formation.  The latter method gained in popularity in recent years due to its native inclusion of spatial and kinetic information in addition to mass and formation history.   Various physical prescriptions such as star formation, stellar feedback, and gas flows are layered on top of these merger trees and are represented as sets of coupled ordinary differential equations.  The output of these models are typically compared to the observable properties of low-redshift galaxies. We refer to \cite{2006RPPh...69.3101B} and \cite{2015ARA&A..53...51S} for a more general discussion on SAMs of galaxy formation.

This approach has been used several times in the past to address the chemical evolution of galaxies as a function of their mass (e.g., \citealt{2013MNRAS.435.3500Y,2016ApJS..222...22C,2017MNRAS.464.3812F}) as well as the chemical evolution of specific galaxies such as the Milky Way (e.g., \citealt{2014ApJ...783..132K,2016ApJ...820...71C}), local dwarf spheroidal and ultrafaint galaxies (e.g., \citealt{2013MNRAS.434..471R,2013MNRAS.429..725S,2015MNRAS.446.4220R}), and the stellar halo \citep[e.g.,][]{2006ApJ...641....1T,2010ApJ...708.1398T}. The strength of SAMs is their low computational cost that facilitates the exploration of parameter spaces and physical processes, which has been done in recent years using a variety of statistically-robust methods including both Bayesian and frequentist Markov Chain Monte Carlo, as well as Bayesian Emulators \citep[e.g.,][]{2012MNRAS.421.1779L,2013MNRAS.431.3373H,2017MNRAS.466.2418R}.  But their weakness is the treatment of those physical processes, which are more abstract and simplified compared to the physics-rich simulations described above. 
Since the physical processes are not directly modelled and resolved, it can be difficult to trust the outputs of SAMs and to understand the degeneracies between their parameterss.

Observationally, metal-poor stars found in dwarf galaxies and in the halo of the Milky Way and Andromeda are Local Group tracers of the first generations of star formation.  The study of these old stellar populations, often called ``Galactic Archaeology,'' provides a glimpse into the distant past by assuming that (1) metal-poor stars are temporally and nucleosynthetically close to the first galaxies that formed in the early universe, and thus can be used to infer their properties; and (2) stellar halos and dwarf galaxies are relatively clean remnants of 
the early hierarchical assembly era of structure formation, with ultra-faint galaxies being particularly pristine environments for the study of the early universe  \citep{2009ApJ...693.1859B,2015ARA&A..53...51S,2015MNRAS.453.1503B,2017MNRAS.469L..83W}.  However, confronting observations of metal-poor stars with theoretical models results in several distinct challenges.  Ultra-faint dwarf satellite galaxies in particular are a challenging environment to model in the same simulations as the massive galaxies that they orbit due to their small mass and shallow potential wells, and thus the huge dynamical ranges required.  Their stellar populations are particularly susceptible to formation environment and to being removed from the galaxies via tidal harassment, particularly if poorly resolved spatially and/or in terms of dark matter particle mass.

At present, it is not possible to run physics-rich simulations of a Milky Way-type galaxy that include the dynamic range in space and mass required to adequately resolve the formation of ultra-faint dwarfs.  A different theoretical approach is thus necessary.  The obvious solution is to use SAMs, which provide the necessary dynamic range.  But the fundamental challenge with applying SAMs to this problem is that their physical prescriptions are often calibrated by comparison to observations of massive and low-redshift galaxies ($L_{\rm gal} \gtrsim L_{\rm MW}$, $z_{\rm obs} \lesssim 1$), though in recent years this has been extended to substantially higher redshifts and to lower-mass galaxies \citep[see, e.g.,][]{2015MNRAS.451.2663H,2016MNRAS.462.3854L,2017MNRAS.466.2418R}.  The applicability of the physical prescriptions and input parameters that are chosen is therefore questionable for extremely high-redshift, very low-mass galaxy formation \citep[e.g.,][]{2006MNRAS.370..645B,2009MNRAS.400.1527K,2010MNRAS.407.2017B}.  In addition, direct observations of high-redshift galaxies are limited.

There is a path forward that will allow us to more reliably use semi-analytic models as theoretical tools for Galactic Archaeology and for the exploration of galaxy formation and evolution in the early universe.  Rather than calibrating SAMs with galaxy observations at low redshift, we propose to calibrate them using physics-rich galaxy formation simulations and low-metallicity stellar populations observed in the Local Group dwarfs.  The theoretical aspect of this calibration can use merger trees derived directly from the selected hydrodynamic simulations.  Detailed analysis of the properties of these calculations can provide valuable insights into the robustness of the SAM prescriptions and lead to the creation of new prescriptions that are more appropriate for low-mass, high-redshift galaxies.  Observations of the ultra-faint dwarf galaxies can calibrate the models in a complementary way, by helping to resolve model parameter degeneracies.

This paper presents the first step in this process, building on our expertise in both semi-analytical modeling of galaxy formation and in physics-rich cosmological simulations \citep{2012MNRAS.427..311W,2014MNRAS.442.2560W,2015ApJ...802..123C,2015ApJ...807L..12O,2016ApJ...824...82C,2016ApJ...833...84X,2017ApJ...835..128C,2017nuco.confb0203C}.  We present a new SAM for galaxy formation and chemical evolution, \gammac, and compare it to the cosmological hydrodynamic simulation of high-redshift galaxy formation used in \citet{2012MNRAS.427..311W,2014MNRAS.442.2560W}. 

The outline of this paper is as follows.  In Section~\ref{sec:def_exp}, we precisely define the experiment pursued in this paper.   In Section~\ref{sec:hydrosims}, we summarize the \enzo\ simulations used as a baseline for comparison with our SAM, and in Section~\ref{sec:chemcode}, we introduce \gammac, our semi-analytic chemical evolution code. In Section~\ref{sec:comparison}, we compare the predictions of the two types of models, and in Sections~\ref{sec:numixing} and~\ref{sec:degeneracy}, we discuss extensions to the baseline model that must be made to improve agreement and degeneracies in the SAM parameters, respectively.  We then discuss some limitations and implications of this work in Section~\ref{sec:discussion}, and summarize our results in Section~\ref{sec:conclusion}.

%% file: def_experiment.tex
In this paper, we experiment with the use of multiphysics cosmological
simulations as a tool to calibrate semi-analytic models of galaxy
formation for  galaxies in the early universe that should be
observable by the James Webb Space Telescope (M$_{\rm vir} \sim
10^9$~M$_\odot$, $z \sim 7-8$), although it may require
gravitationally-lensed fields equivalent to the Hubble Frontier Fields
to do so \citep[see, e.g.,][]{2015ApJ...807L..12O}.  We determine the formation history for the most massive
galaxy in our simulation at the calculation's stopping redshift
($z=7.29$) by using many simulation outputs to create a merger tree
that encompasses all dark matter halos at earlier times which could
plausibly form stars, and then use this merger tree as an input for
our semi-analytic model.  This eliminates the galaxy's growth history
as a source of uncertainty in our inter-model comparison, and allows
us to focus on the physical prescriptions used in the semi-analytic
model as well as the model's other input parameters.  We then ask two
fundamental questions:

\begin{enumerate}

\item Using our standard semi-analytic galaxy formation model (which
is similar to other models of its type), how closely are we able to
match the galaxy's star formation history and metallicity distribution
function, as measured at the final redshift in the simulation?

\item What key qualities of the multiphysics simulations are absent
from the approximations made in the semi-analytical model, and how
does a more complete incorporation of these qualities affect the
predictions made by the model?

\end{enumerate}

The first question is important because the star formation history and MDF
are two of the primary observable properties of low-mass
galaxies (e.g., Local Group dwarf galaxies), with more detailed
physical properties either being difficult to determine or possibly
contaminated by interaction with other, more massive galaxies (e.g.,
structural information about the stellar populations can be difficult
to infer due to small numbers of stars, and can be modified by tidal
harassment from the central galaxy).  We
examine this question in Section~\ref{sec:comparison}.

The second question is important because the level of abstraction in
the multiphysics simulations is much different  than in the
semi-analytic models, and thus it is possible to find emergent
properties of these simulations that can inform the semi-analytic
models.  Put somewhat differently, multiphysics simulations emulate
the universe at a relatively fine level of granularity by modeling
dark matter dynamics, hydrodynamics, radiation transport, the heating
and cooling of gas, and star formation and feedback in many individual
resolution elements that are much smaller than the scale of individual
galaxies.  This allows phenomena like starbursts, accretion of gas,
and galactic winds to develop naturally via interactions of these
individual physical processes.  Semi-analytic models of galaxy
formation, on the other hand, typically model galaxy evolution by
describing physical phenomena such as star formation or galactic winds
using sets of ordinary differential equations that depend on bulk
galaxy quantities such as the mass of gas and stars in the galaxy and
the virial mass of the dark matter halo where the baryons reside.
Given these different levels of abstraction, a close examination of
the multiphysics simulations of low-mass galaxies may show physical
behaviors that are not captured in the standard semi-analytic models
(in particular, behaviors relating to the rate of mixing of
metal-enriched gas, the relatively shallow potential wells of small
dark matter halos, and the rapid evolution of cosmological structure
in the early universe) that can be used to increase the physical
realism, and thus predictive capabilities, of the semi-analytic model.
We examine this question in Sections~\ref{sec:numixing}
and~\ref{sec:degeneracy}.

It is important to acknowledge that the interpretation of the
experiment undertaken in this paper is complicated by the fact that we
are using one theoretical framework (multiphysics simulations) to
inform a second, substantially different theoretical framework
(semi-analytic models).  Both of these theoretical frameworks have
significant implicit and explicit assumptions built into them that may
impact their realism and predictive power, and thus comparing them may
introduce systematic errors.
In addition, it is dangerous to assume
that multiphysics simulations (our baseline for comparison) behave in
a way that is actually ``true.''  (Note that these ideas are explored more thoroughly in
Section~\ref{sec:discussion}).  However, in the absence of an
abundance of observational data of low-mass, high redshift galaxies,
this is the only practical means of improving the physical accuracy of
our semi-analytic models.

%% file: hydro_sims.tex
\subsection{The Enzo Code} \label{sec:enzo_code}

The simulation described in Section~\ref{sec:enzo_sims} is calculated using the
\enzo\ adaptive mesh refinement code \citep{2014ApJS..211...19B}.  \enzo\
uses an N-body adaptive particle-mesh solver to model the dark matter
dynamics, and solves the equations of hydrodynamics using the
second-order-accurate piecewise parabolic method (PPM) and the HLLC
Riemann solver \citep{Woodward84,Bryan97b,Bryan95,toro-1997}.
  \enzo\ uses the Berger and
Colella block-structured adaptive mesh refinement scheme in
Cartesian coordinates 
\citep{Berger89}.  This simulation also includes \enzo's
nine-species non-equilibrium chemistry and cooling model, which
follows species of H, He, and H$_2$ \citep{abel97,anninos97} and
includes the
H$_2$ cooling rates from \citet{2008MNRAS.388.1627G}.  It also uses the \texttt{Moray} radiation transport package
\citep{Wise11_Moray} and ``star
particles'' to represent both individual Population III stars and
ensembles of metal-enriched stars.

\subsection{Simulations} \label{sec:enzo_sims}

The simulation used in this paper is the ``RP'' simulation described
in great detail in \citet[][with this calculation hereafter referred to as the ``W12'' simulation]{2012MNRAS.427..311W,2014MNRAS.442.2560W}, but we repeat the
most important details here.  The calculation uses a simulation box
that is 1 Mpc (comoving) on a side with a $256^3$ root grid and
$256^3$ dark matter particles.  This gives a dark matter mass
resolution of $1{,}840$~M$_\odot$, which is enough to resolve halos with
masses exceeding $\simeq 2 \times 10^5$~M$_\odot$.  The initial
conditions were generated at $z=130$ using the \texttt{grafic} package
\citep{2001ApJS..137....1B} and the seven-year WMAP best-fit
parameters \citep{2011ApJS..192...18K}: $\Omega_{\rm M} = 0.266$,
$\Omega_\Lambda = 0.734$, $\Omega_{\rm b} = 0.0449$, $h=0.71$, $\sigma_8 =
0.81$, and $n=0.963$.

The simulation was evolved from $z=140$ to $z=7.29$ using the physics
described in Section~\ref{sec:enzo_code} and a maximum of $N_{\rm ref}
= 12$ levels of refinement, giving a comoving resolution of
0.95 pc at the highest level.  Cells were refined based on dark matter and baryon mass in a
super-Lagrangian (i.e., increasingly aggressive) manner, and always
resolve the Jeans length by a minimum of 4 cells to avoid artificial
fragmentation during the collapse of gas clouds
\citep{Truelove98}.  Both Population III and metal-enriched
stars were formed, with Pop III stars having a top-heavy IMF and
metal-enriched stars having a more standard Galactic-like Salpeter
IMF.  Gas collapsing in cells with metallicities $[\textrm{Z/H}]$\footnote{[Z/H] $\equiv$ log$_{10}$(Z/H) $-$ log$_{10}$(Z/H)$_\odot$} $< -4$
form a single Population III star whose mass is drawn from a top-heavy
IMF that is a power-law above 100~M$_\odot$ and is exponentially
damped below that mass; more metal-enriched gas forms a star particle
representing an ensemble of metal-enriched stars.  All star particles
feedback radiation using the {\sc Moray} radiation transport algorithm
and mass, metals, and thermal energy through supernova explosions.

\subsection{Analysis} \label{sec:enzo_analysis}

The key data product required from a cosmological simulation for use in
semi-analytic models is  a halo merger tree.  We generate this by
using the \texttt{Rockstar} phase-space halo finder
\citep{2013ApJ...762..109B} on the simulation's dark
matter particles at every simulation data output, which produces a
halo catalog composed of all of
the gravitationally-bound dark matter halos and sub-halos in the
simulation.  We then used the
\texttt{consistent-trees} tool \citep{2013ApJ...763...18B} to create a
gravitationally self-consistent merger tree for every halo in the
simulation at $z=7.29$ (the last simulation data output). To load the output from
\texttt{consistent-trees} and step through the halos in a given tree,
we use the \texttt{ytree}\footnote{\url{http://ytree.readthedocs.io/}}
code \citep{ytree}, an extension of the
\texttt{yt}\footnote{\url{http://yt-project.org/}} analysis toolkit
\citep{2011ApJS..192....9T} designed for the ingestion and manipulation of merger tree data
from multiple sources.  The \texttt{ytree} code provides a Python
interface for merger tree-like data structures, allowing the user to
easily traverse from a halo to its ancestors or descendents, to access
field data with symbolic units for a partial or whole tree, to add new
fields resulting from further analysis, and to save a tree or group of
trees to optimized, reloadable format. 

In this work, we use the definitions provided by \texttt{Rockstar} for
basic halo properties, such as position, virial radius
($R_\mathrm{vir}$), and virial mass ($M_\mathrm{vir}$).  These are
described in detail in \citet{2013ApJ...762..109B}, but we present
them briefly here.  A halo's position is defined as the center of mass
of a central subgroup of N particles for which the Poisson error,
expressed as $\sigma/\sqrt{\rm N}$, is minimized.
Virial masses and radii are defined as the properties of a sphere
within which the average density is equal to the threshold overdensity
given in \citet{1998ApJ...495...80B}.

\subsection{Separating Star Particles} \label{sec:enzo_ssp}

Star particles found inside the virial radius of the main galaxy at the final redshift can have different 
origins.  They might have formed in the main galaxy or in the infalling satellite galaxies, but they could
also have been ejected from neighbouring galaxies.  Since we aim to reproduce the evolution of individual 
galaxies with \gammac\ using their merger tree, we need to separate and tag each star particle
in order to recover their specific star formation history and metallicity distribution function. 

\begin{figure*}
\begin{tabular}{c}
\hspace*{0.01cm} \includegraphics[width=7.in]{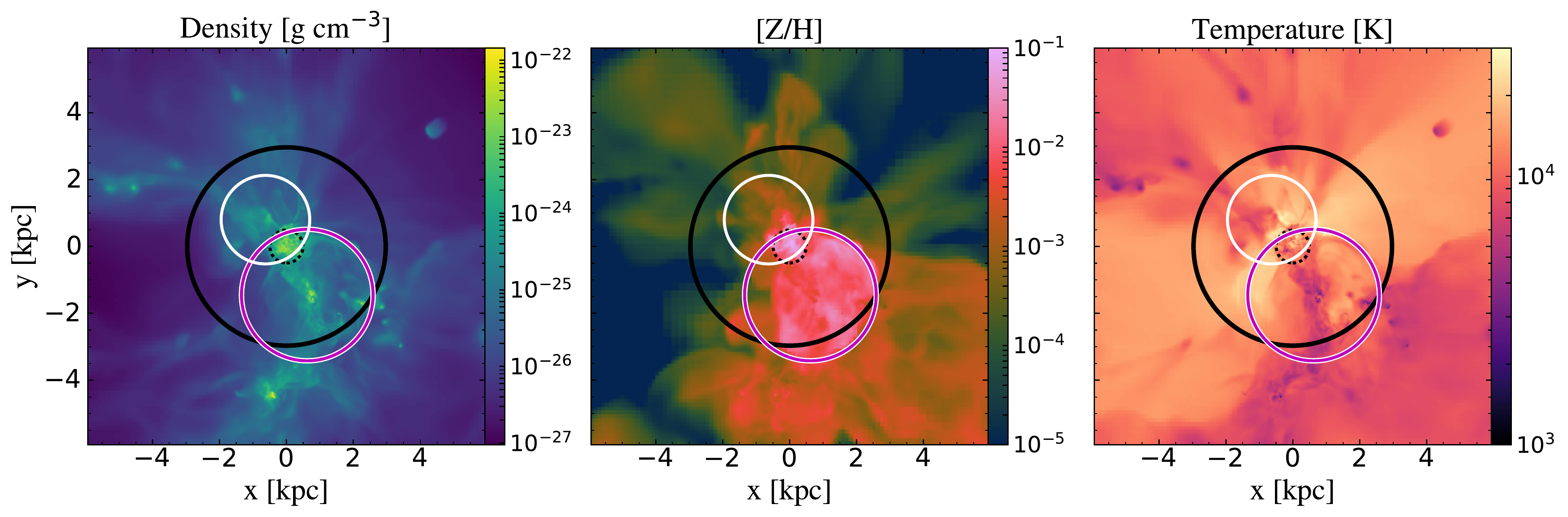} \vspace*{-0.15cm}\\
\includegraphics[width=7.0in]{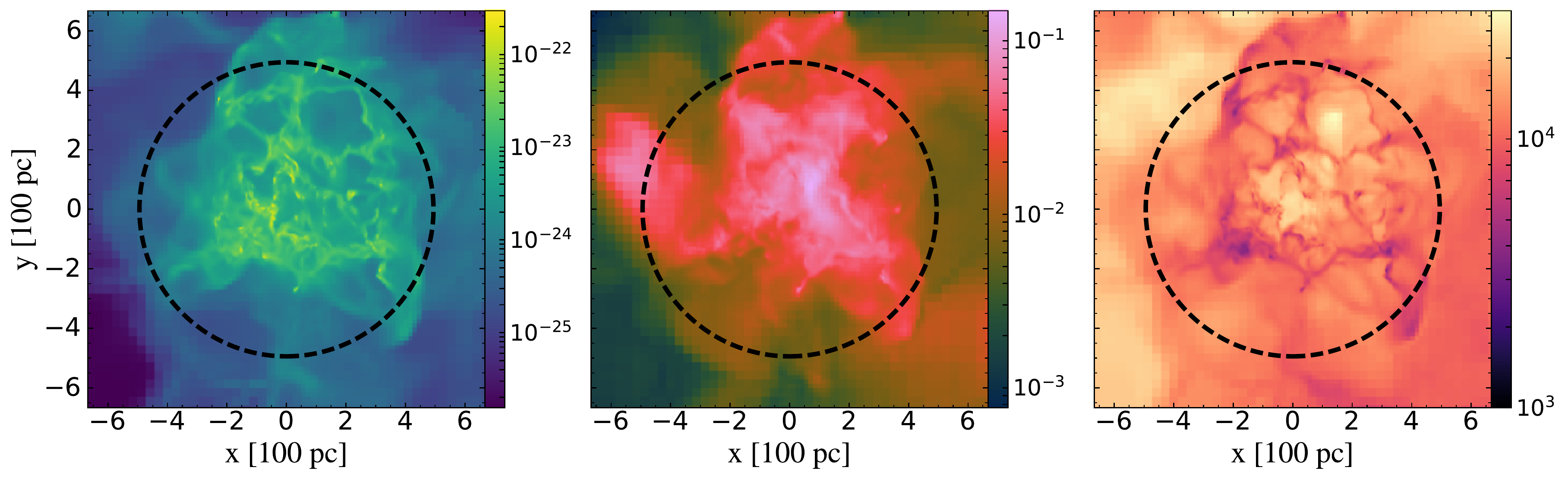}
\end{tabular}
\caption{
  Snapshots at $z=7.45$ of the density (left column),
  metallicity (middle column), and temperature (right column) for the
  most massive galaxy in the simulation.  All panels consist of
  projections weighted by density.  Values for metallicities are
  normalized to $Z_\odot=0.014$.  \textbf{Top row:} The solid
  and dashed black circles trace the virial and inner radius of the
  main halo, respectively.  The radius of the inner region is set to a
  sixth of the virial radius.  The solid pink and white circles trace
  the virial radius of two satellite halos that have not merged by the
  end of the simulation.  \textbf{Bottom row:} Same as in the top
  panels, but zoomed on the inner region. By the end of the simulation,
  the main halo and its two satellites have a total (dark matter plus baryon) mass of $5.45\times10^8$\,M$_\odot$,
  and $1.30\times10^8$ (pink) and $1.11\times10^8$ (white) M$_\odot$, respectively.}
\label{fig_hydro_final_state} 
\end{figure*}

Using \texttt{ytree}, we first identify the most massive progenitors of the main and satellite halos and record their virial radius
$R_\mathrm{vir}$ and the coordinate of their center of mass $\vec{r}_0$ as a function of redshift.  Then, for each redshift
from the highest to the lowest, we identify all star particles that formed between the previous and current redshift.
For each \textit{new} star particle $j$, we calculate its normalized distance from the center of each halo $i$,
\begin{equation}
d_{j,i} = \frac{|\vec{r}_j-\vec{r}_{0,i}|}{R_{\mathrm{vir},i}},
\end{equation}
where all quantities vary as a function of redshift, and $\vec{r}_j$ is the current position of the particle $j$.
Each star particle is then associated with the halo showing the minimum $d_{j,i}$ value.  Using this approach, we found
that the most massive progenitors of the main halo and its most massive satellite (pink circle in Figure~\ref{fig_hydro_final_state})
allow to recover 97.4\,\% the stellar content found inside the main halo at the final redshift, which is sufficient 
for the purpose of this work.

\begin{figure}
\center
\includegraphics[width=3.3in]{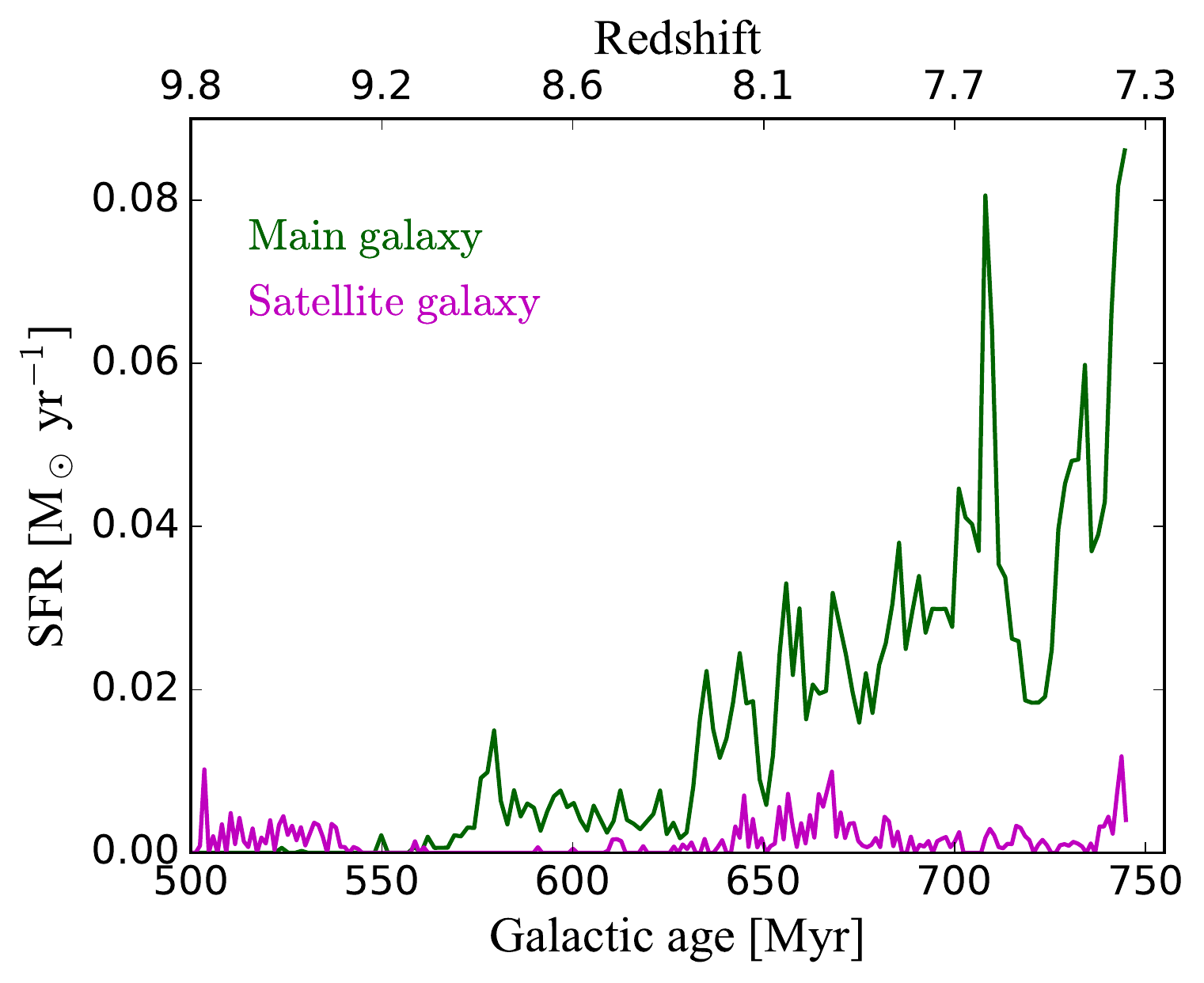}
\caption{Star formation history of the target halo (green) and its most
  massive satellite halo (pink, also shown as the pink circles in
  Figure~\ref{fig_hydro_final_state}). The total integrated stellar mass formed
  in the target and satellite halo at the end of the simulation at $z=7.29$ is $3.58\times10^6$ and $4.60\times10^5$\,M$_\odot$, respectively.
  }
\label{fig_hydro_SFH}
\end{figure}

\subsection{Galaxy Behaviour} \label{sec:gal_behavior}

\begin{figure}
\center
\includegraphics[width=3.3in]{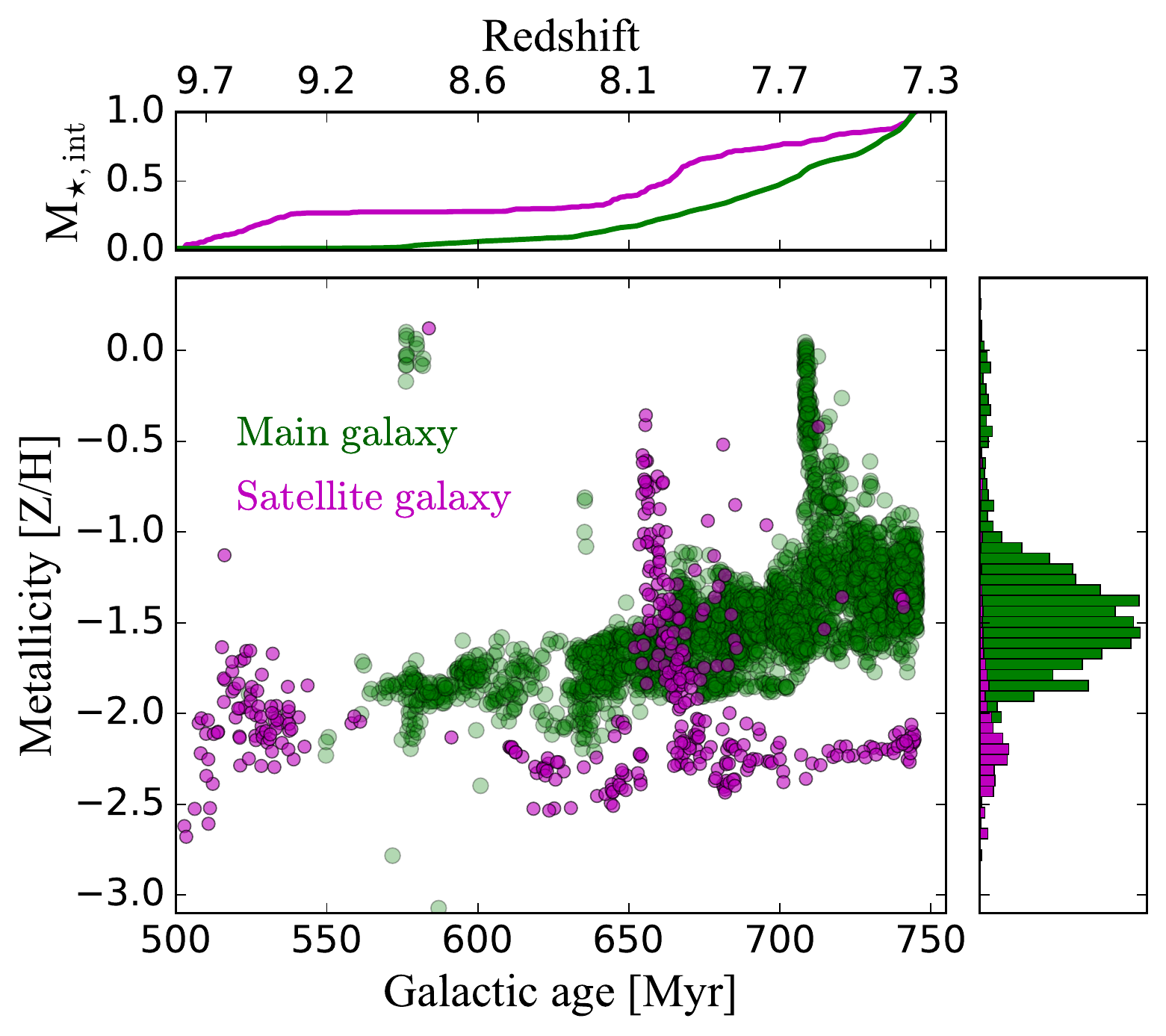}
\caption{\textbf{Main panel:} Metallicity (in units of log Solar
  metallicity) of all stellar populations
  as a function of their formation time (or galactic age) for the target
  halo (green) and its most massive satellite halo (pink).
  \textbf{Top panel:} Integrated stellar mass formed as function of
  time, scaled to the total stellar mass in each halo at $z=7.29$.  \textbf{Left panel:} Metallicity distribution functions of
  all stellar populations.}
\label{fig_hydro_age_Z}
\end{figure}

\begin{figure*}
\center
\includegraphics[width=7.in]{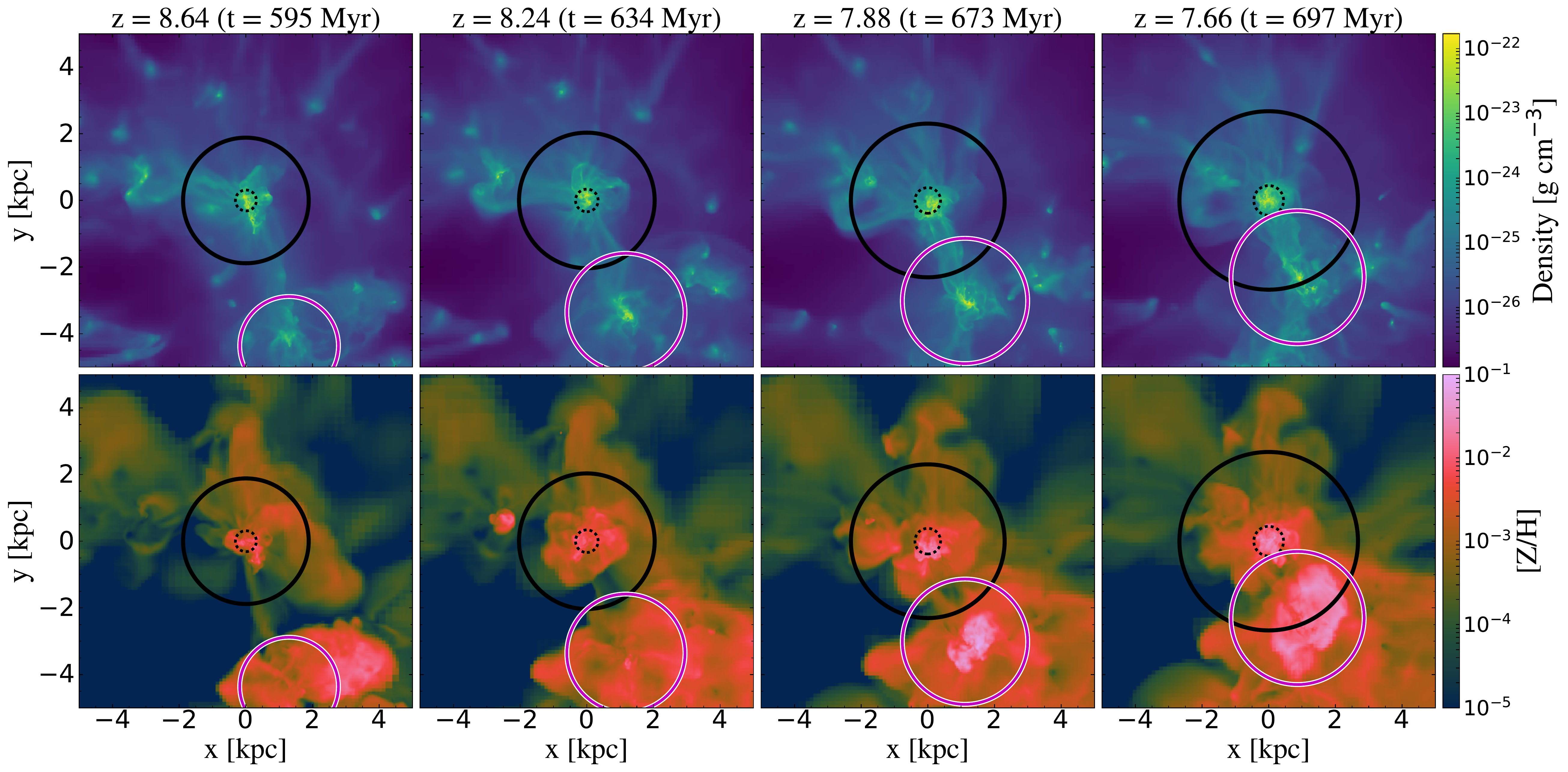}
\caption{Evolution of the density (top row) and metallicity (bottom
  row) as a function of galactic age (from left to right) for the most massive
  galaxy in the simulation (the ``target galaxy''), shown as
  density-weighted projections.  The radius of the solid and dotted black circles represent
  the virial radius of the target halo and a sixth of that radius, respectively.  The radius of the pink
  circle represents the virial radius of the most massive satellite.}
\label{fig_video}
\end{figure*}

In this paper, we focus our attention on the most massive galaxy in
the simulation -- the same galaxy targeted in the W12 simulation.
This galaxy, henceforth referred to as the ``target galaxy,'' has a
range of interesting physical features.
Figures~\ref{fig_hydro_final_state} through~\ref{fig_video} show some
of the large-scale properties of the target halo, which are also
generally representative of high-redshift galaxy formation.  Note that
by the end of the simulation, the main halo and its two satellites
have a total (dark matter plus baryon) mass of
$5.45\times10^8$\,M$_\odot$, and $1.30\times10^8$ and $1.11\times10^8$\,M$_\odot$, respectively.

Figure~\ref{fig_hydro_final_state} shows density-weighted projections at
$z=7.45$ of the density field, metallicity,
and temperature for the target galaxy, at two physical scales: out to
roughly twice the virial radius of the target galaxy, and in the
central star-forming region of the same galaxy.  
These images were chosen to be at this redshift rather than the final
redshift to
highlight the complexity of the baryonic structure at all scales
within the galaxy.  Clear filamentary structure and substantial
inhomogeneity in metallicity can be seen in the halo outside of the
star-forming region, with the regions of highest metallicity relating
to infalling satellite halos rather than metal-enriched galactic
outflows (although those are also present, but correlated more
strongly with high temperatures due to the accompanying radiation from
massive stars).

The central star-forming region has gas at a wide
variety of densities and temperatures, and displaying a range of
metallicities.  While this is in some ways analogous to the
interstellar media in galaxies like the Milky Way, the range of
observed temperatures is much smaller due to the substantially lower
virial temperature of the halo (tens of thousands of Kelvin rather than
millions of Kelvin).  Broadly speaking, there are several different
dense regions of star formation that have differing metallicities,
rather than a single molecular-cloud like structure that one would
expect in, e.g., Population III star formation 
\citep{2002Sci...295...93A,2002ApJ...564...23B,2007ApJ...654...66O,2009Sci...325..601T}.  
In addition, evidence of
stellar feedback can be seen in the hot, low density, metal-enriched
gas in the central regions that are spatially adjacent to star-forming
regions.

Figure~\ref{fig_hydro_SFH} shows the star formation history (SFH) of
both the target halo and its most massive satellite halo,
smoothed on a 2\,Myr time scale.  
All stars included in this figure are formed in the main progenitor halos of the target and satellite galaxies (see Section~\ref{sec:enzo_ssp}).
Overall, the SFH of the target halo displays a steadily increasing trend with time,
although both halos show substantial variability in their star formation rate
on short time scales.  This variability is consistent with other simulations of
high-redshift galaxy formation 
\citep[e.g., the Renaissance simulations,][]{2014ApJ...795..144C,2015ApJ...807L..12O,2016ApJ...833...84X}, 
and with the idea that low-mass galaxies have shallow potential wells with easily-disrupted
star formation.  The total integrated stellar mass formed in the target and satellite halos 
at the end of the simulation is $3.58\times10^6$ and $4.60\times10^5$\,M$_\odot$, respectively.

Figure~\ref{fig_hydro_age_Z} shows the metallicity of each ``star
particle'' (i.e., each specific particle tracing the formation of a
parcel of stars) in the target halo and its most massive satellite as
a function of formation time, with sub-panels displaying the
cumulative star formation history and the metallicity distribution
function of the stars in each galaxy.  As with
Figure~\ref{fig_hydro_SFH}, this includes stars formed in all of the
progenitor halos of each of the two galaxies at a given point in time.
This plot displays several notable features.  First, while the
metallicity of stars formed in the target halo and its progenitors
trends upward, there is substantial variation in stellar metallicity at any
given time.  This is due to the variation in metallicity of star-forming regions within
the galaxy due to non-uniform mixing.  There are ``spikes'' in Figure~\ref{fig_hydro_age_Z}
corresponding to relatively extremely metal-rich star formation at $t \simeq 575$\,Myr and
$720$\,Myr in the main galaxy.  Those stellar populations formed out of 
relatively unmixed gas containing a large fraction of nearby supernova ejecta.
Their formation is likely triggered by the short cooling timescales of the hot metal-rich
gas phase (see Section~\ref{sect:disc_future}), rather then by galaxy merger events.

The satellite halo (purple dots) does not display the same trend of increasing metallicity with time -
rather, its progenitors are more metal-rich at early times, and have
approximately constant stellar metallicity for 200\,Myr afterward with
the exception of a large metallicity spike at $t \simeq 660$\,Myr.  This may
be related to the different formation histories of these objects, and
to their difference in size -- metallicity is determined by a wide
variety of factors, including the production of metal-enriched gas by
supernovae, the outflow of metal-enriched gas from the halo driven by
supernovae, and the inflow of gas of different metallicity from the
cosmic web and infalling satellites (with the infalling satellites
possibly having highly metal-enriched gas in this particular instance, as can be seen in
Figure~\ref{fig_hydro_final_state}).

Figure~\ref{fig_video} shows snapshots of the baryon density and
metallicity of the target galaxy at $z=8.64$, 8.24, 7.88, and 7.66,
with the frame of the image extending to roughly three virial
radii.  As with Figure~\ref{fig_hydro_final_state}, this sequence of
images shows the complexity of the circumgalactic environment for the
target galaxy and its satellites.  The target galaxy is accreting
matter from roughly three different filaments, with a tremendous
amount of metal entering into the circumgalactic environment due to
the approach of the massive satellite (bottom-right corner), which is surrounded by a large
cloud of metal-enriched gas originating from a recent burst of
supernovae.  This burst is associated with the metal-rich stellar populations
shown in Figure~\ref{fig_hydro_age_Z} (the pink metallicity spike at $t \simeq 660$\,Myr).
The formation of these populations in the satellite galaxy occurs before it enters the
virial radius of the target galaxy, which can be seen in the two middle columns of Figure~\ref{fig_video}.

%% file: gamma.tex
\begin{figure}
\center
\includegraphics[width=3.3in]{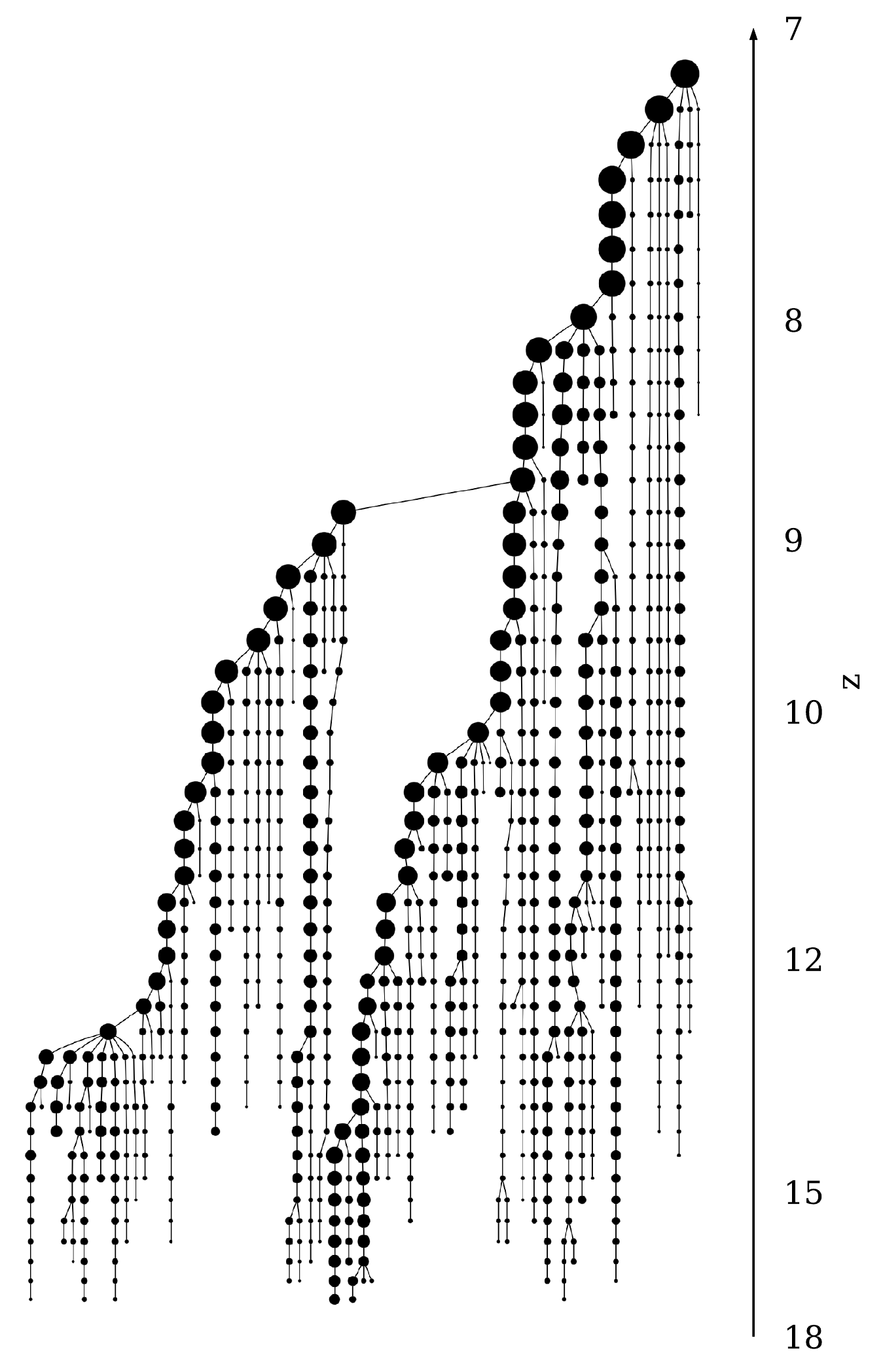}
\caption{Merger tree of the most massive galaxy in the \protect\cite{2012MNRAS.427..311W} simulation.
Every progenitor halo, or tree-node, is represented by a circle where its size is proportional to
the logarithm of its virial mass. The mass assembly of the galaxy can be followed in 
time in the upward direction up to $z=7.29$, the final redshift of
the simulation.  A series of individual tree-nodes connected by a line along the redshift axis shows 
the mass evolution of a halo, but taken at different snapshots in time. 
Tree-nodes that are descending from more than one progenitor
represent mergers. Only halos with masses above $10^5$\,M$_\odot$ are
shown.}
\label{fig_merger_tree}
\end{figure}

In this section we present \gammac\ (Galaxy Assembly with Merger trees for Modeling Abundances), a semi-analytic chemical evolution code that accounts for the mass assembly history of galaxies using merger trees extracted from cosmological simulations. The novelty of \gammac\ is its connection with nuclear astrophysics.  It represents the end-point of our open-source JINA-NuGrid chemical evolution pipeline (\citealt{2017nuco.confb0203C}).  All of our codes are available online\footnote{\url{http://github.com/becot85/JINAPyCEE}}.

\subsection{GAMMA} \label{subsec_gamma} 

The first step in using \gammac\ is to traverse the entire halo merger tree and re-organize the tree-nodes in order to feed the properties of all galaxy mergers into \gammac. In this work, we use the merger tree of the most massive galaxy of the W12 simulation shown in Figure~\ref{fig_merger_tree}.  Each tree-node refers to a unique snapshot of a halo at a specific redshift in the simulation.  In \texttt{ytree}, even if a halo does not experience any merger during a certain redshift interval, the tree-nodes associated with this specific halo will still have unique entries and halo identification numbers in the merger tree.  This is a consequence of the way that merger trees are created from halo catalogs (which are unique at each redshift), and is useful for obtaining accurate halo growth rates.

\begin{figure*}
\includegraphics[width=7.in]{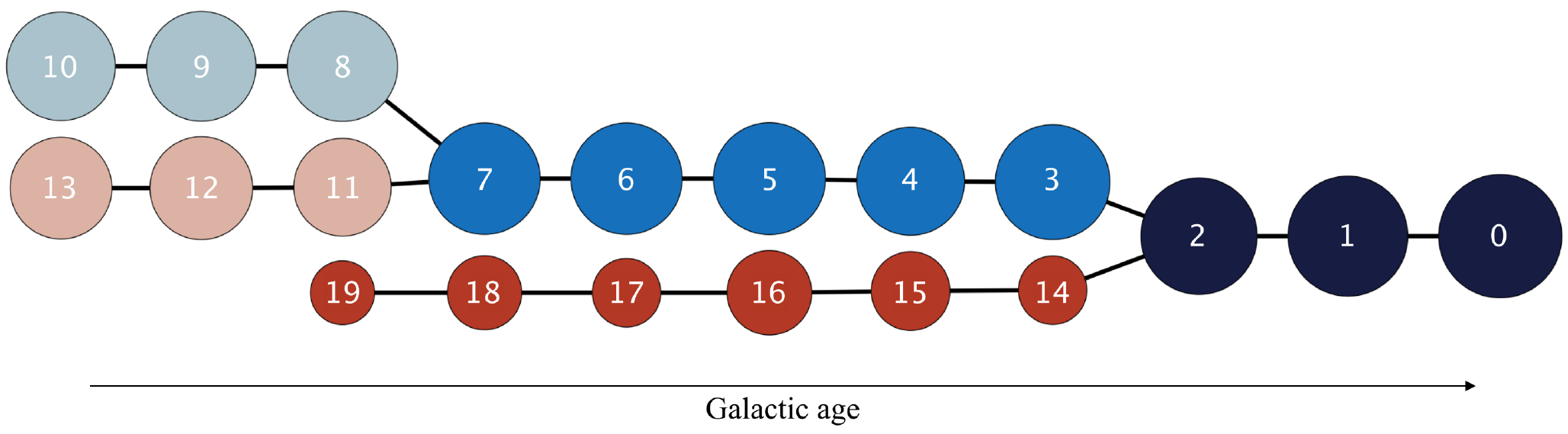}
\caption{Illustration of the terminology used in this work.  Each circle represents a halo at a particular snapshot, a tree-node in the merger tree, where the radius of that circle scales with $\log{M_\mathrm{vir}}$.  Tree-nodes connected together by the same color are defined as a ``branch,'' and refer to an isolated building-block galaxy. These individual galaxies are simulated with \omegap\ (Section~\ref{sect_omegap}) while their interaction with each other and the hierarchical assembly of the target galaxy is orchestrated by \gammac\ (Section~\ref{subsec_gamma}). The number listed on each tree-node refers to its depth-first order when sorting the merger tree (see also Figure~1 in \protect\citealt{2013AN....334..691R}).  In this case, the most massive progenitors correspond to the tree-node segment labelled from 0 to 10.}
\label{fig_tree_demo}
\end{figure*}

While traversing the merger tree, we identify all tree-nodes that are the starting point of a new \textit{branch}, which is defined as a segment in the merger tree (or a series of interconnected tree-nodes) where no merger is occurring, although the halo may still grow by accretion from its surroundings.  For each tree-node, we then move forward in time and identify all interconnected tree-nodes until we encounter a merger, which represents the end point of a branch.  \gammac\ will then consider each branch as an isolated galaxy that is not interacting with surrounding galaxies.  Figure~\ref{fig_tree_demo} illustrates this terminology using the same type of graph as Figure~\ref{fig_merger_tree}, highlighted to emphasize individual branches.

Once all branches have been identified, \gammac\ creates every galaxy in chronological order, from the highest to the lowest redshift, using the \omegap\ code (see Section~\ref{sect_omegap}).  Each galaxy receives the time-dependent properties of their associated branch as an input (e.g., dark matter mass growth) and is evolved until it encounters a merger with one or more galaxies.  When a new galaxy is the result of a merger, \gammac\ combines the stellar and gaseous components of all progenitor galaxies involved in the merger and feeds this information into \omegap\ for the initial conditions of the new galaxy.  During the process, stars formed in the progenitor galaxies stay active in the new galaxy, meaning that they will continue to inject metals and energy into their new environment.

In the present version of \gammac, we do not account for ram pressure and tidal stripping processes of satellite galaxies that can occur when entering the dark matter halo of a central galaxy (see e.g., \citealt{2017arXiv170503018S}).  This can modify the amount of gas and metals present in the CGM of the central galaxies in our models before the merger event between the satellite and central galaxies (see Section~\ref{sec:tot_mass} for a discussion).  This could also reduce the hot gas reservoir of satellite galaxies and therefore reduce their star formation rate.

\subsection{OMEGA+} \label{sect_omegap}

\omegap\ is a galaxy evolution code that consists of a star-forming region, called the cold gas reservoir, which also is assumed to contain the stellar population in the galaxy.  This is surrounded by a hot gas reservoir filling the dark matter halo of the host galaxy.  For the sake of clarity, in this work we refer to the  region containing cold gas and stars as the ``galaxy'' and the reservoir of hot gas surrounding it as the ``circumgalactic medium'', or CGM. The star-forming region is simulated using the galactic chemical evolution code \omega\ \citep[One-zone Model for the Evolution of GAlaxies;][]{2017ApJ...835..128C}.  Given an input star formation history, this latter code calculates the chemical mixture of the galactic gas (i.e., cold gas reservoir) as a function of time by accounting for the contribution of multiple stellar populations as well as the presence of galactic inflows and outflows.

The role of \omegap\ is to interact with \omega\ at each timestep in order to control the rates of inflow, outflow, and star formation (see equations below).  Within this framework, \omega\ is thus only used to calculate the mass and energy returned by all stellar populations as a function of their initial mass, age, and metallicity.  We refer to Section~\ref{sect_yields} for more details on the composition of the mass ejected by each stellar population.  Using \omegap\ provides two main advantages compared to using \omega\ alone: the star formation rate can be self-calculated from the balance between galactic inflows and stellar feedback, and the metals ejected outside the galaxy are mixed with the CGM reservoir and can be recycled into the star-forming region via galactic inflows.

We next present the set of equations that drive the evolution of our galaxy model, which represents one branch of the merger tree.  Since we plan to calibrate our model for the high redshift universe, some of our parametrizations are exploratory.

\subsubsection{Initial Conditions} \label{sect_ini}

Each galaxy is embedded in a virialized system defined by

\begin{equation}
\label{eq_virr}
V^2_\mathrm{vir}=\frac{GM_\mathrm{vir}}{R_\mathrm{vir}},
\end{equation}
where $G$ is the gravitational constant and $V_\mathrm{vir}$, $M_\mathrm{vir}$, and $R_\mathrm{vir}$ are the virial velocity, mass, and radius, respectively (see Section~\ref{sec:enzo_analysis}).  $M_\mathrm{vir}$, which includes the dark matter and baryonic mass, and $R_\mathrm{vir}$ are provided by GAMMA as input parameters.  The virial temperature $T_\mathrm{vir}$ of the system is defined by (\citealt{2001PhR...349..125B})

\begin{equation}
k_{\rm B} T_\mathrm{vir}=\frac{1}{2}\mu m_{\rm H} V^2_\mathrm{vir}
\end{equation}
and is assumed to be the temperature of the CGM gas filling the dark matter halo (\citealt{1991ApJ...379...52W}).  In this last equation, $k_{\rm B}$ is the Boltzmann constant, $\mu$ the mean molecular weight (which can range from 0.6 to 1.2), and $m_{\rm H}$ the mass of a hydrogen atom.  $T_\mathrm{vir}$ is usually used to calculate the cooling timescale of the CGM.  However, for the high-redshift galaxy considered in this work, this timescale is too short to be used in the gas circulation process (see Section~\ref{sect_gal_in}). 

A \textit{primordial} galaxy within the context of this paper refers to a galaxy that is not the result of a galaxy merger, but rather forms directly out of virializing dark matter particles or sub-resolution dark matter halos.  In that case, the gas fraction of the system is set to the universal baryonic fraction and all the of gas is deposited in the CGM, assuming a primordial composition,

\begin{equation}
\label{eq_m_cgm_ini}
M_\mathrm{CGM}(t=0) = \frac{\Omega_{b,0}}{\Omega_0}M_\mathrm{vir}.
\end{equation}
As in the W12 simulation, the cosmological parameters adopted in this work come from \cite{2011ApJS..192...18K}, with the relevant parameters being $\Omega_{b,0}=0.0449$, $\Omega_\Lambda=0.734$, and $h=0.71$.
In Equation~(\ref{eq_m_cgm_ini}), $t=0$ refers to the formation time of the considered galaxy (or branch), and not the beginning of the cosmological simulation.  The actual simulation time is traced by \gammac\ in order to orchestrate the formation of the different building-block galaxies.

If a galaxy is not primordial but is the result of a merger, \omegap\ then uses the input conditions provided by \gammac\ regarding existing stellar populations and the mass and composition of the gas components. Such initial conditions are obtained by summing the components of each progenitor system (i.e., all galactic gas components are combined together).

\subsubsection{Overall Gas Circulation} \label{sect_over_circ}

\begin{figure}
\center
\includegraphics[width=3.3in]{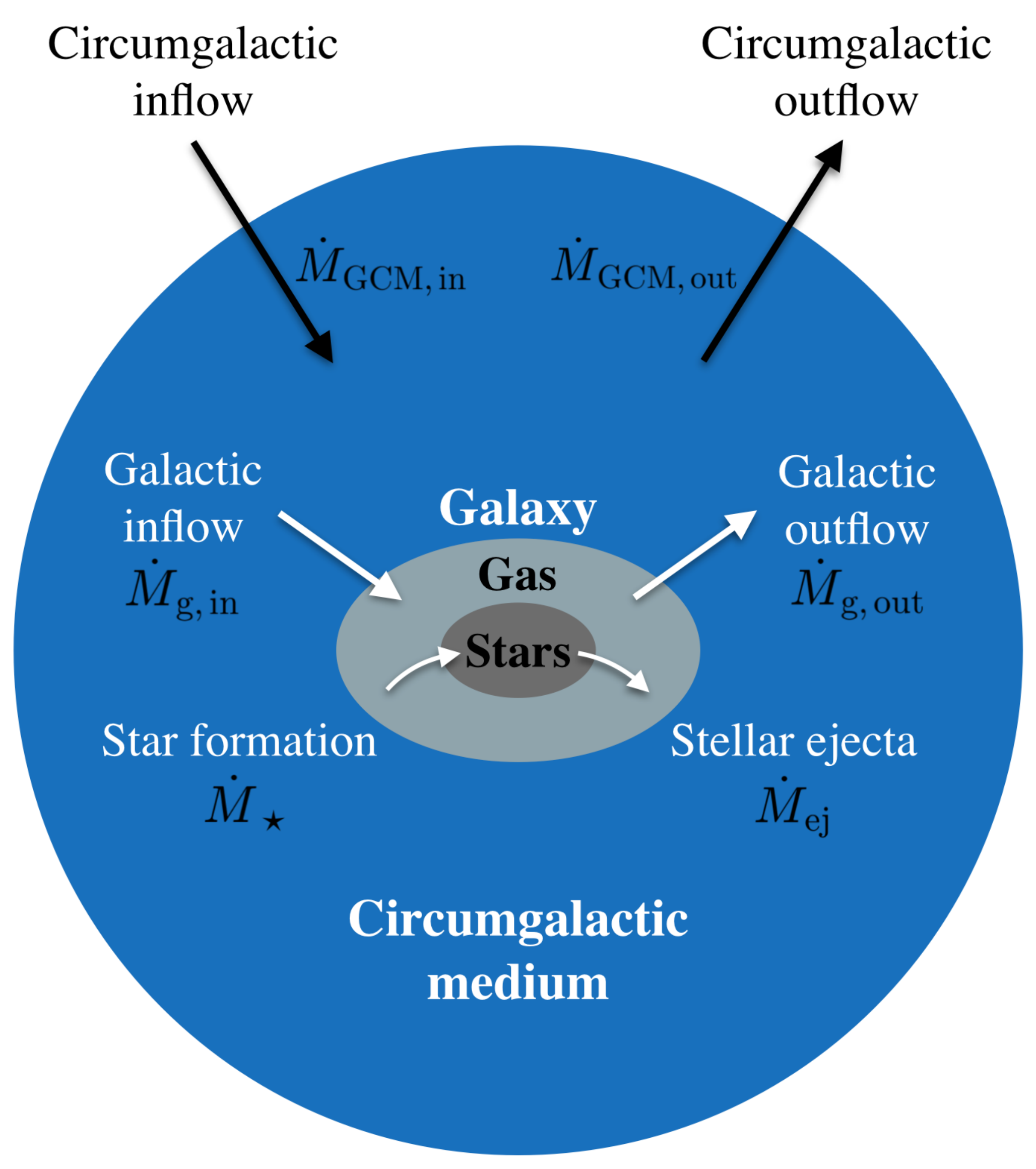}
\caption{Structure overview of \omegap.  The cold star-forming gas (the galaxy) is located at the center of a hot
gas reservoir (the circumgalactic medium) filling the host dark matter halo.  The arrows show the different mass
transfer processes between the different gas components, as described in Section~\ref{sect_omegap}. Within this
framework, the region outside the circumgalactic medium is called the external medium. \gammac, defined in
Section~\ref{subsec_gamma}, uses \omegap\ on top of each branch of the merger tree.}
\label{fig_sam_cartoon}
\end{figure}

Here we present the main equations describing the mass exchange between the different components of \omegap.
The time evolution of $M_\mathrm{gas}$, the mass of the cold gas reservoir (galactic gas), is defined by the following differential equation,

\begin{equation}
\dot{M}_{\rm gas} = \dot{M}_{\rm g,in}  + \dot{M}_{\rm ej} - \dot{M}_\star - \dot{M}_{\rm g,out},
\label{eq_main_gal}
\end{equation}
where the four terms on the right-hand side are the inflow rate from the CGM into the galaxy ($\dot{M}_{\rm g,in}$), the combined mass-loss rate of all stars ($\dot{M}_{\rm ej}$), the star formation rate ($\dot{M}_\star$), and the outflow rate from the galaxy into the CGM ($\dot{M}_{\rm g,out}$).  While the magnitude of the star formation rate drives how much metal mass is ejected by stars, the galactic inflows typically
dilute the metallicity of the galactic gas (\citealt{2017ASSL..430..221F}).  We refer to \cite{2012MNRAS.421...98D} for an analytical model where all the physical processes stated above are in equilibrium.  The time evolution of $M_\mathrm{CGM}$, the mass of the hot gas reservoir (CGM), is defined by

\begin{equation}
\dot{M}_{\rm CGM}= \dot{M}_{\rm CGM,in} + \dot{M}_{\rm g,out} - \dot{M}_\mathrm{g,in} - \dot{M}_{\rm CGM,out},
\label{eq_main_cgm}
\end{equation}
where $\dot{M}_{\rm CGM,in}$ is the inflow rate from the external medium into the CGM, and $\dot{M}_{\rm CGM,out}$ is the outflow rate from the CGM into the external medium. All terms presented in Equations~(\ref{eq_main_gal}) and (\ref{eq_main_cgm}) evolve as a function of time and are described in more details below (see also Figure~\ref{fig_sam_cartoon}).  

\subsubsection{Circumgalactic Inflows} \label{sect_cgm_in}

The CGM can increase its mass by accreting gas from the external medium.  Since we do not follow the evolution of this external
gas reservoir with \gammac, the mass gained by the CGM during a timestep is based on how much dark matter has been accreted.  Assuming 
a universal baryonic fraction, the mass accretion rate from the external medium to the CGM is defined by

\begin{equation}
\label{eq_cgm_in}
\dot{M}_{\rm CGM,in} = \dot{M}_\mathrm{DM}\left(\frac{\Omega_{\rm M,0}}{\Omega_{\rm b,0}}-1\right)^{-1},
\end{equation}
where the growth rate of the dark matter mass $\dot{M}_\mathrm{DM}$ is extracted from the merger tree and provided by \gammac.  The mass
accreted is assumed to have a primordial composition.   Note that this is not entirely representative of the hydrodynamic simulation, as
metals ejected outside the virial radius can be re-accreted at a later time and metals ejected by surrounding
galaxies can be introduced in the halo of the considered galaxy (see also \citealt{2017MNRAS.470.4698A}).

If the dark matter mass decreases during a timestep, we calculate the fraction of dark matter lost and use that
fraction to remove gas from the CGM in the same proportion.  We do not use Equation~(\ref{eq_cgm_in}) in this
case because the gas fraction in the CGM does not always reflect the universal baryonic ratio ($\Omega_{b,0}/\Omega_{\rm M,0}$).
Indeed, CGM outflows can reduce the gas fraction by expelling gas beyond the virial radius (see Section~\ref{sect_cgm_out}).

\subsubsection{Galactic Inflows} \label{sect_gal_in}

The rate at which the CGM gas is introduced inside the galactic gas is defined by

\begin{equation}
\label{eq_m_g_in}
\dot{M}_{\rm g,in} = \frac{M_\mathrm{CGM}}{\tau_\mathrm{inflow}},
\end{equation}
where $\tau_\mathrm{inflow}$ represents the inflow timescale.  We refer to \cite{2011MNRAS.416..660L} for
more details on the inflow prescriptions typically used in semi-analytic models. The time needed
for the hot gas to be transferred in the central galaxy depends on how fast gas can
cool (cooling timescale, $\tau_\mathrm{cool}$) and how fast gas can physically travel
from the CGM to the galaxy (free-fall timescale, $\tau_\mathrm{ff}$).  Those timescales
are defined as (\citealt{1991ApJ...379...52W})

\begin{equation}
\tau_\mathrm{cool}=\frac{3k_BT}{2n_e\Lambda(T,Z)},
\end{equation}
\begin{equation}
\tau_\mathrm{ff}=0.1H^{-1}_0(1+z)^{-3/2},
\label{eq_t_ff}
\end{equation}
where $n_e$ and $H_0$ are the electron number density and the current Hubble parameter, respectively.  The cooling function $\Lambda$ depends on the temperature $T$ and metallicity $Z$ of the gas reservoir, here the CGM.
Equation~(\ref{eq_t_ff}) actually refers to 10\,\% the Hubble time in an Einstein-de Sitter universe, which is a reasonable approximation for the dynamical timescale of a halo.  In our model, however, this equation is only used to provide a default setup for the redshift-dependent gas inflow timescale, which is modified by exploratory parameters as described below.

We used \texttt{Grackle}\footnote{\url{http://grackle.readthedocs.io/en/grackle-3.0/}} (\citealt{2017MNRAS.466.2217S}) to calculate the average cooling timescale of the CGM, assuming it is heated to $T_\mathrm{vir}$.  For halos forming the bulk of the final stellar mass in the W12 simulation, we found that $\tau_\mathrm{cool}$ is always shorter than $\tau_\mathrm{ff}$ because of the high gas density at high redshift.
For the purpose of this paper, we therefore do not use $\tau_\mathrm{cool}$ to calculate the inflow timescale and do not need to track the density and temperature of the CGM.  We define the inflow timescale of cooling gas as

\begin{equation}
\tau_\mathrm{inflow} = 0.1H^{-1}_0C_\tau (1+z)^{-3\gamma_\mathrm{ff}/2},
\end{equation}
\begin{equation}
\label{eq_inflow_norm}
C_\tau = 10H_0\tau_{\mathrm{inflow},0}(1+z_f)^{3\gamma_\mathrm{ff}/2},
\end{equation}
where $\tau_{\mathrm{inflow},0}$ is the gas inflow timescale of the final galaxy (the resulting most massive branch in the merger tree)
at the end of the simulation at redshift $z_f$.  When using \gammac, the value of $\tau_{\mathrm{inflow},0}$ is the same for all \omegap\ instances,
so that each building-block galaxies considered in the merger tree evolves following the same set of equations.
The $\gamma_\mathrm{ff}$ parameter has been introduced to explore different redshift dependencies, with the goal to provide
maximum flexibility to compare our SAM with hydrodynamic simulations.  When $\gamma_\mathrm{ff}=1$, the gas inflow timescale
scales linearly with the free-fall timescale. The impact of $\tau_{\mathrm{inflow},0}$ and $\gamma_\mathrm{ff}$ on the temporal
evolution of $\tau_\mathrm{inflow}$ and on the predicted SFH and MDF
is presented in Section~\ref{sect:comp_inflow} and shown in the second row of Figure~\ref{fig_param_1}.

\subsubsection{Star Formation} \label{sect_SF}

The star formation in \omegap\ depends linearly on the mass of the cold gas reservoir (e.g., \citealt{2001MNRAS.328..726S,2006RPPh...69.3101B}):

\begin{equation}
\label{eq_SFR}
\dot{M}_\star = \frac{\epsilon_\star}{\tau_\star}M_\mathrm{gas}=f_\star M_\mathrm{gas},
\end{equation}
where $\epsilon_\star$ is the dimensionless star formation efficiency (SFE) and $\tau_\star$ is the star formation timescale. We combine these two last parameter into a single SFE parameter ($f_\star$) which has
units of yr$^{-1}$.  Here we allow $f_\star$ to vary as a function of the dark matter halo mass of the galaxy and define the SFE as

\begin{equation}
\label{eq_sfe_yr}
f_\star = f_{\star,0}\left(\frac{M_\mathrm{DM}}{M_\mathrm{DM,0}}\right)^{\gamma_\star},
\end{equation} 
where $f_{\star,0}$ and $M_\mathrm{DM,0}$ are the SFE and dark matter mass
of the final galaxy at the end of the simulation. The former is a free parameter while the latter given by \gammac.  The $\gamma_\star$ parameter controls the power-law dependence of the star formation efficiency on the size of the dark matter halo, $M_\mathrm{DM}$, and should be considered to be an exploratory parameter.  As for $\gamma_\mathrm{ff}$ (see Section~\ref{sect_gal_in}), the value of $\gamma_\star$ is the same for all \omegap\ instances when using \gammac. The impact of $f_{\star,0}$ and $\gamma_\star$ on the temporal evolution of $f_\star$ and on the predicted SFH and MDF
is presented in Section~\ref{sect:comp_SFE} and shown in the first row of Figure~\ref{fig_param_1}.

\subsubsection{Galactic Outflows} \label{sect_gal_out}

In our model, galactic outflows (the mass ejected from the galaxy into the CGM) are driven
by the mechanical energy released by massive stars.  The outflow rate is therefore based
on the mass-loading factor defined by \citet{2005ApJ...618..569M}:

\begin{equation}
\label{eq_eta}
\eta_\mathrm{gal}=\frac{\dot{M}_\mathrm{g,out}}{\dot{M}_\star}.
\end{equation}
According to \cite{2005ApJ...618..569M},

\begin{equation}
\label{eq_eta_M05}
\eta_\mathrm{gal}\propto v_\mathrm{out}^{-\gamma_\eta},
\end{equation}
where $v_\mathrm{out}$ is the velocity of the gas contained in the outflow.
In \cite{2005ApJ...618..569M}, $\gamma_\eta=1$ for momentum-driven outflows
and $\gamma_\eta=2$ for energy-driven outflows.  We note that $\gamma_\eta$
can be larger than 2 for low-mass dwarf galaxies (\citealt{2015MNRAS.454.2691M}).
\cite{2005ApJ...621..227M} found that the outflow velocity of galaxies correlates with their rotation velocities,
which related to the velocity $V_\mathrm{vir}$ of the virialized
systems (see \citealt{2015ARA&A..53...51S}).  By replacing $v_\mathrm{out}$ by
$V_\mathrm{vir}$ in Equation~(\ref{eq_eta_M05}), and by substituting $V_\mathrm{vir}$ with
Equation~(\ref{eq_virr}), the mass-loading factor becomes dependent on $M_\mathrm{vir}$
and $R_\mathrm{vir}$.  Using the relation between $R_\mathrm{vir}$, $M_\mathrm{vir}$, and 
redshift $z$ found in \cite{1991ApJ...379...52W}, the mass-loading factor can be expressed as
follow (see derivation in \citealt{2017ApJ...835..128C})


\begin{equation}
\eta_\mathrm{gal}=C_\eta M^{-\gamma_\eta/3}_\mathrm{DM}(1+z)^{-\gamma_\eta/2},
\label{eq_eta_gal_main}
\end{equation}
\begin{equation}
\label{eq_C_eta}
C_\eta=\eta_{\mathrm{gal},0}M^{\gamma_\eta/3}_\mathrm{DM,0}(1+z_f)^{\gamma_\eta/2},
\end{equation}
where $\eta_{\mathrm{gal},0}$ is a free parameter representing the mass-loading factor of the final
galaxy at the end the simulation.

As with $\gamma_\mathrm{ff}$ (see Section~\ref{sect_gal_in}),
$\gamma_\eta$ is also set as a free parameter for which its value is the same for all \omegap\ instances.
We refer to \cite{2016ApJ...819...29B} for alternative approaches to model galactic outflows.
The impact of $\eta_{\mathrm{gal},0}$ and $\gamma_\eta$ on the temporal
evolution of $\eta_\mathrm{gal}$ and on the predicted SFH and MDF
is presented in Section~\ref{sect:comp_gmlf} and shown in the third row of Figure~\ref{fig_param_1}.
In this work, Equations~(\ref{eq_eta_gal_main}) and (\ref{eq_C_eta}) and their free parameters represent
an exploratory parametrization to allow flexibility to reproduce the high-redshift hydrodynamic simulation.
In future work we plan to use the latter simulation to derive analytical prescriptions to be used in SAMs,
as done in \cite{2012MNRAS.421.3522H} and \cite{2015MNRAS.454.2691M}.

At a given time $t_i$, associated with the $i^\mathrm{th}$ timestep, the mass of the galactic outflow
driven by the $i^\mathrm{th}$ SSP is defined by

\begin{equation}
\label{eq_Migout}
M^i_\mathrm{g,out}=\eta_\mathrm{gal}(t_i)\dot{M}_\star(t_i)\Delta t_i,
\end{equation}
where $\Delta t_i$ is the duration of the $i^\mathrm{th}$ timestep\footnote{As for {\texttt{SYGMA}} and \omega,
the timesteps in \omegap\ and \gammac\ can vary arbitrarily as a function of time. See \url{https://github.com/NuGrid/NuPyCEE/blob/master/DOC/Capabilities/Timesteps_size_management.ipynb}}.  This outflowing mass is then
distributed in time following the evolution of the mechanical luminosity
$L_\star(t)$ released by the stars of the $i^\mathrm{th}$ SSP.  In \omegap, $L_\star(t)$
is re-normalized such that 

\begin{equation}
\int_{0}^{\infty}L_\star(t)\mathrm{d}t = 1.
\end{equation}
Integrating $L_\star(t)$ over a certain time interval thus represents the fraction
of the total outflowing mass (see Equation~\ref{eq_Migout}) lost at each timestep.
In our code, $L_\star(t)$ can be of any form.  However, in this paper, we use a
constant luminosity from 4 to 20\,Myr in order to be consistent with the SNe feedback
implemented in \cite{2012MNRAS.427..311W}.  Below 4\,Myr and above 20\,Myr, $L_\star(t)$ is set
to zero.

At any time $t$, the total mass ejected from the galaxy over a time $\Delta t$ is obtained by summing
the contribution of all SSPs,

\begin{equation}
M_\mathrm{g,out}(t) = \sum_i^{N_\mathrm{SSP}} M^i_\mathrm{g,out}\int_{t-t_i}^{t-t_i+\Delta t}L_\star(t)\mathrm{d}t.
\label{eq_delayed_outflow}
\end{equation}
In this last equation, $t-t_i$ represents the age of the $i^\mathrm{th}$ SSP. The summation assumes a
cumulative process in the sense that twice as much energy will lead to twice as much outflowing gas.  In other
words, no nonlinear interaction is assumed between the different SSPs that contribute to an outflow.

\subsubsection{Circumgalactic Outflows} \label{sect_cgm_out}

The mass introduced into the CGM by galactic outflows can generate an excess
of energy.  When this happens, a fraction of the hot gas contained in the CGM can
be expelled outside the host dark matter halo (e.g., \citealt{2006MNRAS.365...11C}).
Because this feedback process is driven by galactic outflows, which are driven
by stellar feedback, we use the mass-loading factor definition to describe the CGM
outflow rate, as in equation~(\ref{eq_eta}),

\begin{equation}
\eta_\mathrm{CGM}=\frac{\dot{M}_\mathrm{CGM,out}}{\dot{M}_\star}.
\end{equation}
We then assume that the CGM outflow rate is proportional to the galactic outflow rate
through the free parameter $f_\eta$.  The CGM outflow rate is then
given by

\begin{equation}
\label{eq_m_cgm_out}
\dot{M}_\mathrm{CGM,out} = f_\eta\eta_\mathrm{gal}\dot{M}_\star.
\end{equation}
In the current version of \omegap, when a CGM outflow is generated, the mass lost
is not kept in a separate reservoir to be reincorporated later on in the virialized system -- rather, it is assumed that the cooling time of this gas is long enough that it is greater than the current Hubble time in the model, and thus will not return to the halo on a time scale that is relevant to the calculation.
The impact of $f_\eta$ on the predicted SFH and MDF
is presented in Section~\ref{sect:comp_cmlf} and shown in the last row of Figure~\ref{fig_param_1}.

\subsection{Stellar Yields} \label{sect_yields}

Because of their connection with the JINA-NuGrid chemical evolution pipeline (\citealt{2017nuco.confb0203C}),
\omegap\ and \gammac\ automatically have access to the complete NuPyCEE\footnote{\url{http://github.com/NuGrid/NuPyCEE}} stellar yields 
library.  In practice, a simulation with \gammac\ can include as many elements as desired, including
the 280 isotopes available with NuGrid yields (\citealt{2016ApJS..225...24P,2017arXiv170908677R}).  In addition,
an arbitrary number of enrichment sources can be included such as compact binary mergers, Type~Ia 
supernovae, neutrino-driven winds in core-collapse SNe, and any other sources that can be modelled
using a set of yields and a delay-time distribution function\footnote{\url{https://github.com/NuGrid/NuPyCEE/blob/master/DOC/Capabilities/Delayed_extra_sources.ipynb}}.

However, in this work, because we want to be consistent with the W12 simulation, we do
not follow individual species -- we only follow the total mass of metals. Throughout this paper,
each SSP will eject 25\,\% of its original mass via SNe.  As in W12, the metallicity
in mass fraction of all stellar ejecta is $Z=0.02$.

%% file: input_parameter_impact.tex
\begin{figure*}
\includegraphics[width=7.in]{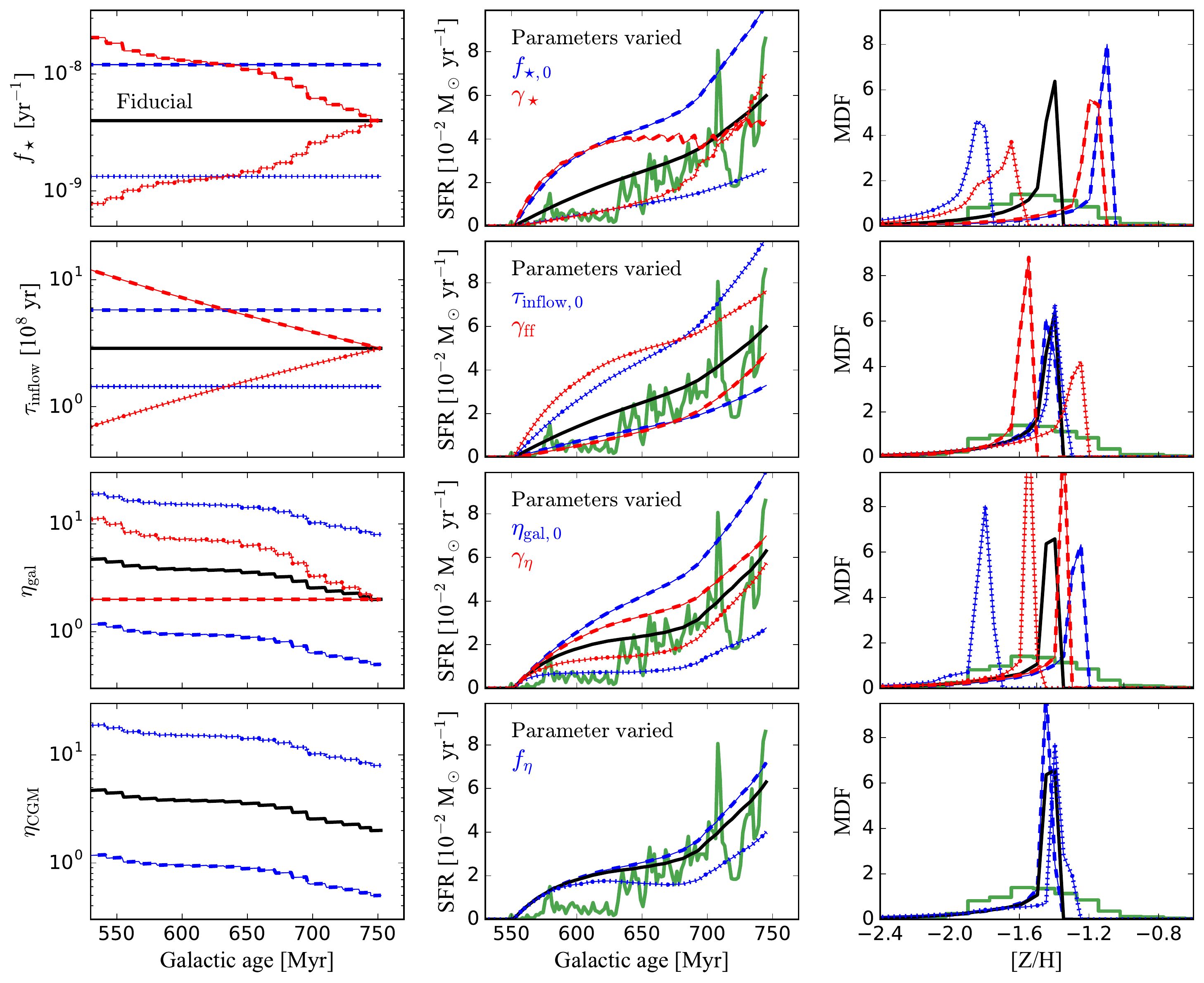}
\caption{Impact of our input parameters (left column) on the predicted
star formation history (middle column) and metallicity distribution
function (right column) using \gammac\ with the target galaxy's merger
tree.  The green lines in the middle and right columns display the
physical values extracted from the hydrodynamic simulation.  In each
row, only one key physical quantity has been modified (left panel),
and the different line styles are used to keep track of each case.
The fiducial case is shown as the black solid line.
The physical quantities shown in the three first top rows are
controlled by two parameters, and their individual impact is
highlighted using different colors (see Table~\ref{tab_param_fig5}).
\textbf{Top row:} Star formation efficiency ($f_\star$,
Equation~\ref{eq_sfe_yr}).  Variation of $f_{*,0}$ is shown by the
blue lines, and $\gamma_*$ by the red lines.  \textbf{Second row:} Gas inflow
timescale ({\small $\tau_\mathrm{inflow}$}, Equation~\ref{eq_inflow_norm}) used
to calculate the rate at which the hot gas reservoir (CGM) is
introduced in the star-forming region.  Variation of
{\small $\tau_{\mathrm{inflow},0}$} is represented by the blue lines, and
$\gamma_\mathrm{ff}$ by the red lines.  \textbf{Third row:} Mass-loading factor of the
galaxy ({\small $\eta_\mathrm{gal}$}, Equation~\ref{eq_C_eta}) representing the
mass transfer of gas from the star-forming region to the hot gas
reservoir.  Variation of {\small $\eta_{\mathrm{gal},0}$} is shown by the blue
lines, and $\gamma_\eta$ by the red lines.  \textbf{Bottom row:} Mass-loading
factor of the hot gas reservoir ({\small $\eta_\mathrm{CGM}$},
Equation~\ref{eq_m_cgm_out}) representing the mass transfer of gas
from the hot gas reservoir to the external medium (outside the virial
radius). We note that none of the metallicity distribution functions produced
in this parameter exploration match the broad distribution extracted from the
simulations. We explain the reason for this disagreement in Section
\ref{sec:numixing}.
}
\label{fig_param_1}
\end{figure*}

In this section we combine \gammac\ with the merger tree of the most
massive galaxy found in the W12 calculation and compare its
predictions with the equivalent quantities extracted from the
hydrodynamic simulations.  The goal of this section is to show
variations of the model parameters impact our predictions, rather than
tuning \gammac\ to best match the results obtained by the W12
calculation (which is shown in Section~\ref{sec:degeneracy}).  In the
following subsections we discuss Figure~\ref{fig_param_1}, which
focuses on variations in fundamental physical quantities such as the
star formation efficiency and the gas inflow timescale rather than on
specific input parameters. This is because different parameterizations
can be used in different studies, while fundamental processes such as
star formation and gas circulation are inherent to most chemical
evolution and semi-analytic models.  The input parameters used in this
section to control those physical quantities are summarized in
Table~\ref{tab_param_fig5}.

When interpreting the following subsections, it is important to note
that the predictions shown in the top two and bottom two rows of
Figure~\ref{fig_param_1} assumed different fiducial values for the
parameters (these are also indicated in Table~\ref{tab_param_fig5}).
This is to motivated by two reasons. First, to facilitate the building of intuition, since the impact of
different parameters are best visualized in different \textit{galaxy
evolution regimes}.  And second, we want to highlight that, under certain
circumstances, different physical quantities can in principle modify the predicted
SFR and MDF in a similar way.

To show the impact of the star
formation efficiency (SFE), we adopt weak galactic outflows to
minimize their impact on the predictions.  Otherwise, when
increasing the SFE and reducing the mass of galactic gas, outflows
would start to significantly alter the star formation history by
generating episodic behaviours, which are less trivial to interpret in
the context of this section (but see Section~\ref{sec:SF_Z} and
Figure~\ref{fig_degeneracy}).  With stronger outflows, previous studies shown that the SFE has little
impact on the SFR of galaxies due to a balance between star formation and feedback
(e.g., \citealt{2010MNRAS.402.1536S,2011MNRAS.417..950H,2011MNRAS.416.1566L}).
Our target galaxy, however, only efficiently formed stars for $\sim$\,200\,Myr and did 
not reach an equilibrium state by the end of the simulation. The large variations shown
in this section caused by the SFE should not be taken as a general case. To show the impact
of galactic and CGM outflows (bottom two rows of
Figure~\ref{fig_param_1}), they need to be strong.  Otherwise, if
they are too weak relative to the amount of gas available, not enough
gas would be removed from the galaxy and the CGM to significantly
alter the star formation history and the metallicity distribution
function.


Some quantities seen in Figure~\ref{fig_param_1} show discrete steps in 
their temporal profiles.  This is because in the most massive galaxy found
in the simulation, there is typically at least one merger with a non-star-forming
halo between data outputs.  This means we do not always have access
to the time-evolution of the properties of halos (e.g., $M_\mathrm{vir}$)
within a brach of the merger tree.  When this occurs, we input constant
halo properties in \omegap, which in turn fixes parameters that depend
on $M_\mathrm{vir}$ (e.g., $f_\star$, $\eta_\mathrm{gal}$) to a constant value.

\subsection{Star Formation Efficiency} \label{sect:comp_SFE}

The top row of Figure~\ref{fig_param_1} shows the impact of the star
formation efficiency (SFE; $f_\star$) adopted in \gammac\ (left panel)
on the predicted star formation history (SFH; middle panel) and
metallicity distribution function (MDF; right panel).  The magnitude
of the SFE is controlled by the $f_{\star,0}$ parameter
(Equation~\ref{eq_sfe_yr}).  A higher SFE generates higher SFRs and
increases the total stellar mass formed by the end of the simulation
(see blue lines).  Overall, the shape of the temporal profile of the
SFH stays similar when the SFE is scaled up and down by a constant
factor.  The main role of the SFE in a chemical evolution code is to
modify the stellar-to-gas mass ratio in the galactic component.  By
increasing the SFE, the metals ejected by stars are thus deposited in
greater concentration relative to the mass of gas and the peak of the
MDF is pushed to higher metallicities, and vice-versa.

\begin{deluxetable*}{lllll}
\tablewidth{0pc}
\tablecaption{Input parameters used in \gammac\ to produce Figure~\ref{fig_param_1}. \label{tab_param_fig5}}
\tablehead{ \colhead{\multirow{2}{*}{Parameter}} & \colhead{\multirow{2}{*}{Description}} & \colhead{Values in top} & \colhead{Values in bottom} & \colhead{Modified in row} \\
\colhead{} & \colhead{} & \colhead{rows ($f_\star$, $\tau_\mathrm{inflow}$)} & \colhead{rows ($\eta_\mathrm{gal}$, $\eta_\mathrm{CGM}$)} & \colhead{(from top to bottom)} }
\startdata
$f_{\star,0}$ & Final SFE value at $z_f$ [10$^{-9}$\,yr$^{-1}$] & (1.3, \textbf{4.0}, 12.0) & \textbf{6.0} & \multirow{2}{*}{1 ($f_\star$, Equation~\ref{eq_sfe_yr})} \\
$\gamma_\star$ & DM mass dependency for the SFE & ($-1.0$, \textbf{0.0}, 1.0) & \textbf{0.0} & \\
$\tau_{\mathrm{inflow},0}$ & Final inflow timescale at $z_f$ [$\tau_{\mathrm{ff},0}$] & (2.5, \textbf{5.0}, 10.0) & \textbf{3.5} & \multirow{2}{*}{2 ($\tau_\mathrm{inflow}$, Equation~\ref{eq_inflow_norm})} \\
$\gamma_\mathrm{ff}$ & Redshift dependency for the inflow timescale & ($-4.0$, \textbf{0.0}, 4.0) & \textbf{0} &  \\
$\eta_{\mathrm{gal},0}$ & Final galactic mass-loading factor  at $z_f$ & \textbf{1.0} & (0.5, \textbf{2.0}, 8.0) & \multirow{2}{*}{3 ($\eta_\mathrm{gal}$, Equation~\ref{eq_C_eta})} \\
$\gamma_\eta$ & Galactic mass-loading power-law index & \textbf{0.0} & (0.0, \textbf{2.0}, 4.0) & \\
$f_\eta$ & CGM mass-loading factor scaling [$\eta_\mathrm{gal}$] & \textbf{0.0} & (0.25, \textbf{1.0}, 4.0) & 4 ($\eta_\mathrm{CGM}$, Equation~\ref{eq_m_cgm_out}) \\
\enddata
\tablecomments{We used different sets of input parameter
values for the top and bottom panels of Figure~\ref{fig_param_1} to
better highlight the impact of each parameter.  The fiducial values are shown in
boldface while the ones used when a parameter is varied are shown in
parenthesis. SFE stands for star formation efficiency, DM for dark matter, $z_f$ for final
redshift, and CGM for circumgalactic medium.
}
\end{deluxetable*}

The shape of the SFH is affected by the shape of the temporal profile
of the SFE (see red lines).  The time-dependence of this latter
quantity originates from its relation with the dark matter halo mass
of the host galaxy.  This relation is controlled by the $\gamma_\star$
parameter in the form of a power law (Equation~\ref{eq_sfe_yr}).  A
SFE that decreases with time tends to form more stars at early time
and less stars at later time compared to a constant SFE.  In our case,
this generates an initial star formation burst followed by a
relatively flat SFR (red dashed line).  On the other hand, a SFE that
increases with time tends to generate an ever-increasing SFR (red
dotted line) since gas gets increasingly efficient in turning into
stars.

\subsection{Inflow Timescale} \label{sect:comp_inflow}

The second row of Figure~\ref{fig_param_1} shows the impact of the
inflow timescale $\tau_\mathrm{inflow}$, which sets the gas transfer
rate from the CGM into the galactic component
(Equation~\ref{eq_m_g_in}).  The magnitude of the timescale is set by
the $\tau_{\mathrm{inflow},0}$ parameter, while the slope of its
temporal profile is controlled by $\gamma_\mathrm{ff}$
(Equation~\ref{eq_inflow_norm}).  A lower $\tau_\mathrm{inflow}$
increases the SFR of the galaxy because the CGM gas is introduced more
rapidly in the galactic component to fuel the star formation (see blue
lines), and vice-versa.  The MDF is similar from one run to another
because the shape of the inflow timescale and the SFE value were kept
unchanged.  Although the total mass of stars and gas scale with
$\tau_\mathrm{inflow}$, the stellar-to-gas mass ratio and thus the
metal concentration remain the same (see Section~\ref{sect:comp_SFE}).

As for the SFE, modifying the slope of the inflow timescale modifies
the shape of the SFH (red lines).  Adopting an increasing trend 
as a function of time for $\tau_\mathrm{inflow}$
will tend to form more stars at early times and less stars at later
times in comparison to a constant $\tau_\mathrm{inflow}$, and
vice-versa.  The MDF is also affected by the shape of
$\tau_\mathrm{inflow}$, since it modifies the final mass of gas
present in the galactic component at the end of the simulation.  For
example, for two galaxies that will form in total the same amount of
stars (blue and red dashed lines), the galaxy with a decreasing
$\tau_\mathrm{inflow}$ will have to form more stars at later time
compared to the one with a constant $\tau_\mathrm{inflow}$, in order
to compensate for the lower SFR at early time.  This late star
formation enhancement requires a large amount of inflowing gas.  As a
result, the galaxy with a decreasing $\tau_\mathrm{inflow}$ as a function of time will end
up with more gas in the galactic component compared to the galaxy with
a constant $\tau_\mathrm{inflow}$, which dilutes the metallicity and
pushes the peak of the MDF to lower metallicities.

\subsection{Galactic Mass-Loading Factor} \label{sect:comp_gmlf}

The third row of Figure~\ref{fig_param_1} shows the impact of the
galactic mass-loading factor, $\eta_\mathrm{gal}$, which sets the mass
transfer from the galaxy to the CGM via galactic outflows
(Equation~\ref{eq_eta}).  This quantity scales with
$\eta_{\mathrm{gal},0}$ while its temporal profile is defined by
$\gamma_\eta$ (Equation~\ref{eq_C_eta}). When $\eta_\mathrm{gal}$ is
increased by a constant factor, the SFR is reduced because more gas is
ejected from the galaxy (blue lines).  Recall that the star formation
history is self-generated with \omegap. With \omega, the SFH is an
input parameter and galactic outflows are balanced by galactic inflows
in order to sustain the desired SFH
(\citealt{2017ApJ...835..128C}). With \omegap, inflows and outflows
are not directly connected and the SFH can thus be modified by
$\eta_\mathrm{gal}$.

Galactic outflows are efficient mechanisms to remove metals from
galaxies.  A stronger outflow will systematically shift the peak of
the MDF to lower metallicities, and vice-versa (see also
\citealt{2017ApJ...835..224A}).  Modifying the shape of
$\eta_\mathrm{gal}$ affects the shape of the SFH because it changes
the way the mass of gas inside the galaxy is evolving with time.

\subsection{Circumgalactic Mass-Loading Factor} \label{sect:comp_cmlf}

The bottom row of Figure~\ref{fig_param_1} shows the impact of the CGM
mass-loading factor, $\eta_\mathrm{CGM}$, which sets the gas transfer
rate from the CGM to the external medium (which one could think of as
the truly intergalactic medium).  It is directly related to
$\eta_\mathrm{gal}$ via the proportionality constant $f_\eta$
(Equation~\ref{eq_m_cgm_out}).  Because the galactic inflow rate
depends on the mass of the CGM (Equation~\ref{eq_m_g_in}), the
strength of a CGM outflow regulates the mass of the galactic gas and
thus the star formation history.  In Figure~\ref{fig_param_1}, the SFH
eventually saturates when $\eta_\mathrm{CGM}$ is decreased (blue
dashed line) since the CGM outflow rate becomes negligible relative to
the total mass of the CGM.

In terms of the scaling of the inflow timescale, changing the CGM
outflow rate by a constant factor does not significantly shift the
peak value of the MDF.  This is because the stellar-to-gas mass ratio
in the galactic component does not depend on the amount of inflowing
gas, as long as the temporal profile (not the magnitude) of the
inflows and star formation efficiency are kept unchanged (see
Section~\ref{sect:comp_inflow}).

%% file: non_uniform_mixing.tex
As seen in Figure~\ref{fig_param_1}, no combinations of input parameters can generate a MDF with \gammac\ that is as broad as is seen in the hydrodynamic simulation.  This is because each building-block galaxy in \gammac\ is represented by a uniformly-mixed one-zone model (\omega).  This implies that the star-forming gas has only one metallicity value per timestep, which can be seen as the mass-weighted mean metallicity of what should be a non-uniformly-mixed medium. In \gammac, all stars formed during a timestep in a given branch therefore have the same initial metallicity, as opposed to in hydrodynamic simulations that show star formation in a range of metallicities at a given point in time in the target galaxy and its largest satellite.  Uniformly-mixed models tend to create a narrow MDF with a sharp break at its high-metallicity end when a galaxy is in an active accretion phase (\citealt{2017ApJ...837..183W}), which is the case here.

Figure~\ref{fig_hydro_age_Z} shows that the dispersion of metallicity in the W12 simulation is stochastic and depends on time and on the characteristics of the particular progenitor galaxy.  If we ignore the largest dispersions caused by triggered star formation, as shown in Section~\ref{sec:gal_behavior} (a reasonable approximation, because very little stellar mass is produced in these events), the dispersion in the main galaxy increases as a function of time (green dots) while it decreases in the smaller satellite galaxy (pink dots). We recall that the MDFs of metal-poor stars observed today in real galaxies represent the combination of several low-mass galaxies that likely had different enrichment histories.  

\begin{figure}
\center
\includegraphics[width=3.3in]{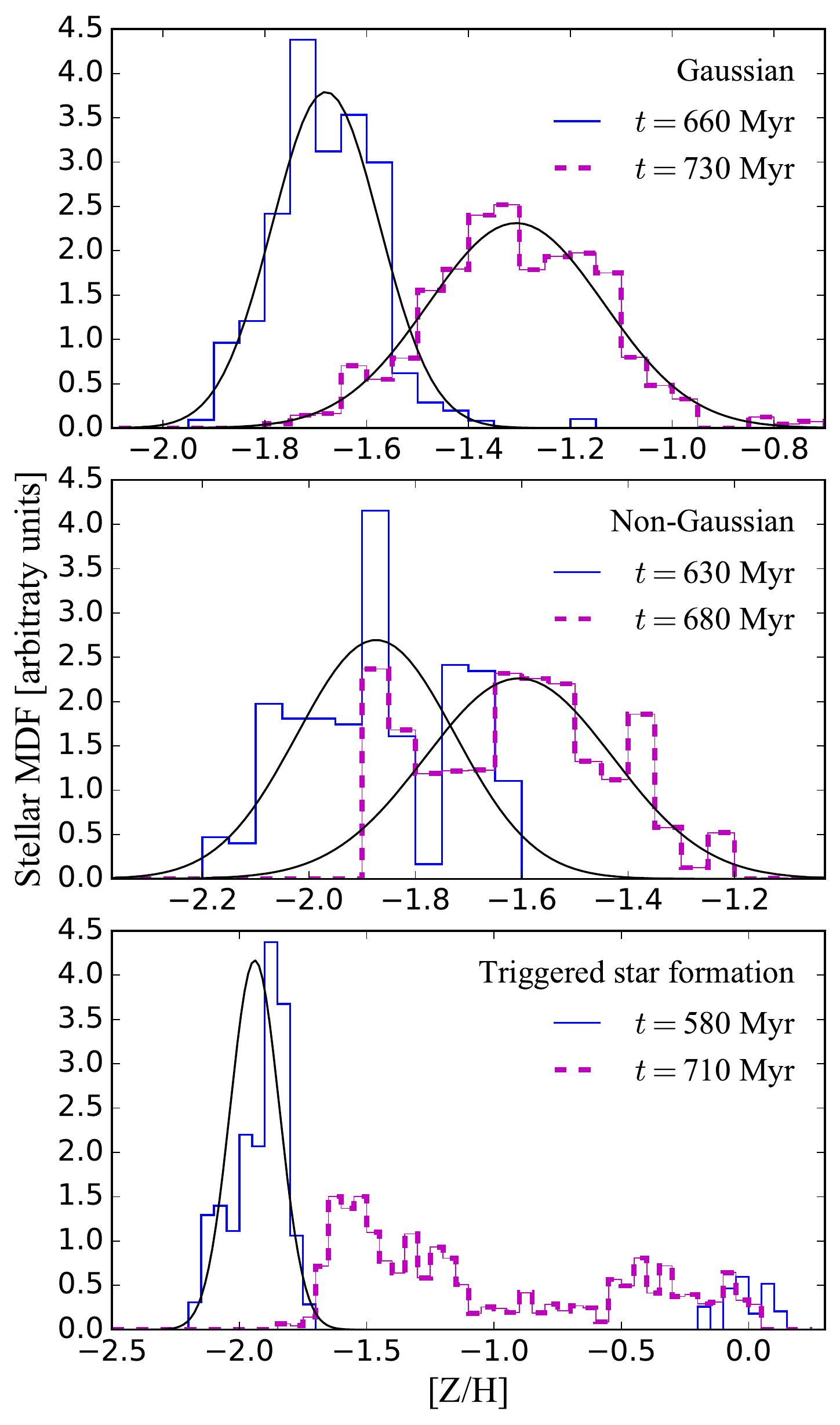}
\caption{Examples of metallicity distribution functions (histograms) of all stars formed during a certain time interval in the target galaxy.  The middle value of the time intervals, which has a range of 10\,Myr in all cases, is labelled in each panel.  The solid black curves represent Gaussian fits. There is no GAMMA prediction in this figure. \textbf{Top panel:} Distributions that are well described by a Gaussian function. \textbf{Middle panel:} Distributions that are not well described by a Gaussian function. This could be due, however, to insufficient sampling. \textbf{Bottom panel:} Distributions that include a triggered star-formation process suddenly generating high-metallicity stars compared to the average gas metallicity (see Figure~\ref{fig_hydro_age_Z}).}
\label{fig_MDF_t}
\end{figure}

The histograms in Figure~\ref{fig_MDF_t} show the stellar metallicity distributions taken at different times, as indicated on the figure, for the main galaxy.  Some distributions can be represented by Gaussian functions with different standard deviations (upper panel). This justifies the process of convolving the MDF with a Gaussian function in order to reproduce observations with simple models (e.g., \citealt{2003PASA...20..189F,2012nuco.confE.227P,2016MNRAS.460.2238V,2016MNRAS.463.3755C}). However, at earlier times, the metallicity distributions have more complex shapes that are not well reproduced by individual Gaussians (middle panel).  This deviation from a Gaussian distribution could, however, be caused by insufficient sampling due to the low number of stellar populations formed.  When triggered star formation events occur, the contribution of distinct stellar populations can be seen in the metallicity dispersion profile (bottom panel).

\begin{figure}
\center
\includegraphics[width=3.3in]{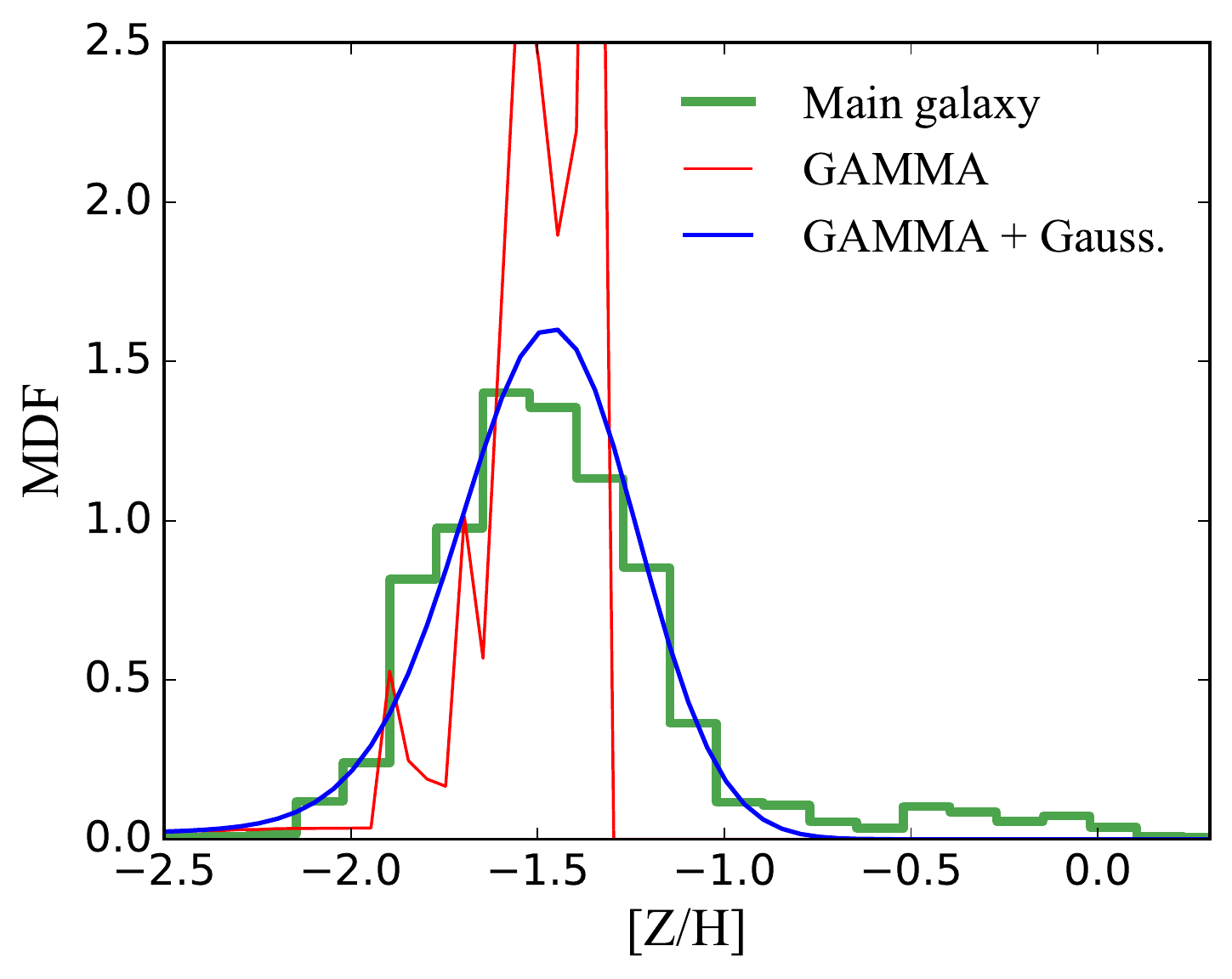}
\caption{Metallicity distribution function (MDF) predicted by \gammac\ compared to the one extracted from the hydrodynamic simulation (\blue{W12}, green line) for the main galaxy.  The red line represents
the raw MDF while the blue line shows the result of convoluting this MDF with Gaussian functions having a standard deviation of 0.2 (see Section~\ref{sec:numixing}). The long high-metallicity
tail seen in the \blue{W12} simulation originates from triggered star-formation events (see Figure~\ref{fig_hydro_age_Z}). The \gammac\ model presented in this figure corresponds to the blue
model in Figure~\ref{fig_degeneracy} (see Section~\ref{sec:degeneracy}).}
\label{fig_MDF_gaussian}
\end{figure}

To induce spread in the MDF predicted by \gammac, we post-process our results by convolving the MDF of all building-block galaxies by a Gaussian function with a standard deviation of 0.2, a value extracted from the hydrodynamic simulation at late times.  As seen in Figure~\ref{fig_MDF_gaussian}, this convolution process significantly improves the agreement between the semi-analytic and hydrodynamic approaches at the final redshift.  Although more cases need to be investigated before deriving a general prescription for SAMs, our results suggests that non-uniform mixing can be captured by simple models in a post-processing manner.  However, when modeling the chemical evolution of different elements, the situation becomes more complex because different abundance ratios detected in metal-poor stars have different dispersions depending on the targeted elements (e.g., \citealt{2013ApJ...778...56C,2014AJ....147..136R,2015ARA&A..53..631F,2017MNRAS.466.2474H}).  Deriving analytic prescriptions for the non-uniform mixing of individual elements is not possible with the W12 simulation, since only the overall metallicity was tracked rather than individual species abundances.

In spite of our ability to reproduce the MDF of the hydrodynamic simulation at the final redshift, we cannot capture the triggered star formation events predicted by this simulation. These stochastic events create the long high-metallicity tail in the MDF (see Figure~\ref{fig_hydro_age_Z}), which is absent in the MDF predicted by \gammac\ (see Figure~\ref{fig_MDF_gaussian}).

%% file: reproducing_hydro_sims.tex
In this section we apply the non-uniform mixing prescription
constructed in Section~\ref{sec:numixing} and attempt to reproduce the
\textit{observable quantities} from the W12 simulation that are
presented in Section~\ref{sec:gal_behavior}. The ``best fit'' models
shown in this section have been found by tuning input parameters by
hand.  This is not a statistically rigorous approach, and is intended
to be the first step of a long-term model calibration and optimization
process that will be improved by using statistical tools such as
Markov Chain Monte Carlo calculations, which are already part of our
chemical evolution pipeline (see \citealt{2017nuco.confb0203C}).

Figure~\ref{fig_degeneracy} shows a comparison between the W12
simulation and \gammac, with panels $a$\,-\,$c$ showing three particular
parameterizations of $f_\star$, $\tau_\mathrm{inflow}$, and $\eta_\mathrm{gal}$ that are
quite dissimilar from each other (see also Table~\ref{tab_param_best}).
However, all those parameterizations do a reasonable job
of reproducing the observable properties of the target halo (panels $d$\,-\,$f$). 
We examine these in more detail in the following subsections.

\subsection{Star Formation and Metallicity} \label{sec:SF_Z}

\begin{figure*}
\includegraphics[width=7.05in]{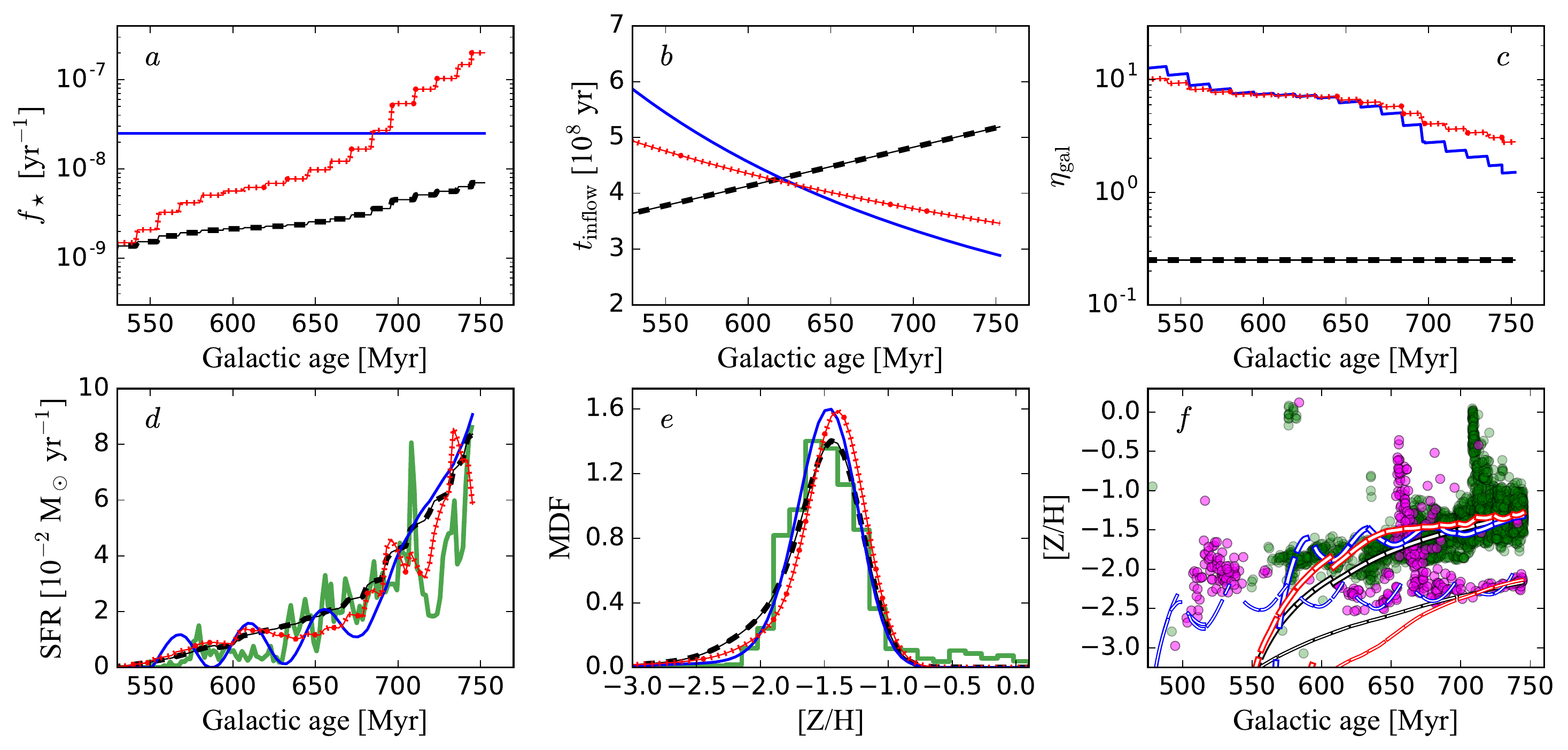}
\caption{Comparison between the hydrodynamic simulation of
  \protect\cite{2012MNRAS.427..311W} and the semi-analytic code
  \gammac\ using the merger trees extracted from that simulation.
  \textbf{Top row:} The solid blue, dashed black, and red dotted lines
  represent three different \gammac\ models using different
  assumptions for the evolution of the star formation efficiency
  (panel~$a$), the gas inflow timescale (panel~$b$), and the galactic
  mass-loading factor (panel~$c$), as presented in Table~\ref{tab_param_best}.  These three panels only show
  the predictions for the target galaxy, and do not explicitly target
  its satellite.  \textbf{Bottom row:} Panels
  $d$ and $e$ show star formation history and metallicity distribution
  function (MDF) predicted by \gammac\ compared to the ones predicted
  by the hydrodynamic simulation (green thick line), for the target
  galaxy.  Panel $f$ shows the age-metallicity relationship.  For
  \gammac, each color is associated to a specific model as in the
  other panels but the target and satellite galaxies are represented by
  thick and thin line segments (see Section~\ref{sect:age_Z}),
  respectively. For the hydrodynamic simulation, the target and
  satellite galaxies are represented by green and pink dots,
  respectively.}
\label{fig_degeneracy}
\end{figure*}

Panels $d$ and $e$ of Figure~\ref{fig_degeneracy} shows the star
formation history and metallicity distribution functions predicted by
\gammac\ using three the different sets of input parameters
presented in Table~\ref{tab_param_best} (blue, red, and black lines).  
We cannot reproduce the bursty nature
and the relatively small oscillation periods of the star formation
history, but our three models are still able to reproduce the global
trend of the hydrodynamic simulation (green thick line in each panel).
The oscillating behaviour in the SFR our blue model prior $\sim$\,700\,Myr
is discussed in Section~\ref{sect:disc_limitations_gamma}.
The MDFs predicted by our models are very similar to each other,
although the adopted SFE and gas circulation processes are
substantially different from one model to another (panels $a$, $b$,
and $c$).  This is because the input parameters can modify the MDF in
a similar manner (see Section~\ref{sec:comparison}). It is therefore
possible to vary those parameters in such as way that the impact of
their variations is cancelled for these particular observables.  This
results is complementary to the work of \citealt{2017ApJ...835..128C}
who showed that only fitting the stellar abundance ratios of a real
galaxy is insufficient to understand its evolution, since those
observations can numerically be reproduced in multiple ways (i.e.,
there is significant degeneracy between these model parameters).

\subsection{Age-Metallicity Relationship} \label{sect:age_Z}

Panel $f$ of Figure~\ref{fig_degeneracy} shows the age-metallicity
relationship predicted by \gammac\ for the main galaxy (thick lines)
and for the satellite galaxy (thin lines) against the W12 data (green
and pink dots).  Each line segment represents a building-block galaxy
in the merger tree (or a branch).  The segments tend to align and form
continuous lines because most of the mergers are minor and involve
low-mass halos that do not form stars. Although the merger tree of the
main galaxy is complex (see Figure~\ref{fig_merger_tree}), this galaxy
represents a relatively simple case because star formation
predominantly occurs in the most massive progenitors at $z \lesssim 9$
(see Figure~\ref{fig_hydro_SFH}).

To match the satellite galaxy with each \gammac\ model, we used the
same input parameters as for the main galaxy (panels $a$, $b$, and $c$
of Figure~\ref{fig_degeneracy}), but we scaled down the magnitude of
the star formation efficiency by a factor of 5.  The slight
differences seen in the low-metallicity end of the main galaxy's MDFs
predicted by our models (panel $e$) are more visible when looking at
the age-metallicity relationship (panel $f$).  Although the models are
consistent with each others at late time ($t>700$\,Myr), they are
divergent at earlier time.  For the main galaxy, the black model best fits
the W12 data between 650 and 700\,Myr, while only the blue model is
able reproduce the stellar metallicity at $\sim575$\,Myr.  For the
satellite galaxy, only the blue model can match the W12 data between
600 and 700\,Myr.  Its predictions at $\sim525$\,Myr are too low
compared to W12, but the other models are at least an order of
magnitude lower.

\begin{deluxetable}{ccccc}
\tablewidth{0pc}
\tablecaption{Input parameters used for the three \gammac\ models shown Figure~\ref{fig_degeneracy}. \label{tab_param_best}}
\tablehead{ \colhead{\multirow{2}{*}{Parameter}} &  \multicolumn{3}{c}{Model} & \colhead{\multirow{2}{*}{Units}} \\ 
                   \colhead{} & \colhead{Blue} & \colhead{Red} & \colhead{Black} & \colhead{}}
\startdata
$f_{\star,0}$ & 2.5 & 20.0 & 0.7 &10$^{-8}$\,yr$^{-1}$ \\
$\gamma_\star$ & 0.0 & 3.0 & 1.0 & -- \\
$\tau_{\mathrm{inflow},0}$ & 5.0 & 6.0 & 9.0 & $\tau_\mathrm{ff}$ \\
$\gamma_\mathrm{ff}$ & -2.0 & -1.0 & 1.0 & -- \\
$\eta_{\mathrm{gal},0}$ & 1.5 & 2.8 & 0.25 & --  \\
$\gamma_\eta$ & 5.0 & 3.0 & 0.0 & -- \\
$f_\eta$ & 0.0 & 0.0 & 0.0 & $\eta_\mathrm{gal}$ \\
\enddata
\end{deluxetable}

Overall, the blue model is our current best-fit
model because of its fast early enrichment that allows to reproduce
(although not perfectly) the age-metallicity relationship at all time, for both the
main and the satellite galaxies.
However, our predictions for the early stages of the age-metallicity relationship
is likely affected by the current lack of pre-enrichment from Population III stars in \gammac,
which is present in W12 (see Section~\ref{sect:disc_limitations_gamma}).

\subsection{CGM Outflows} \label{sec:tot_mass}
Figure~\ref{fig_m_tot} presents the total mass of gas and the total mass of metals
found within the virial radius of the main galaxy as a function of time.  As shown in
Table~\ref{tab_param_best}, our best-fit \gammac\ models do not include large-scale
CGM outflows ($f_\eta=0.0$), meaning that all the gas is retained inside
the virial radius.  According to Figure~\ref{fig_m_tot}, this assumption generates a 
good match with the quantity extracted from the hydrodynamic simulation (top thick
green line) for the entire period of active star formation ($t\gtrsim600$\,Myr).
This high baryonic retention, however, is likely to change if the simulation were to continue
to lower redshifts ($z<7$).  Indeed, cosmological zoom-in simulations
have shown that low-mass galaxies lose a significant fraction of their
baryon by $z=0$ (e.g., \citealt{2015MNRAS.454.2691M}).

In Figure~\ref{fig_m_tot}, variations can be seen between the simulation and the \gammac\ models in between
400 and 600\,Myr of galactic evolution.  At 400\,Myr, in the simulation,
a Pop~III explodes and triggers a short burst of star formation which generates
a CGM outflow while the main galaxy is $\sim$\,50 times less massive than
at the end of the simulation.  This event is not considered in our \gammac\ models,
which is why the total mass inside the halo in the simulation tends to be lower than
our models during that period of time (see upper lines in Figure~\ref{fig_m_tot}). 
The metal enrichment caused by this event can be seen as a bump in the mass of
metals at $\sim$\,420\,Myr (bottom thick green line).  After this time, the mass
of metals decreases because mass is lost outside the virial radius.  We note 
that the stellar mass formed during this event is negligible compared to the
total mass formed during the active star formation period (see Figure~\ref{fig_hydro_SFH}).

Although a GCM outflow is generated at early time, the mass lost eventually
falls back and returns into the galactic halo after $\sim$\,600\,Myr of evolution.
This early starburst event provides a metallicity floor of [Z/H]\,$\sim$\,$-3$ at
the onset of the active star formation period where our \gammac\
models start to form stars (see discussion in Section~\ref{sect:disc_limitations_gamma}).
This could affect the metal-poor end of the MDF predicted by our models (see
panel~$e$ of Figure~\ref{fig_degeneracy}).  The slight increase in the mass of metals
seen in the simulation at $\sim$\,700\,Myr compared to our model predictions is due
to the infall of metal-rich gas accompanying the most massive satellite galaxy (see
pink circle in Figure~\ref{fig_video}). We recall that since this satellite did not merge
before the end of the simulation, it is not include in the merger tree of the main galaxy
and it's impact on the mass of metals is thus not accounted in the \gammac\ models
for the main galaxy.

%% file: discussion.tex
In this section we discuss the limitations and uncertainties of this work, highlight
the advantages of our approach, and describe the next steps required 
in this research to better constrain our semi-analytic model, \gammac.

\begin{figure}
\center
\includegraphics[width=3.3in]{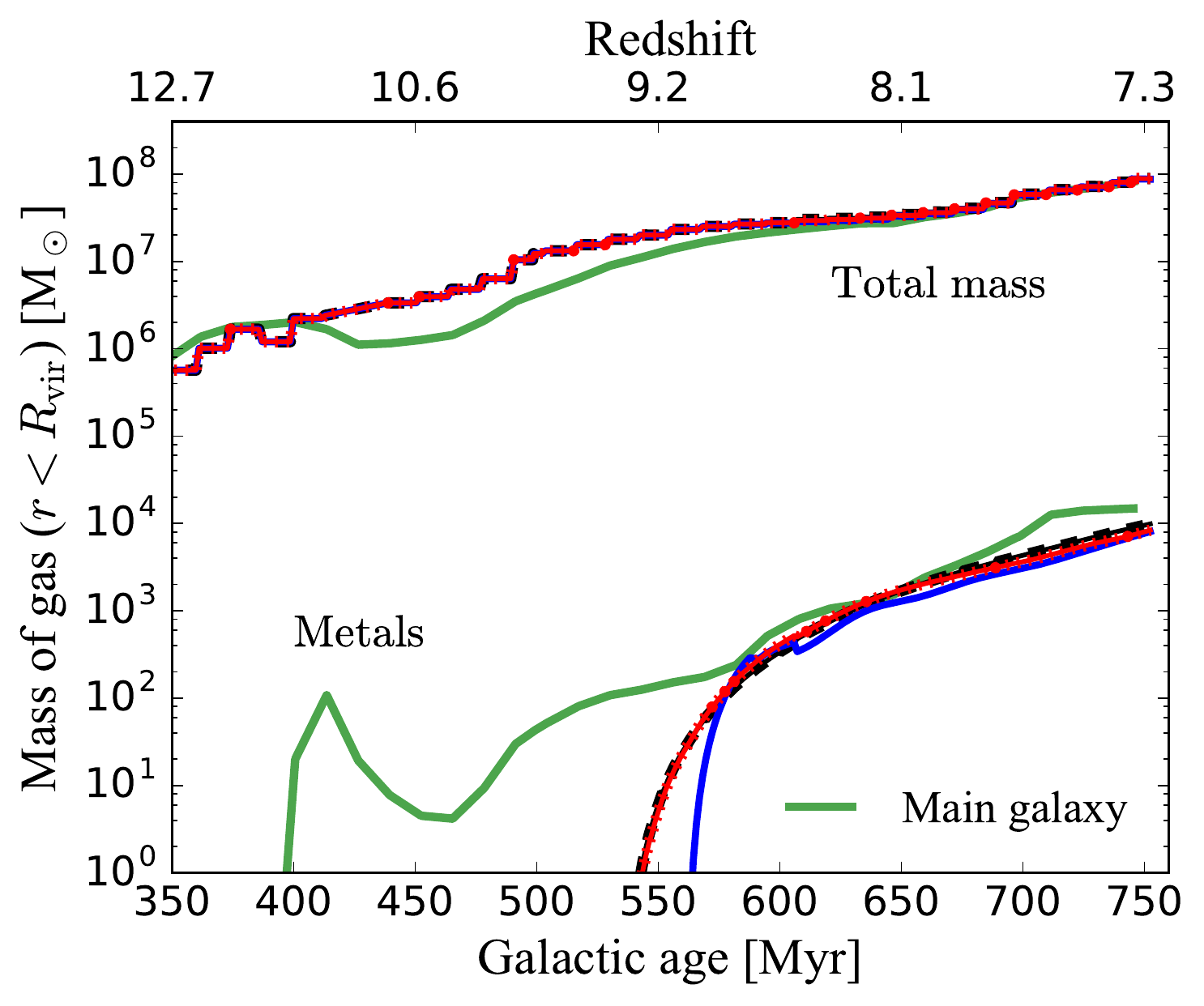}
\caption{Total mass of gas (top lines) and mass of metals (bottom lines) present inside the virial
radius of the main galaxy as a function of time.  The thick green line represents the values extracted
from the hydrodynamic simulation, while the solid blue, dashed black, and red dotted lines represent
the three different \gammac\ models presented in Figure~\ref{fig_degeneracy}.}
\label{fig_m_tot}
\end{figure}

\subsection{Modeling Uncertainties} \label{sect:disc_modeling_uncertainties}

One of the most important sources of uncertainty in this project is
the use of hydrodynamical simulations as ``ground truth.''  The W12
calculation is complex and includes a wide range of physics acting
over a large span of spatial and temporal scales.  A particular
concern is the subgrid model for star formation and feedback used in
this simulation -- these subgrid models are, in general, the largest
source of uncertainty in current-generation cosmological simulations
of galaxy formation \citep[see,
e.g.,][]{2012MNRAS.423.1726S,2017ARA&A..55...59N}.

\begin{figure}
\center
\includegraphics[width=\columnwidth]{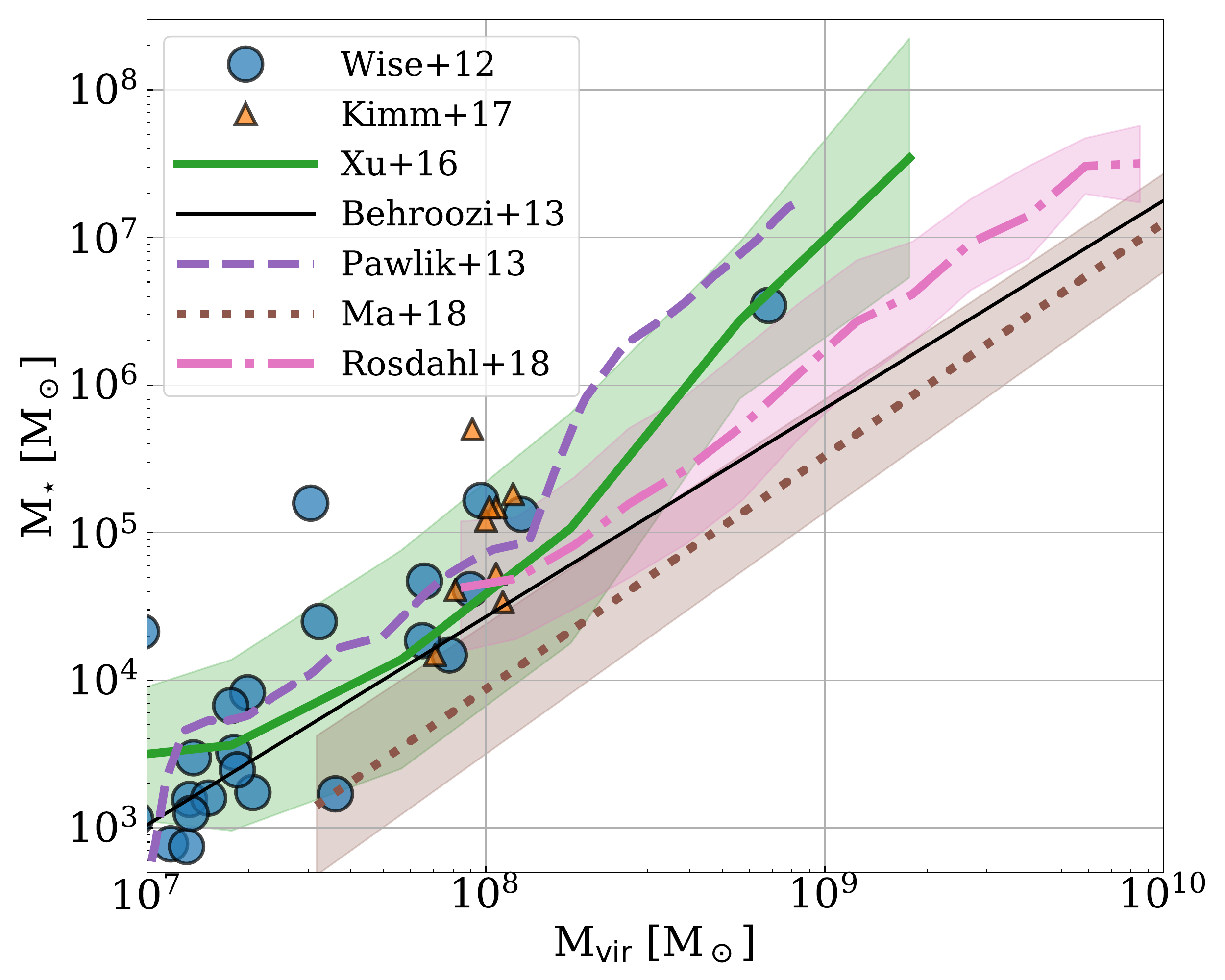}
\caption{Comparison of the stellar mass growth as a function of halo
  mass between several simulations (see legend).  The shaded areas
  depict 1-$\sigma$ errors.  The thin black line shows the abundance
  matching results at $z=0$ extrapolated to low masses
  \citep{2013ApJ...770...57B}.  The stellar mass of the galaxy
  analyzed here is close to the median relation in the Renaissance
  Simulations \citep{2016ApJ...833...84X}, is consistent with the
  SPHINX simulations \citep{2018arXiv180107259R}, and is an order of
  magnitude higher than the FIRE-2 simulations
  \citep{2017arXiv170606605M}.}
\label{fig:compare-mstar}
\end{figure}

Figure~\ref{fig:compare-mstar} compares the stellar mass --
  halo mass relation from several ``first galaxies'' simulations
  \citep{2013ApJ...767...59P, 2016ApJ...833...84X,
    2017MNRAS.466.4826K, 2017arXiv170606605M, 2018arXiv180107259R}.
  The galaxy studied in this work has a stellar mass nearly equal to
  the median found in the ``Normal region'' of the Renaissance
  simulations.  In addition to the W12 and Renaissance Simulations,
  the simulations of \citet{2017MNRAS.466.4826K} include Population
  III star formation.  All simulations except
  \citet{2017arXiv170606605M} solve the radiative transfer equation.
  At halo masses $M_{\rm vir} \la 10^8~{\rm M}_\odot$, all of the
  simulation results are consistent with each other.  Above this mass,
  the relation slope starts to increase in W12, the Renaissance
  Simulations, and \citet{2013ApJ...767...59P}, whereas the SPHINX
  \citep{2018arXiv180107259R} and FIRE-2 simulations
  \citep{2017arXiv170606605M} relation slope remains basically
  unchanged.  The stellar masses from the SPHINX simulations are
  consistently a factor of 3--6 higher than the FIRE-2 simulations.  The
  largest W12 galaxy is a factor of five higher than the median of the
  SPHINX simulations, however it should be noted that the Renaissance
  and SPHINX simulations' stellar mass -- halo mass relation are
  consistent to 1-$\sigma$ over their overlapping halo mass range
  $10^8 - 10^{9.25}~{\rm M}_\odot$.

Furthermore, while the W12 calculation stops at a relatively low
redshift that is observable by the Hubble Space Telescope ($z=7.29$),
even the most massive galaxy in this simulation is too dim to be
  seen by HST, even if strongly lensed.  One important consequence of
this is that the simulation we are using to validate and improve the
\gammac\ semi-analytical model is constrained by observations only
indirectly, and thus will represent a significant source of
uncertainty when \gammac\ is ultimately used to make predictions about
the relevant galaxy populations.

That said, there is significant
utility in using the W12 simulation as a \textit{case study}, to highlight possible
missing or inaccurate physical prescriptions in \gammac.  In this way,
this project is quite successful, as we have identified several key
areas where the \gammac\ model could be improved. For example,
Section~\ref{sec:numixing} highlights the need to examine star-forming
regions in more detail, and to implement more realistic models for
nonuniform mixing that are motivated by the hydrodynamical
simulations.  These models will be refined in future efforts by
comparison to additional galaxies in the W12 simulations, and in more
recent calculations like the Renaissance Simulations
\citep{2015ApJ...807L..12O}.  Similarly, as future generations of
physics-rich cosmological simulations add improved subgrid models for
the ISM and star formation and feedback, this will result in updated
physical prescriptions in our semi-analytic models.

One significant strength of directly comparing two different
theoretical techniques, rather than using observations to calibrate
the semi-analytical model, is that our validation methodology is
unaffected by nuclear astrophysics and observational uncertainties.
This is particularly important in terms of uncertainties relating to
stellar modeling \citep[see, e.g.,][]{2018ApJS..234...19F}.  Observationally, old stellar populations present a
challenge when converting from integrated stellar spectra or color-color
diagrams to metallicities and star formation rates, given
uncertainties in stellar synthesis models and degeneracies between
stellar age and metallicity \citep[see, e.g.,][]{2009ApJ...699..486C,2010ApJ...708...58C,2010ApJ...712..833C}.  In the type of comparison we have
undertaken, the hydrodynamical simulation provides these quantities
with no uncertainty (although undoubtedly with inaccuracy, as described
above and in the cited references).  From a theoretical standpoint, stellar modeling is affected
by uncertainties in nuclear reaction rates and in the physical
prescriptions put into stellar evolution models
(e.g., \citealt{2007ApJ...664.1033Y,2009ApJ...702.1068T,2015MNRAS.447.3115J,2017arXiv170903144D,2017MNRAS.469.1752N}).  These both impose
significant systematic uncertainties when comparing a hydrodynamical
simulation or semi-analytical model to observations.  When doing a
model-to-model comparison such as we have undertaken, however, we can
use the same underlying stellar models in both types of calculation
and thus eliminate these sources of systematic uncertainty.

\subsection{Current Limitations of GAMMA} \label{sect:disc_limitations_gamma}

The current version of \gammac\ is limited in several ways in terms of its modeling assumptions.
We did not include the contribution of Pop~III stars,
which probably affects our treatment of
the lowest-metallicity stars in our models \citep{2003ApJ...596L.135B,2012ApJ...745...50W,2016arXiv161100759G}.
A single PopIII star explosion can create a metallicity floor of [Z/H]\,$\sim$\,$-3$ in the W12 simulation.
One of the next steps is to include primordial stellar populations using a stochastic formation process similar to
the one adopted in W12.  We counted 13 Pop~III remnants in the main target halo at the end of the
simulation, which only represents $\sim$\,0.03\% of the total stellar mass formed.  The last Pop~III star
exploded at $t=508$\,Myr ($z=9.7$).

Another limitation of \gammac\ is its oversimplification of the gas phases. While 
the hydrodynamic simulation clearly shows multiphase interstellar
and circumgalactic media with highly variable metallicity, our SAM
only works with averaged quantities.  The complexity of the interstellar medium should be included in \gammac, but we first
need to analyze the hydrodynamic simulation in more detail (see Section~\ref{sect:disc_future}).
We aim to improve \gammac\ step-by-step where each new implementation is fully motivated and
needed, as for the non-uniform mixing prescription (see Section~\ref{sec:numixing}).

Our SAM is also limited in terms of its analytical nature.  Even if its complexity is increased, it
will always provide a simplified representation of galaxies. For example,
the stochastic (bursty) features and the rapid oscillation periods seen in the SFH of the W12
simulation is unlikely to be reproduced by \gammac.  Although the latter can generate oscillating behaviours
(see the blue line in panel~$d$ of Figure~\ref{fig_degeneracy}), they are not driven
by localized individual star formation bursts within the interstellar medium of the galaxy.
They are rather caused by a periodic transfer of gas between the galaxy and the CGM
(see, e.g., \citealt{2007ApJ...667..170S,2008MNRAS.386.2227Q,2015ApJ...802..123C,2015MNRAS.454.2691M}).
In certain circumstances, because our models account for a delay between the formation
of stars and the release of energy by SNe (see Equation~\ref{eq_delayed_outflow}),
stars can accumulate enough \textit{potential} energy to completely empty the galactic
gas component.  When the galactic outflow is launched, the galactic gas is transferred
into the CGM and the star formation process is momentarily stopped, until part of the CGM gas falls
back into the galaxy to start a new cycle.

Finally, we do not include yet the impact of an ionizing ultraviolet radiation background (UVB),
which can prevent or limit star formation in dwarf galaxies after the reionization (e.g., \citealt{2008MNRAS.390..920O,2017arXiv170606605M}).
Currently, if we were to run \gammac\ beyond the reionization period down to redshift zero,
our models would overestimate the stellar mass formed in low-mass galaxies.  This UVB will be included in future work.
We note that the W12 and Renaissance simulations include an H2-dissociating (Lyman-Werner)
radiation background, but not an ionizing background.  The ionizing radiation field is directly calculated
by the radiation transport solver.  Low-mass halos ($M_\mathrm{vir} < 10^7$\,M$_\odot$) are
photo-evaporated from radiation, especially if they are near a galaxy.
\\
\\
\subsection{Why Calibrate with Hydrodynamic Simulations?} \label{sect:disc_why}

Semi-analytic galaxy formation models targeting the nearby
low-redshift universe primarily use observations for calibration, but
some practitioners
also use hydrodynamic simulations
to inform them and to constrain their analytical prescriptions
\citep[e.g.,][]{2007MNRAS.377...63C,2010MNRAS.406..729S,2010MNRAS.407..632S,
2012MNRAS.419.3200H,2012MNRAS.421.3579N,2014MNRAS.441.2058M,2016MNRAS.461.3457G,
2017arXiv170908647M}.  This approach is even more necessary when SAMs are targeting
the early universe because the observational data available to constrain models are 
significantly less abundant and reduced in quality compared to the low-redshift universe.

The degeneracy highlighted in Section~\ref{sec:degeneracy} suggests that it is
unlikely that this type of modeling can constrain the physical processes which
occur inside dwarf spheroidal and ultra-faint galaxies by only looking
at their star formation history and stellar metallicity distribution functions.
Technically, the observed age-metallicity relationship could break down
our degeneracy (see Figure~\ref{fig_degeneracy}), but this relationship requires the determination of stellar ages which are
uncertain by a factor of a few Gyr for the oldest stars \citep{2014A&A...562A..71B,2014A&A...565A..89B}. In addition,
the typical time bin for the star formation history in observed dwarf spheroidal galaxies is $\sim1$\,Gyr
\citep{2009ARA&A..47..371T,2012A&A...539A.103D,2012A&A...544A..73D,2014A&A...572A..10D,2014ApJ...789..147W},
which is a reflection of these uncertainties.  This is large enough to
hide the stochastic and bursty nature of the SFH at $z\gtrsim7$

At high redshift, observations are thus limited in their ability to constrain the physical
processes implemented in SAMs. Adding stellar abundances in the list of
constraints could help, but other constraints outside the realm of chemical evolution (e.g.,
gas fraction, star formation efficiency, gas flow rates) are the most valuable to break the degeneracy
(\citealt{2017ApJ...835..128C}).  However, the Galactic halo and dwarf spheroidal and ultra-faint
galaxies do not form stars anymore and cannot directly probe the physical conditions
that led to their formation.  In spite of the imperfection of cosmological hydrodynamic
simulations (see Section~\ref{sect:disc_modeling_uncertainties}), they still represent the best
option to calibrate SAMs, since they are adequately resolved and provide crucial
data that are otherwise unknown with observations.

\subsection{Future Directions} \label{sect:disc_future}

The next step of this project is to extract further constraints from the W12 simulation.
An important quantity to extract is the amount of gas
involved in the star formation process, as it will allow us to
constrain the star formation efficiency in \gammac\
(see Figure~\ref{fig_degeneracy}).  To create a consistent base for the comparison, we need to 
define the star-forming region in the hydrodynamic simulation.  This
reservoir needs to include different gas phases in order to be representative of the 
simplified and averaged nature of the galactic gas in our SAM.  However, the 
task is non-trivial since the spatial distribution of stars in the targeted high-redshift galaxy is complex and cannot
be represented by a sphere or a rotating disk as in \cite{2017arXiv170908647M}.
While the spatial distribution and the geometry of the star-forming region is
not accounted for in our \gammac\ models, they are important from the 
point of view of extracting the \textit{correct} quantities from the hydrodynamic simulation.

Phase-space cuts (e.g., temperature, density, and metallicity thresholds) could be used
to isolate the geometry of the star-forming region as a function of time, but we first need
to better understand the metal recycling process within that central region.
Indeed, we noticed that the hot metal-rich gas phase containing most the SNe ejecta has
a very-short cooling timescale ($\lesssim1$\,Myr), which is below the time-resolution limit
of our data outputs.  At this stage, it is not clear whether this hot metal-rich gas efficiently
mix with its surrounding before forming new stars.  Deeper analysis is required, which
possibly includes re-running parts of the simulation with higher temporal resolution to study
the gas dynamics in more details.

In addition of isolating the mass of the star-forming region, inflowing and outflowing gas
fluxes going through the star-forming region should be extracted from the simulation.  These
quantities would help to constrain the gas circulation process adopted in our SAM.
Once the comparison framework between our semi-analytical models and
high-redshift cosmological simulations is established, our 
plan is to repeat our experiment with the other galaxies found in the W12
simulation.  The goal will be to derive general prescriptions that are
valid for a wide range of galaxy masses. On the longer term, we will
apply our methodology and inform \gammac\ using the Renaissance simulations
\citep{2015ApJ...807L..12O,2016ApJ...833...84X} in order to address a greater
variety of galaxy formation environments in the early universe.


%% file: conclusion.tex
In this paper we presented \gammac, our new semi-analytic galaxy formation model designed to address the origin of metal-poor stars and to reconstruct the chemical evolution history of low-mass galaxies in the early universe.  A critical goal of this project is to calibrate our model by comparing its predictions with the ones extracted from high-redshift cosmological hydrodynamic simulations. In this paper, we targeted the most massive galaxy in the \citet[][referred to as W12]{2012MNRAS.427..311W} simulation, extracted its mass assembly history (merger tree), and used it as an input in \gammac\ to provide a consistent setup for comparing the semi-analytical and hydrodynamical approaches.

We found that \gammac\ is able to reproduce the global trends predicted by the W12 simulation for the star formation history (SFH), the metallicity distribution function (MDF), and late stages of the age-metallicity relationship. However, there are degeneracies between the input parameters of \gammac\ and it is not possible to constrain the star formation efficiency and the baryonic circulation processes using these constraints alone.  Additional constraints such as the mass of gas involved in the star formation process need to be extracted from the hydrodynamic simulation to break the degeneracy (see Section~\ref{sect:disc_future}).

Non-uniform mixing in the interstellar medium causes a broadening in the MDF, which can be emulated with \gammac\ when convolving its MDF with Gaussian functions having a standard deviation of $\sim$\,0.2, a prescription motivated by the metallicity distributions of newly-formed stars seen in the hydrodynamic simulation at late times.  However, some other features seen in the W12 simulation cannot be captured by \gammac. This includes the stochasticity of the SFH and the sudden star formation events triggered by supernova explosions, which generate a non-negligible high-metallicity tail in the MDF.  In addition, infalling satellite galaxies have active star formation, and tend to have outflows of metal-enriched gas while simultaneously falling into the halo of the targeted galaxy.  This interaction between the satellite galaxies and the circumgalactic medium of the central galaxy is not include in the current version of our SAM.


The long-term goal of this project is to improve \gammac\ to a point where it can provide valuable insights into the formation and evolution of local dwarf spheroidal and ultra-faint galaxies as well as the Galactic halo. This requires the derivation of galaxy evolution prescriptions that are general enough to be applicable to a wide range of galaxy masses and formation environments.  To do so, we plan to repeat our experiment with more galaxies in the W12 simulation and ultimately extend our methodology to the Renaissances simulations \citep{2015ApJ...807L..12O,2016ApJ...833...84X}.

%% file: ms.bbl
\begin{thebibliography}{}
\providecommand\natexlab[1]{#1}
\providecommand\JournalTitle[1]{#1}

\bibitem[{{Abel} {et~al.}(1997){Abel}, {Anninos}, {Zhang}, \&
  {Norman}}]{abel97}
{Abel}, T., {Anninos}, P., {Zhang}, Y., \& {Norman}, M.~L. 1997,
  \href{http://dx.doi.org/10.1016/S1384-1076(97)00010-9}{\JournalTitle{New
  Astronomy}, 2, 181}

\bibitem[{{Abel} {et~al.}(2002){Abel}, {Bryan}, \&
  {Norman}}]{2002Sci...295...93A}
{Abel}, T., {Bryan}, G.~L., \& {Norman}, M.~L. 2002,
  \href{http://dx.doi.org/10.1126/science.295.5552.93}{\JournalTitle{Science},
  295, 93}

\bibitem[{{Agertz} \& {Kravtsov}(2016)}]{2016ApJ...824...79A}
{Agertz}, O., \& {Kravtsov}, A.~V. 2016,
  \href{http://dx.doi.org/10.3847/0004-637X/824/2/79}{\JournalTitle{\apj}, 824,
  79}

\bibitem[{{Andrews} {et~al.}(2017){Andrews}, {Weinberg}, {Sch{\"o}nrich}, \&
  {Johnson}}]{2017ApJ...835..224A}
{Andrews}, B.~H., {Weinberg}, D.~H., {Sch{\"o}nrich}, R., \& {Johnson}, J.~A.
  2017,
  \href{http://dx.doi.org/10.3847/1538-4357/835/2/224}{\JournalTitle{\apj},
  835, 224}

\bibitem[{{Angl{\'e}s-Alc{\'a}zar} {et~al.}(2017){Angl{\'e}s-Alc{\'a}zar},
  {Faucher-Gigu{\`e}re}, {Kere{\v s}}, {Hopkins}, {Quataert}, \&
  {Murray}}]{2017MNRAS.470.4698A}
{Angl{\'e}s-Alc{\'a}zar}, D., {Faucher-Gigu{\`e}re}, C.-A., {Kere{\v s}}, D.,
  {et~al.} 2017,
  \href{http://dx.doi.org/10.1093/mnras/stx1517}{\JournalTitle{\mnras}, 470,
  4698}

\bibitem[{{Anninos} {et~al.}(1997){Anninos}, {Zhang}, {Abel}, \&
  {Norman}}]{anninos97}
{Anninos}, P., {Zhang}, Y., {Abel}, T., \& {Norman}, M.~L. 1997,
  \href{http://dx.doi.org/10.1016/S1384-1076(97)00009-2}{\JournalTitle{New
  Astronomy}, 2, 209}

\bibitem[{{Atek} {et~al.}(2014){Atek}, {Richard}, {Kneib}, {Clement}, {Egami},
  {Ebeling}, {Jauzac}, {Jullo}, {Laporte}, {Limousin}, \&
  {Natarajan}}]{2014ApJ...786...60A}
{Atek}, H., {Richard}, J., {Kneib}, J.-P., {et~al.} 2014,
  \href{http://dx.doi.org/10.1088/0004-637X/786/1/60}{\JournalTitle{\apj}, 786,
  60}

\bibitem[{{Barkana} \& {Loeb}(2001)}]{2001PhR...349..125B}
{Barkana}, R., \& {Loeb}, A. 2001,
  \href{http://dx.doi.org/10.1016/S0370-1573(01)00019-9}{\JournalTitle{\physrep},
  349, 125}

\bibitem[{{Baugh}(2006)}]{2006RPPh...69.3101B}
{Baugh}, C.~M. 2006,
  \href{http://dx.doi.org/10.1088/0034-4885/69/12/R02}{\JournalTitle{Reports on
  Progress in Physics}, 69, 3101}

\bibitem[{{Beckwith} {et~al.}(2006){Beckwith}, {Stiavelli}, {Koekemoer},
  {Caldwell}, {Ferguson}, {Hook}, {Lucas}, {Bergeron}, {Corbin}, {Jogee},
  {Panagia}, {Robberto}, {Royle}, {Somerville}, \&
  {Sosey}}]{2006AJ....132.1729B}
{Beckwith}, S.~V.~W., {Stiavelli}, M., {Koekemoer}, A.~M., {et~al.} 2006,
  \href{http://dx.doi.org/10.1086/507302}{\JournalTitle{\aj}, 132, 1729}

\bibitem[{{Behroozi} {et~al.}(2013{\natexlab{a}}){Behroozi}, {Wechsler}, \&
  {Conroy}}]{2013ApJ...770...57B}
{Behroozi}, P.~S., {Wechsler}, R.~H., \& {Conroy}, C. 2013{\natexlab{a}},
  \href{http://dx.doi.org/10.1088/0004-637X/770/1/57}{\JournalTitle{\apj}, 770,
  57}

\bibitem[{{Behroozi} {et~al.}(2013{\natexlab{b}}){Behroozi}, {Wechsler}, \&
  {Wu}}]{2013ApJ...762..109B}
{Behroozi}, P.~S., {Wechsler}, R.~H., \& {Wu}, H.-Y. 2013{\natexlab{b}},
  \href{http://dx.doi.org/10.1088/0004-637X/762/2/109}{\JournalTitle{\apj},
  762, 109}

\bibitem[{{Behroozi} {et~al.}(2013{\natexlab{c}}){Behroozi}, {Wechsler}, {Wu},
  {Busha}, {Klypin}, \& {Primack}}]{2013ApJ...763...18B}
{Behroozi}, P.~S., {Wechsler}, R.~H., {Wu}, H.-Y., {et~al.} 2013{\natexlab{c}},
  \href{http://dx.doi.org/10.1088/0004-637X/763/1/18}{\JournalTitle{\apj}, 763,
  18}

\bibitem[{{Bensby} {et~al.}(2014){Bensby}, {Feltzing}, \&
  {Oey}}]{2014A&A...562A..71B}
{Bensby}, T., {Feltzing}, S., \& {Oey}, M.~S. 2014,
  \href{http://dx.doi.org/10.1051/0004-6361/201322631}{\JournalTitle{\aap},
  562, A71}

\bibitem[{{Bergemann} {et~al.}(2014){Bergemann}, {Ruchti}, {Serenelli},
  {Feltzing}, {Alves-Brito}, {Asplund}, {Bensby}, {Gruyters}, {Heiter},
  {Hourihane}, {Korn}, {Lind}, {Marino}, {Jofre}, {Nordlander}, {Ryde},
  {Worley}, {Gilmore}, {Randich}, {Ferguson}, {Jeffries}, {Micela},
  {Negueruela}, {Prusti}, {Rix}, {Vallenari}, {Alfaro}, {Allende Prieto},
  {Bragaglia}, {Koposov}, {Lanzafame}, {Pancino}, {Recio-Blanco}, {Smiljanic},
  {Walton}, {Costado}, {Franciosini}, {Hill}, {Lardo}, {de Laverny}, {Magrini},
  {Maiorca}, {Masseron}, {Morbidelli}, {Sacco}, {Kordopatis}, \& {Tautvai{\v
  s}ien{\.e}}}]{2014A&A...565A..89B}
{Bergemann}, M., {Ruchti}, G.~R., {Serenelli}, A., {et~al.} 2014,
  \href{http://dx.doi.org/10.1051/0004-6361/201423456}{\JournalTitle{\aap},
  565, A89}

\bibitem[{{Berger} \& {Colella}(1989)}]{Berger89}
{Berger}, M.~J., \& {Colella}, P. 1989, \JournalTitle{J. Comp. Phys.}, 82, 64

\bibitem[{{Bertschinger}(2001)}]{2001ApJS..137....1B}
{Bertschinger}, E. 2001,
  \href{http://dx.doi.org/10.1086/322526}{\JournalTitle{\apjs}, 137, 1}

\bibitem[{{Bland-Hawthorn} \& {Gerhard}(2016)}]{2016ARA&A..54..529B}
{Bland-Hawthorn}, J., \& {Gerhard}, O. 2016,
  \href{http://dx.doi.org/10.1146/annurev-astro-081915-023441}{\JournalTitle{\araa},
  54, 529}

\bibitem[{{Bovill} \& {Ricotti}(2009)}]{2009ApJ...693.1859B}
{Bovill}, M.~S., \& {Ricotti}, M. 2009,
  \href{http://dx.doi.org/10.1088/0004-637X/693/2/1859}{\JournalTitle{\apj},
  693, 1859}

\bibitem[{{Bower} {et~al.}(2006){Bower}, {Benson}, {Malbon}, {Helly}, {Frenk},
  {Baugh}, {Cole}, \& {Lacey}}]{2006MNRAS.370..645B}
{Bower}, R.~G., {Benson}, A.~J., {Malbon}, R., {et~al.} 2006,
  \href{http://dx.doi.org/10.1111/j.1365-2966.2006.10519.x}{\JournalTitle{\mnras},
  370, 645}

\bibitem[{{Bower} {et~al.}(2010){Bower}, {Vernon}, {Goldstein}, {Benson},
  {Lacey}, {Baugh}, {Cole}, \& {Frenk}}]{2010MNRAS.407.2017B}
{Bower}, R.~G., {Vernon}, I., {Goldstein}, M., {et~al.} 2010,
  \href{http://dx.doi.org/10.1111/j.1365-2966.2010.16991.x}{\JournalTitle{\mnras},
  407, 2017}

\bibitem[{{Boylan-Kolchin} {et~al.}(2015){Boylan-Kolchin}, {Weisz}, {Johnson},
  {Bullock}, {Conroy}, \& {Fitts}}]{2015MNRAS.453.1503B}
{Boylan-Kolchin}, M., {Weisz}, D.~R., {Johnson}, B.~D., {et~al.} 2015,
  \href{http://dx.doi.org/10.1093/mnras/stv1736}{\JournalTitle{\mnras}, 453,
  1503}

\bibitem[{{Bromm} {et~al.}(2002){Bromm}, {Coppi}, \&
  {Larson}}]{2002ApJ...564...23B}
{Bromm}, V., {Coppi}, P.~S., \& {Larson}, R.~B. 2002,
  \href{http://dx.doi.org/10.1086/323947}{\JournalTitle{\apj}, 564, 23}

\bibitem[{{Bromm} {et~al.}(2003){Bromm}, {Yoshida}, \&
  {Hernquist}}]{2003ApJ...596L.135B}
{Bromm}, V., {Yoshida}, N., \& {Hernquist}, L. 2003,
  \href{http://dx.doi.org/10.1086/379359}{\JournalTitle{\apjl}, 596, L135}

\bibitem[{{Bryan} \& {Norman}(1997)}]{Bryan97b}
{Bryan}, G.~L., \& {Norman}, M.~L. 1997, \JournalTitle{ArXiv Astrophysics
  e-prints}, \href{http://arxiv.org/abs/astro-ph/9710187}{{\sffamily
  astro-ph/9710187}}

\bibitem[{{Bryan} \& {Norman}(1998)}]{1998ApJ...495...80B}
---. 1998, \href{http://dx.doi.org/10.1086/305262}{\JournalTitle{\apj}, 495,
  80}

\bibitem[{{Bryan} {et~al.}(1995){Bryan}, {Norman}, {Stone}, {Cen}, \&
  {Ostriker}}]{Bryan95}
{Bryan}, G.~L., {Norman}, M.~L., {Stone}, J.~M., {Cen}, R., \& {Ostriker},
  J.~P. 1995, \JournalTitle{Comp. Phys. Comm}, 89, 149

\bibitem[{{Bryan} {et~al.}(2014){Bryan}, {Norman}, {O'Shea}, {Abel}, {Wise},
  {Turk}, {Reynolds}, {Collins}, {Wang}, {Skillman}, {Smith}, {Harkness},
  {Bordner}, {Kim}, {Kuhlen}, {Xu}, {Goldbaum}, {Hummels}, {Kritsuk}, {Tasker},
  {Skory}, {Simpson}, {Hahn}, {Oishi}, {So}, {Zhao}, {Cen}, {Li}, \& {Enzo
  Collaboration}}]{2014ApJS..211...19B}
{Bryan}, G.~L., {Norman}, M.~L., {O'Shea}, B.~W., {et~al.} 2014,
  \href{http://dx.doi.org/10.1088/0067-0049/211/2/19}{\JournalTitle{\apjs},
  211, 19}

\bibitem[{{Bustard} {et~al.}(2016){Bustard}, {Zweibel}, \&
  {D'Onghia}}]{2016ApJ...819...29B}
{Bustard}, C., {Zweibel}, E.~G., \& {D'Onghia}, E. 2016,
  \href{http://dx.doi.org/10.3847/0004-637X/819/1/29}{\JournalTitle{\apj}, 819,
  29}

\bibitem[{{Cattaneo} {et~al.}(2007){Cattaneo}, {Blaizot}, {Weinberg}, {Kere{\v
  s}}, {Colombi}, {Dav{\'e}}, {Devriendt}, {Guiderdoni}, \&
  {Katz}}]{2007MNRAS.377...63C}
{Cattaneo}, A., {Blaizot}, J., {Weinberg}, D.~H., {et~al.} 2007,
  \href{http://dx.doi.org/10.1111/j.1365-2966.2007.11597.x}{\JournalTitle{\mnras},
  377, 63}

\bibitem[{{Chen} {et~al.}(2014){Chen}, {Wise}, {Norman}, {Xu}, \&
  {O'Shea}}]{2014ApJ...795..144C}
{Chen}, P., {Wise}, J.~H., {Norman}, M.~L., {Xu}, H., \& {O'Shea}, B.~W. 2014,
  \href{http://dx.doi.org/10.1088/0004-637X/795/2/144}{\JournalTitle{\apj},
  795, 144}

\bibitem[{{Cohen} {et~al.}(2013){Cohen}, {Christlieb}, {Thompson}, {McWilliam},
  {Shectman}, {Reimers}, {Wisotzki}, \& {Kirby}}]{2013ApJ...778...56C}
{Cohen}, J.~G., {Christlieb}, N., {Thompson}, I., {et~al.} 2013,
  \href{http://dx.doi.org/10.1088/0004-637X/778/1/56}{\JournalTitle{\apj}, 778,
  56}

\bibitem[{{Conroy} \& {Gunn}(2010)}]{2010ApJ...712..833C}
{Conroy}, C., \& {Gunn}, J.~E. 2010,
  \href{http://dx.doi.org/10.1088/0004-637X/712/2/833}{\JournalTitle{\apj},
  712, 833}

\bibitem[{{Conroy} {et~al.}(2009){Conroy}, {Gunn}, \&
  {White}}]{2009ApJ...699..486C}
{Conroy}, C., {Gunn}, J.~E., \& {White}, M. 2009,
  \href{http://dx.doi.org/10.1088/0004-637X/699/1/486}{\JournalTitle{\apj},
  699, 486}

\bibitem[{{Conroy} {et~al.}(2010){Conroy}, {White}, \&
  {Gunn}}]{2010ApJ...708...58C}
{Conroy}, C., {White}, M., \& {Gunn}, J.~E. 2010,
  \href{http://dx.doi.org/10.1088/0004-637X/708/1/58}{\JournalTitle{\apj}, 708,
  58}

\bibitem[{{C{\^o}t{\'e}} {et~al.}(2015){C{\^o}t{\'e}}, {Martel}, \&
  {Drissen}}]{2015ApJ...802..123C}
{C{\^o}t{\'e}}, B., {Martel}, H., \& {Drissen}, L. 2015,
  \href{http://dx.doi.org/10.1088/0004-637X/802/2/123}{\JournalTitle{\apj},
  802, 123}

\bibitem[{{C{\^o}t{\'e}} {et~al.}(2017{\natexlab{a}}){C{\^o}t{\'e}}, {O'Shea},
  {Ritter}, {Herwig}, \& {Venn}}]{2017ApJ...835..128C}
{C{\^o}t{\'e}}, B., {O'Shea}, B.~W., {Ritter}, C., {Herwig}, F., \& {Venn},
  K.~A. 2017{\natexlab{a}},
  \href{http://dx.doi.org/10.3847/1538-4357/835/2/128}{\JournalTitle{\apj},
  835, 128}

\bibitem[{{C{\^o}t{\'e}} {et~al.}(2017{\natexlab{b}}){C{\^o}t{\'e}}, {Ritter},
  {Herwig}, {O'Shea}, {Pignatari}, {Silvia}, {Jones}, \&
  {Fryer}}]{2017nuco.confb0203C}
{C{\^o}t{\'e}}, B., {Ritter}, C., {Herwig}, F., {et~al.} 2017{\natexlab{b}},
  \href{http://dx.doi.org/10.7566/JPSCP.14.020203}{in 14th International
  Symposium on Nuclei in the Cosmos (NIC2016), ed. S.~{Kubono}, T.~{Kajino},
  S.~{Nishimura}, T.~{Isobe}, S.~{Nagataki}, T.~{Shima}, \& Y.~{Takeda}},
  020203

\bibitem[{{C{\^o}t{\'e}} {et~al.}(2016{\natexlab{a}}){C{\^o}t{\'e}}, {Ritter},
  {O'Shea}, {Herwig}, {Pignatari}, {Jones}, \& {Fryer}}]{2016ApJ...824...82C}
{C{\^o}t{\'e}}, B., {Ritter}, C., {O'Shea}, B.~W., {et~al.} 2016{\natexlab{a}},
  \href{http://dx.doi.org/10.3847/0004-637X/824/2/82}{\JournalTitle{\apj}, 824,
  82}

\bibitem[{{C{\^o}t{\'e}} {et~al.}(2016{\natexlab{b}}){C{\^o}t{\'e}}, {West},
  {Heger}, {Ritter}, {O'Shea}, {Herwig}, {Travaglio}, \&
  {Bisterzo}}]{2016MNRAS.463.3755C}
{C{\^o}t{\'e}}, B., {West}, C., {Heger}, A., {et~al.} 2016{\natexlab{b}},
  \href{http://dx.doi.org/10.1093/mnras/stw2244}{\JournalTitle{\mnras}, 463,
  3755}

\bibitem[{{Crosby} {et~al.}(2016){Crosby}, {O'Shea}, {Beers}, \&
  {Tumlinson}}]{2016ApJ...820...71C}
{Crosby}, B.~D., {O'Shea}, B.~W., {Beers}, T.~C., \& {Tumlinson}, J. 2016,
  \href{http://dx.doi.org/10.3847/0004-637X/820/1/71}{\JournalTitle{\apj}, 820,
  71}

\bibitem[{{Croton} {et~al.}(2006){Croton}, {Springel}, {White}, {De Lucia},
  {Frenk}, {Gao}, {Jenkins}, {Kauffmann}, {Navarro}, \&
  {Yoshida}}]{2006MNRAS.365...11C}
{Croton}, D.~J., {Springel}, V., {White}, S.~D.~M., {et~al.} 2006,
  \href{http://dx.doi.org/10.1111/j.1365-2966.2005.09675.x}{\JournalTitle{\mnras},
  365, 11}

\bibitem[{{Croton} {et~al.}(2016){Croton}, {Stevens}, {Tonini}, {Garel},
  {Bernyk}, {Bibiano}, {Hodkinson}, {Mutch}, {Poole}, \&
  {Shattow}}]{2016ApJS..222...22C}
{Croton}, D.~J., {Stevens}, A.~R.~H., {Tonini}, C., {et~al.} 2016,
  \href{http://dx.doi.org/10.3847/0067-0049/222/2/22}{\JournalTitle{\apjs},
  222, 22}

\bibitem[{{Dav{\'e}} {et~al.}(2012){Dav{\'e}}, {Finlator}, \&
  {Oppenheimer}}]{2012MNRAS.421...98D}
{Dav{\'e}}, R., {Finlator}, K., \& {Oppenheimer}, B.~D. 2012,
  \href{http://dx.doi.org/10.1111/j.1365-2966.2011.20148.x}{\JournalTitle{\mnras},
  421, 98}

\bibitem[{{de Boer} {et~al.}(2014){de Boer}, {Tolstoy}, {Lemasle}, {Saha},
  {Olszewski}, {Mateo}, {Irwin}, \& {Battaglia}}]{2014A&A...572A..10D}
{de Boer}, T.~J.~L., {Tolstoy}, E., {Lemasle}, B., {et~al.} 2014,
  \href{http://dx.doi.org/10.1051/0004-6361/201424119}{\JournalTitle{\aap},
  572, A10}

\bibitem[{{de Boer} {et~al.}(2012{\natexlab{a}}){de Boer}, {Tolstoy}, {Hill},
  {Saha}, {Olszewski}, {Mateo}, {Starkenburg}, {Battaglia}, \&
  {Walker}}]{2012A&A...544A..73D}
{de Boer}, T.~J.~L., {Tolstoy}, E., {Hill}, V., {et~al.} 2012{\natexlab{a}},
  \href{http://dx.doi.org/10.1051/0004-6361/201219547}{\JournalTitle{\aap},
  544, A73}

\bibitem[{{de Boer} {et~al.}(2012{\natexlab{b}}){de Boer}, {Tolstoy}, {Hill},
  {Saha}, {Olsen}, {Starkenburg}, {Lemasle}, {Irwin}, \&
  {Battaglia}}]{2012A&A...539A.103D}
---. 2012{\natexlab{b}},
  \href{http://dx.doi.org/10.1051/0004-6361/201118378}{\JournalTitle{\aap},
  539, A103}

\bibitem[{{deBoer} {et~al.}(2017){deBoer}, {Gorres}, {Wiescher}, {Azuma},
  {Best}, {Brune}, {Fields}, {Jones}, {Pignatari}, {Sayre}, {Smith}, {Timmes},
  \& {Uberseder}}]{2017arXiv170903144D}
{deBoer}, R.~J., {Gorres}, J., {Wiescher}, M., {et~al.} 2017,
  \JournalTitle{ArXiv e-prints},
  \href{http://arxiv.org/abs/1709.03144}{{\sffamily arXiv:1709.03144
  [nucl-ex]}}

\bibitem[{{Fenner} \& {Gibson}(2003)}]{2003PASA...20..189F}
{Fenner}, Y., \& {Gibson}, B.~K. 2003,
  \href{http://dx.doi.org/10.1071/AS02047}{\JournalTitle{\pasa}, 20, 189}

\bibitem[{{Fields} {et~al.}(2018){Fields}, {Timmes}, {Farmer}, {Petermann},
  {Wolf}, \& {Couch}}]{2018ApJS..234...19F}
{Fields}, C.~E., {Timmes}, F.~X., {Farmer}, R., {et~al.} 2018,
  \href{http://dx.doi.org/10.3847/1538-4365/aaa29b}{\JournalTitle{\apjs}, 234,
  19}

\bibitem[{{Finlator}(2017)}]{2017ASSL..430..221F}
{Finlator}, K. 2017, \href{http://dx.doi.org/10.1007/978-3-319-52512-9_10}{in
  Astrophysics and Space Science Library, Vol. 430, Astrophysics and Space
  Science Library, ed. A.~{Fox} \& R.~{Dav{\'e}}}, 221

\bibitem[{{Fontanot} {et~al.}(2017){Fontanot}, {De Lucia}, {Hirschmann},
  {Bruzual}, {Charlot}, \& {Zibetti}}]{2017MNRAS.464.3812F}
{Fontanot}, F., {De Lucia}, G., {Hirschmann}, M., {et~al.} 2017,
  \href{http://dx.doi.org/10.1093/mnras/stw2612}{\JournalTitle{\mnras}, 464,
  3812}

\bibitem[{{Frebel} \& {Norris}(2015)}]{2015ARA&A..53..631F}
{Frebel}, A., \& {Norris}, J.~E. 2015,
  \href{http://dx.doi.org/10.1146/annurev-astro-082214-122423}{\JournalTitle{\araa},
  53, 631}

\bibitem[{{Glover} \& {Abel}(2008)}]{2008MNRAS.388.1627G}
{Glover}, S.~C.~O., \& {Abel}, T. 2008,
  \href{http://dx.doi.org/10.1111/j.1365-2966.2008.13224.x}{\JournalTitle{\mnras},
  388, 1627}

\bibitem[{{Griffen} {et~al.}(2016){Griffen}, {Dooley}, {Ji}, {O'Shea},
  {G{\'o}mez}, \& {Frebel}}]{2016arXiv161100759G}
{Griffen}, B.~F., {Dooley}, G.~A., {Ji}, A.~P., {et~al.} 2016,
  \JournalTitle{ArXiv e-prints},
  \href{http://arxiv.org/abs/1611.00759}{{\sffamily arXiv:1611.00759}}

\bibitem[{{Guo} {et~al.}(2016){Guo}, {Gonzalez-Perez}, {Guo}, {Schaller},
  {Furlong}, {Bower}, {Cole}, {Crain}, {Frenk}, {Helly}, {Lacey}, {Lagos},
  {Mitchell}, {Schaye}, \& {Theuns}}]{2016MNRAS.461.3457G}
{Guo}, Q., {Gonzalez-Perez}, V., {Guo}, Q., {et~al.} 2016,
  \href{http://dx.doi.org/10.1093/mnras/stw1525}{\JournalTitle{\mnras}, 461,
  3457}

\bibitem[{{Henriques} {et~al.}(2015){Henriques}, {White}, {Thomas}, {Angulo},
  {Guo}, {Lemson}, {Springel}, \& {Overzier}}]{2015MNRAS.451.2663H}
{Henriques}, B.~M.~B., {White}, S.~D.~M., {Thomas}, P.~A., {et~al.} 2015,
  \href{http://dx.doi.org/10.1093/mnras/stv705}{\JournalTitle{\mnras}, 451,
  2663}

\bibitem[{{Henriques} {et~al.}(2013){Henriques}, {White}, {Thomas}, {Angulo},
  {Guo}, {Lemson}, \& {Springel}}]{2013MNRAS.431.3373H}
---. 2013, \href{http://dx.doi.org/10.1093/mnras/stt415}{\JournalTitle{\mnras},
  431, 3373}

\bibitem[{{Hirai} {et~al.}(2017){Hirai}, {Ishimaru}, {Saitoh}, {Fujii},
  {Hidaka}, \& {Kajino}}]{2017MNRAS.466.2474H}
{Hirai}, Y., {Ishimaru}, Y., {Saitoh}, T.~R., {et~al.} 2017,
  \href{http://dx.doi.org/10.1093/mnras/stw3342}{\JournalTitle{\mnras}, 466,
  2474}

\bibitem[{{Hirschmann} {et~al.}(2012){Hirschmann}, {Naab}, {Somerville},
  {Burkert}, \& {Oser}}]{2012MNRAS.419.3200H}
{Hirschmann}, M., {Naab}, T., {Somerville}, R.~S., {Burkert}, A., \& {Oser}, L.
  2012,
  \href{http://dx.doi.org/10.1111/j.1365-2966.2011.19961.x}{\JournalTitle{\mnras},
  419, 3200}

\bibitem[{{Hopkins} {et~al.}(2014){Hopkins}, {Kere{\v s}}, {O{\~n}orbe},
  {Faucher-Gigu{\`e}re}, {Quataert}, {Murray}, \&
  {Bullock}}]{2014MNRAS.445..581H}
{Hopkins}, P.~F., {Kere{\v s}}, D., {O{\~n}orbe}, J., {et~al.} 2014,
  \href{http://dx.doi.org/10.1093/mnras/stu1738}{\JournalTitle{\mnras}, 445,
  581}

\bibitem[{{Hopkins} {et~al.}(2011){Hopkins}, {Quataert}, \&
  {Murray}}]{2011MNRAS.417..950H}
{Hopkins}, P.~F., {Quataert}, E., \& {Murray}, N. 2011,
  \href{http://dx.doi.org/10.1111/j.1365-2966.2011.19306.x}{\JournalTitle{\mnras},
  417, 950}

\bibitem[{{Hopkins} {et~al.}(2012){Hopkins}, {Quataert}, \&
  {Murray}}]{2012MNRAS.421.3522H}
---. 2012,
  \href{http://dx.doi.org/10.1111/j.1365-2966.2012.20593.x}{\JournalTitle{\mnras},
  421, 3522}

\bibitem[{{Illingworth} {et~al.}(2013){Illingworth}, {Magee}, {Oesch},
  {Bouwens}, {Labb{\'e}}, {Stiavelli}, {van Dokkum}, {Franx}, {Trenti},
  {Carollo}, \& {Gonzalez}}]{2013ApJS..209....6I}
{Illingworth}, G.~D., {Magee}, D., {Oesch}, P.~A., {et~al.} 2013,
  \href{http://dx.doi.org/10.1088/0067-0049/209/1/6}{\JournalTitle{\apjs}, 209,
  6}

\bibitem[{{Ishigaki} {et~al.}(2015){Ishigaki}, {Kawamata}, {Ouchi}, {Oguri},
  {Shimasaku}, \& {Ono}}]{2015ApJ...799...12I}
{Ishigaki}, M., {Kawamata}, R., {Ouchi}, M., {et~al.} 2015,
  \href{http://dx.doi.org/10.1088/0004-637X/799/1/12}{\JournalTitle{\apj}, 799,
  12}

\bibitem[{{Jones} {et~al.}(2015){Jones}, {Hirschi}, {Pignatari}, {Heger},
  {Georgy}, {Nishimura}, {Fryer}, \& {Herwig}}]{2015MNRAS.447.3115J}
{Jones}, S., {Hirschi}, R., {Pignatari}, M., {et~al.} 2015,
  \href{http://dx.doi.org/10.1093/mnras/stu2657}{\JournalTitle{\mnras}, 447,
  3115}

\bibitem[{{Kim} {et~al.}(2009){Kim}, {Baugh}, {Cole}, {Frenk}, \&
  {Benson}}]{2009MNRAS.400.1527K}
{Kim}, H.-S., {Baugh}, C.~M., {Cole}, S., {Frenk}, C.~S., \& {Benson}, A.~J.
  2009,
  \href{http://dx.doi.org/10.1111/j.1365-2966.2009.15560.x}{\JournalTitle{\mnras},
  400, 1527}

\bibitem[{{Kimm} {et~al.}(2017){Kimm}, {Katz}, {Haehnelt}, {Rosdahl},
  {Devriendt}, \& {Slyz}}]{2017MNRAS.466.4826K}
{Kimm}, T., {Katz}, H., {Haehnelt}, M., {et~al.} 2017,
  \href{http://dx.doi.org/10.1093/mnras/stx052}{\JournalTitle{\mnras}, 466,
  4826}

\bibitem[{{Komatsu} {et~al.}(2011){Komatsu}, {Smith}, {Dunkley}, {Bennett},
  {Gold}, {Hinshaw}, {Jarosik}, {Larson}, {Nolta}, {Page}, {Spergel},
  {Halpern}, {Hill}, {Kogut}, {Limon}, {Meyer}, {Odegard}, {Tucker}, {Weiland},
  {Wollack}, \& {Wright}}]{2011ApJS..192...18K}
{Komatsu}, E., {Smith}, K.~M., {Dunkley}, J., {et~al.} 2011,
  \href{http://dx.doi.org/10.1088/0067-0049/192/2/18}{\JournalTitle{\apjs},
  192, 18}

\bibitem[{{Komiya} {et~al.}(2014){Komiya}, {Yamada}, {Suda}, \&
  {Fujimoto}}]{2014ApJ...783..132K}
{Komiya}, Y., {Yamada}, S., {Suda}, T., \& {Fujimoto}, M.~Y. 2014,
  \href{http://dx.doi.org/10.1088/0004-637X/783/2/132}{\JournalTitle{\apj},
  783, 132}

\bibitem[{{Lacey} {et~al.}(2016){Lacey}, {Baugh}, {Frenk}, {Benson}, {Bower},
  {Cole}, {Gonzalez-Perez}, {Helly}, {Lagos}, \&
  {Mitchell}}]{2016MNRAS.462.3854L}
{Lacey}, C.~G., {Baugh}, C.~M., {Frenk}, C.~S., {et~al.} 2016,
  \href{http://dx.doi.org/10.1093/mnras/stw1888}{\JournalTitle{\mnras}, 462,
  3854}

\bibitem[{{Lagos} {et~al.}(2011){Lagos}, {Lacey}, {Baugh}, {Bower}, \&
  {Benson}}]{2011MNRAS.416.1566L}
{Lagos}, C.~D.~P., {Lacey}, C.~G., {Baugh}, C.~M., {Bower}, R.~G., \& {Benson},
  A.~J. 2011,
  \href{http://dx.doi.org/10.1111/j.1365-2966.2011.19160.x}{\JournalTitle{\mnras},
  416, 1566}

\bibitem[{{Lu} {et~al.}(2011){Lu}, {Kere{\v s}}, {Katz}, {Mo}, {Fardal}, \&
  {Weinberg}}]{2011MNRAS.416..660L}
{Lu}, Y., {Kere{\v s}}, D., {Katz}, N., {et~al.} 2011,
  \href{http://dx.doi.org/10.1111/j.1365-2966.2011.19072.x}{\JournalTitle{\mnras},
  416, 660}

\bibitem[{{Lu} {et~al.}(2012){Lu}, {Mo}, {Katz}, \&
  {Weinberg}}]{2012MNRAS.421.1779L}
{Lu}, Y., {Mo}, H.~J., {Katz}, N., \& {Weinberg}, M.~D. 2012,
  \href{http://dx.doi.org/10.1111/j.1365-2966.2012.20435.x}{\JournalTitle{\mnras},
  421, 1779}

\bibitem[{{Ma} {et~al.}(2017){Ma}, {Hopkins}, {Garrison-Kimmel},
  {Faucher-Gigu{\`e}re}, {Quataert}, {Boylan-Kolchin}, {Hayward}, {Feldmann},
  \& {Kere{\v s}}}]{2017arXiv170606605M}
{Ma}, X., {Hopkins}, P.~F., {Garrison-Kimmel}, S., {et~al.} 2017,
  \JournalTitle{ArXiv e-prints},
  \href{http://arxiv.org/abs/1706.06605}{{\sffamily arXiv:1706.06605}}

\bibitem[{{Martin}(2005)}]{2005ApJ...621..227M}
{Martin}, C.~L. 2005,
  \href{http://dx.doi.org/10.1086/427277}{\JournalTitle{\apj}, 621, 227}

\bibitem[{{Mitchell} {et~al.}(2017){Mitchell}, {Lacey}, {Lagos}, {Frenk},
  {Bower}, {Cole}, {Helly}, {Schaller}, {Gonzalez-Perez}, \&
  {Theuns}}]{2017arXiv170908647M}
{Mitchell}, P.~D., {Lacey}, C.~G., {Lagos}, C.~D.~P., {et~al.} 2017,
  \JournalTitle{ArXiv e-prints},
  \href{http://arxiv.org/abs/1709.08647}{{\sffamily arXiv:1709.08647}}

\bibitem[{{Monaco} {et~al.}(2014){Monaco}, {Benson}, {De Lucia}, {Fontanot},
  {Borgani}, \& {Boylan-Kolchin}}]{2014MNRAS.441.2058M}
{Monaco}, P., {Benson}, A.~J., {De Lucia}, G., {et~al.} 2014,
  \href{http://dx.doi.org/10.1093/mnras/stu655}{\JournalTitle{\mnras}, 441,
  2058}

\bibitem[{{Muratov} {et~al.}(2015){Muratov}, {Kere{\v s}},
  {Faucher-Gigu{\`e}re}, {Hopkins}, {Quataert}, \&
  {Murray}}]{2015MNRAS.454.2691M}
{Muratov}, A.~L., {Kere{\v s}}, D., {Faucher-Gigu{\`e}re}, C.-A., {et~al.}
  2015, \href{http://dx.doi.org/10.1093/mnras/stv2126}{\JournalTitle{\mnras},
  454, 2691}

\bibitem[{{Murray} {et~al.}(2005){Murray}, {Quataert}, \&
  {Thompson}}]{2005ApJ...618..569M}
{Murray}, N., {Quataert}, E., \& {Thompson}, T.~A. 2005,
  \href{http://dx.doi.org/10.1086/426067}{\JournalTitle{\apj}, 618, 569}

\bibitem[{{Naab} \& {Ostriker}(2017)}]{2017ARA&A..55...59N}
{Naab}, T., \& {Ostriker}, J.~P. 2017,
  \href{http://dx.doi.org/10.1146/annurev-astro-081913-040019}{\JournalTitle{\araa},
  55, 59}

\bibitem[{{Neistein} {et~al.}(2012){Neistein}, {Khochfar}, {Dalla Vecchia}, \&
  {Schaye}}]{2012MNRAS.421.3579N}
{Neistein}, E., {Khochfar}, S., {Dalla Vecchia}, C., \& {Schaye}, J. 2012,
  \href{http://dx.doi.org/10.1111/j.1365-2966.2012.20584.x}{\JournalTitle{\mnras},
  421, 3579}

\bibitem[{{Nishimura} {et~al.}(2017){Nishimura}, {Hirschi}, {Rauscher},
  {Murphy}, \& {Cescutti}}]{2017MNRAS.469.1752N}
{Nishimura}, N., {Hirschi}, R., {Rauscher}, T., {Murphy}, A.~S.~J., \&
  {Cescutti}, G. 2017,
  \href{http://dx.doi.org/10.1093/mnras/stx696}{\JournalTitle{\mnras}, 469,
  1752}

\bibitem[{{Okamoto} {et~al.}(2008){Okamoto}, {Gao}, \&
  {Theuns}}]{2008MNRAS.390..920O}
{Okamoto}, T., {Gao}, L., \& {Theuns}, T. 2008,
  \href{http://dx.doi.org/10.1111/j.1365-2966.2008.13830.x}{\JournalTitle{\mnras},
  390, 920}

\bibitem[{{O'Shea} \& {Norman}(2007)}]{2007ApJ...654...66O}
{O'Shea}, B.~W., \& {Norman}, M.~L. 2007,
  \href{http://dx.doi.org/10.1086/509250}{\JournalTitle{\apj}, 654, 66}

\bibitem[{{O'Shea} {et~al.}(2015){O'Shea}, {Wise}, {Xu}, \&
  {Norman}}]{2015ApJ...807L..12O}
{O'Shea}, B.~W., {Wise}, J.~H., {Xu}, H., \& {Norman}, M.~L. 2015,
  \href{http://dx.doi.org/10.1088/2041-8205/807/1/L12}{\JournalTitle{\apjl},
  807, L12}

\bibitem[{{Pawlik} {et~al.}(2013){Pawlik}, {Milosavljevi{\'c}}, \&
  {Bromm}}]{2013ApJ...767...59P}
{Pawlik}, A.~H., {Milosavljevi{\'c}}, M., \& {Bromm}, V. 2013,
  \href{http://dx.doi.org/10.1088/0004-637X/767/1/59}{\JournalTitle{\apj}, 767,
  59}

\bibitem[{{Pignatari} {et~al.}(2016){Pignatari}, {Herwig}, {Hirschi},
  {Bennett}, {Rockefeller}, {Fryer}, {Timmes}, {Ritter}, {Heger}, {Jones},
  {Battino}, {Dotter}, {Trappitsch}, {Diehl}, {Frischknecht}, {Hungerford},
  {Magkotsios}, {Travaglio}, \& {Young}}]{2016ApJS..225...24P}
{Pignatari}, M., {Herwig}, F., {Hirschi}, R., {et~al.} 2016,
  \href{http://dx.doi.org/10.3847/0067-0049/225/2/24}{\JournalTitle{\apjs},
  225, 24}

\bibitem[{{Pilkington} \& {Gibson}(2012)}]{2012nuco.confE.227P}
{Pilkington}, K., \& {Gibson}, B.~K. 2012, in Nuclei in the Cosmos (NIC XII),
  227

\bibitem[{{Quillen} \& {Bland-Hawthorn}(2008)}]{2008MNRAS.386.2227Q}
{Quillen}, A.~C., \& {Bland-Hawthorn}, J. 2008,
  \href{http://dx.doi.org/10.1111/j.1365-2966.2008.13193.x}{\JournalTitle{\mnras},
  386, 2227}

\bibitem[{{Riebe} {et~al.}(2013){Riebe}, {Partl}, {Enke}, {Forero-Romero},
  {Gottl{\"o}ber}, {Klypin}, {Lemson}, {Prada}, {Primack}, {Steinmetz}, \&
  {Turchaninov}}]{2013AN....334..691R}
{Riebe}, K., {Partl}, A.~M., {Enke}, H., {et~al.} 2013,
  \href{http://dx.doi.org/10.1002/asna.201211900}{\JournalTitle{Astronomische
  Nachrichten}, 334, 691}

\bibitem[{{Ritter} \& {C{\^o}t{\'e}}(2016)}]{2016ascl.soft10015R}
{Ritter}, C., \& {C{\^o}t{\'e}}, B. 2016, {NuPyCEE: NuGrid Python Chemical
  Evolution Environment}, Astrophysics Source Code Library,
  \href{http://arxiv.org/abs/1610.015}{{\sffamily ascl:1610.015}}

\bibitem[{{Ritter} {et~al.}(2017){Ritter}, {Herwig}, {Jones}, {Pignatari},
  {Fryer}, \& {Hirschi}}]{2017arXiv170908677R}
{Ritter}, C., {Herwig}, F., {Jones}, S., {et~al.} 2017, \JournalTitle{ArXiv
  e-prints}, \href{http://arxiv.org/abs/1709.08677}{{\sffamily arXiv:1709.08677
  [astro-ph.SR]}}

\bibitem[{{Rodrigues} {et~al.}(2017){Rodrigues}, {Vernon}, \&
  {Bower}}]{2017MNRAS.466.2418R}
{Rodrigues}, L.~F.~S., {Vernon}, I., \& {Bower}, R.~G. 2017,
  \href{http://dx.doi.org/10.1093/mnras/stw3269}{\JournalTitle{\mnras}, 466,
  2418}

\bibitem[{{Roederer} {et~al.}(2014){Roederer}, {Preston}, {Thompson},
  {Shectman}, {Sneden}, {Burley}, \& {Kelson}}]{2014AJ....147..136R}
{Roederer}, I.~U., {Preston}, G.~W., {Thompson}, I.~B., {et~al.} 2014,
  \href{http://dx.doi.org/10.1088/0004-6256/147/6/136}{\JournalTitle{\aj}, 147,
  136}

\bibitem[{{Romano} {et~al.}(2015){Romano}, {Bellazzini}, {Starkenburg}, \&
  {Leaman}}]{2015MNRAS.446.4220R}
{Romano}, D., {Bellazzini}, M., {Starkenburg}, E., \& {Leaman}, R. 2015,
  \href{http://dx.doi.org/10.1093/mnras/stu2427}{\JournalTitle{\mnras}, 446,
  4220}

\bibitem[{{Romano} \& {Starkenburg}(2013)}]{2013MNRAS.434..471R}
{Romano}, D., \& {Starkenburg}, E. 2013,
  \href{http://dx.doi.org/10.1093/mnras/stt1033}{\JournalTitle{\mnras}, 434,
  471}

\bibitem[{{Rosdahl} {et~al.}(2018){Rosdahl}, {Katz}, {Blaizot}, {Kimm},
  {Michel-Dansac}, {Garel}, {Haehnelt}, {Ocvirk}, \&
  {Teyssier}}]{2018arXiv180107259R}
{Rosdahl}, J., {Katz}, H., {Blaizot}, J., {et~al.} 2018, \JournalTitle{ArXiv
  e-prints}, \href{http://arxiv.org/abs/1801.07259}{{\sffamily
  arXiv:1801.07259}}

\bibitem[{{Saro} {et~al.}(2010){Saro}, {De Lucia}, {Borgani}, \&
  {Dolag}}]{2010MNRAS.406..729S}
{Saro}, A., {De Lucia}, G., {Borgani}, S., \& {Dolag}, K. 2010,
  \href{http://dx.doi.org/10.1111/j.1365-2966.2010.16737.x}{\JournalTitle{\mnras},
  406, 729}

\bibitem[{{Scannapieco} {et~al.}(2012){Scannapieco}, {Wadepuhl}, {Parry},
  {Navarro}, {Jenkins}, {Springel}, {Teyssier}, {Carlson}, {Couchman}, {Crain},
  {Dalla Vecchia}, {Frenk}, {Kobayashi}, {Monaco}, {Murante}, {Okamoto},
  {Quinn}, {Schaye}, {Stinson}, {Theuns}, {Wadsley}, {White}, \&
  {Woods}}]{2012MNRAS.423.1726S}
{Scannapieco}, C., {Wadepuhl}, M., {Parry}, O.~H., {et~al.} 2012,
  \href{http://dx.doi.org/10.1111/j.1365-2966.2012.20993.x}{\JournalTitle{\mnras},
  423, 1726}

\bibitem[{{Schaye} {et~al.}(2010){Schaye}, {Dalla Vecchia}, {Booth}, {Wiersma},
  {Theuns}, {Haas}, {Bertone}, {Duffy}, {McCarthy}, \& {van de
  Voort}}]{2010MNRAS.402.1536S}
{Schaye}, J., {Dalla Vecchia}, C., {Booth}, C.~M., {et~al.} 2010,
  \href{http://dx.doi.org/10.1111/j.1365-2966.2009.16029.x}{\JournalTitle{\mnras},
  402, 1536}

\bibitem[{{Schaye} {et~al.}(2015){Schaye}, {Crain}, {Bower}, {Furlong},
  {Schaller}, {Theuns}, {Dalla Vecchia}, {Frenk}, {McCarthy}, {Helly},
  {Jenkins}, {Rosas-Guevara}, {White}, {Baes}, {Booth}, {Camps}, {Navarro},
  {Qu}, {Rahmati}, {Sawala}, {Thomas}, \& {Trayford}}]{2015MNRAS.446..521S}
{Schaye}, J., {Crain}, R.~A., {Bower}, R.~G., {et~al.} 2015,
  \href{http://dx.doi.org/10.1093/mnras/stu2058}{\JournalTitle{\mnras}, 446,
  521}

\bibitem[{{Simpson} {et~al.}(2017){Simpson}, {Grand}, {G{\'o}mez}, {Marinacci},
  {Pakmor}, {Springel}, {Campbell}, \& {Frenk}}]{2017arXiv170503018S}
{Simpson}, C.~M., {Grand}, R.~J.~J., {G{\'o}mez}, F.~A., {et~al.} 2017,
  \JournalTitle{ArXiv e-prints},
  \href{http://arxiv.org/abs/1705.03018}{{\sffamily arXiv:1705.03018}}

\bibitem[{Smith(2018)}]{ytree}
Smith, B. 2018, https://github.com/brittonsmith/ytree: ytree version 2.0.2
  release

\bibitem[{{Smith} {et~al.}(2017){Smith}, {Bryan}, {Glover}, {Goldbaum}, {Turk},
  {Regan}, {Wise}, {Schive}, {Abel}, {Emerick}, {O'Shea}, {Anninos}, {Hummels},
  \& {Khochfar}}]{2017MNRAS.466.2217S}
{Smith}, B.~D., {Bryan}, G.~L., {Glover}, S.~C.~O., {et~al.} 2017,
  \href{http://dx.doi.org/10.1093/mnras/stw3291}{\JournalTitle{\mnras}, 466,
  2217}

\bibitem[{{Somerville} \& {Dav{\'e}}(2015)}]{2015ARA&A..53...51S}
{Somerville}, R.~S., \& {Dav{\'e}}, R. 2015,
  \href{http://dx.doi.org/10.1146/annurev-astro-082812-140951}{\JournalTitle{\araa},
  53, 51}

\bibitem[{{Springel} {et~al.}(2001){Springel}, {White}, {Tormen}, \&
  {Kauffmann}}]{2001MNRAS.328..726S}
{Springel}, V., {White}, S.~D.~M., {Tormen}, G., \& {Kauffmann}, G. 2001,
  \href{http://dx.doi.org/10.1046/j.1365-8711.2001.04912.x}{\JournalTitle{\mnras},
  328, 726}

\bibitem[{{Starkenburg} {et~al.}(2013){Starkenburg}, {Helmi}, {De Lucia}, {Li},
  {Navarro}, {Font}, {Frenk}, {Springel}, {Vera-Ciro}, \&
  {White}}]{2013MNRAS.429..725S}
{Starkenburg}, E., {Helmi}, A., {De Lucia}, G., {et~al.} 2013,
  \href{http://dx.doi.org/10.1093/mnras/sts367}{\JournalTitle{\mnras}, 429,
  725}

\bibitem[{{Stinson} {et~al.}(2007){Stinson}, {Dalcanton}, {Quinn}, {Kaufmann},
  \& {Wadsley}}]{2007ApJ...667..170S}
{Stinson}, G.~S., {Dalcanton}, J.~J., {Quinn}, T., {Kaufmann}, T., \&
  {Wadsley}, J. 2007,
  \href{http://dx.doi.org/10.1086/520504}{\JournalTitle{\apj}, 667, 170}

\bibitem[{{Stringer} {et~al.}(2010){Stringer}, {Brooks}, {Benson}, \&
  {Governato}}]{2010MNRAS.407..632S}
{Stringer}, M.~J., {Brooks}, A.~M., {Benson}, A.~J., \& {Governato}, F. 2010,
  \href{http://dx.doi.org/10.1111/j.1365-2966.2010.16944.x}{\JournalTitle{\mnras},
  407, 632}

\bibitem[{{Tolstoy} {et~al.}(2009){Tolstoy}, {Hill}, \&
  {Tosi}}]{2009ARA&A..47..371T}
{Tolstoy}, E., {Hill}, V., \& {Tosi}, M. 2009,
  \href{http://dx.doi.org/10.1146/annurev-astro-082708-101650}{\JournalTitle{\araa},
  47, 371}

\bibitem[{Toro(1997)}]{toro-1997}
Toro, E.~F. 1997, Riemann solvers and numerical methods for fluid dynamics : a
  practical introduction (Berlin, New York: Springer)

\bibitem[{{Truelove} {et~al.}(1998){Truelove}, {Klein}, {McKee}, {Holliman},
  {Howell}, {Greenough}, \& {Woods}}]{Truelove98}
{Truelove}, J.~K., {Klein}, R.~I., {McKee}, C.~F., {et~al.} 1998,
  \href{http://dx.doi.org/10.1086/305329}{\JournalTitle{\apj}, 495, 821}

\bibitem[{{Tumlinson}(2006)}]{2006ApJ...641....1T}
{Tumlinson}, J. 2006,
  \href{http://dx.doi.org/10.1086/500383}{\JournalTitle{\apj}, 641, 1}

\bibitem[{{Tumlinson}(2010)}]{2010ApJ...708.1398T}
---. 2010,
  \href{http://dx.doi.org/10.1088/0004-637X/708/2/1398}{\JournalTitle{\apj},
  708, 1398}

\bibitem[{{Tur} {et~al.}(2009){Tur}, {Heger}, \&
  {Austin}}]{2009ApJ...702.1068T}
{Tur}, C., {Heger}, A., \& {Austin}, S.~M. 2009,
  \href{http://dx.doi.org/10.1088/0004-637X/702/2/1068}{\JournalTitle{\apj},
  702, 1068}

\bibitem[{{Turk} {et~al.}(2009){Turk}, {Abel}, \&
  {O'Shea}}]{2009Sci...325..601T}
{Turk}, M.~J., {Abel}, T., \& {O'Shea}, B. 2009,
  \href{http://dx.doi.org/10.1126/science.1173540}{\JournalTitle{Science}, 325,
  601}

\bibitem[{{Turk} {et~al.}(2011){Turk}, {Smith}, {Oishi}, {Skory}, {Skillman},
  {Abel}, \& {Norman}}]{2011ApJS..192....9T}
{Turk}, M.~J., {Smith}, B.~D., {Oishi}, J.~S., {et~al.} 2011,
  \href{http://dx.doi.org/10.1088/0067-0049/192/1/9}{\JournalTitle{\apjs}, 192,
  9}

\bibitem[{{Van Der Walt} {et~al.}(2011){Van Der Walt}, {Colbert}, \&
  {Varoquaux}}]{2011arXiv1102.1523V}
{Van Der Walt}, S., {Colbert}, S.~C., \& {Varoquaux}, G. 2011,
  \JournalTitle{ArXiv e-prints},
  \href{http://arxiv.org/abs/1102.1523}{{\sffamily arXiv:1102.1523 [cs.MS]}}

\bibitem[{{Vincenzo} {et~al.}(2016){Vincenzo}, {Matteucci}, {de Boer},
  {Cignoni}, \& {Tosi}}]{2016MNRAS.460.2238V}
{Vincenzo}, F., {Matteucci}, F., {de Boer}, T.~J.~L., {Cignoni}, M., \& {Tosi},
  M. 2016,
  \href{http://dx.doi.org/10.1093/mnras/stw1145}{\JournalTitle{\mnras}, 460,
  2238}

\bibitem[{{Vogelsberger} {et~al.}(2014){Vogelsberger}, {Genel}, {Springel},
  {Torrey}, {Sijacki}, {Xu}, {Snyder}, {Nelson}, \&
  {Hernquist}}]{2014MNRAS.444.1518V}
{Vogelsberger}, M., {Genel}, S., {Springel}, V., {et~al.} 2014,
  \href{http://dx.doi.org/10.1093/mnras/stu1536}{\JournalTitle{\mnras}, 444,
  1518}

\bibitem[{{Weinberg} {et~al.}(2017){Weinberg}, {Andrews}, \&
  {Freudenburg}}]{2017ApJ...837..183W}
{Weinberg}, D.~H., {Andrews}, B.~H., \& {Freudenburg}, J. 2017,
  \href{http://dx.doi.org/10.3847/1538-4357/837/2/183}{\JournalTitle{\apj},
  837, 183}

\bibitem[{{Weisz} \& {Boylan-Kolchin}(2017)}]{2017MNRAS.469L..83W}
{Weisz}, D.~R., \& {Boylan-Kolchin}, M. 2017,
  \href{http://dx.doi.org/10.1093/mnrasl/slx043}{\JournalTitle{\mnras}, 469,
  L83}

\bibitem[{{Weisz} {et~al.}(2014){Weisz}, {Dolphin}, {Skillman}, {Holtzman},
  {Gilbert}, {Dalcanton}, \& {Williams}}]{2014ApJ...789..147W}
{Weisz}, D.~R., {Dolphin}, A.~E., {Skillman}, E.~D., {et~al.} 2014,
  \href{http://dx.doi.org/10.1088/0004-637X/789/2/147}{\JournalTitle{\apj},
  789, 147}

\bibitem[{{White} \& {Frenk}(1991)}]{1991ApJ...379...52W}
{White}, S.~D.~M., \& {Frenk}, C.~S. 1991,
  \href{http://dx.doi.org/10.1086/170483}{\JournalTitle{\apj}, 379, 52}

\bibitem[{{Williams} {et~al.}(1996){Williams}, {Blacker}, {Dickinson}, {Dixon},
  {Ferguson}, {Fruchter}, {Giavalisco}, {Gilliland}, {Heyer}, {Katsanis},
  {Levay}, {Lucas}, {McElroy}, {Petro}, {Postman}, {Adorf}, \&
  {Hook}}]{1996AJ....112.1335W}
{Williams}, R.~E., {Blacker}, B., {Dickinson}, M., {et~al.} 1996,
  \href{http://dx.doi.org/10.1086/118105}{\JournalTitle{\aj}, 112, 1335}

\bibitem[{{Wise} \& {Abel}(2011)}]{Wise11_Moray}
{Wise}, J.~H., \& {Abel}, T. 2011,
  \href{http://dx.doi.org/10.1111/j.1365-2966.2011.18646.x}{\JournalTitle{\mnras},
  414, 3458}

\bibitem[{{Wise} {et~al.}(2012{\natexlab{a}}){Wise}, {Abel}, {Turk}, {Norman},
  \& {Smith}}]{2012MNRAS.427..311W}
{Wise}, J.~H., {Abel}, T., {Turk}, M.~J., {Norman}, M.~L., \& {Smith}, B.~D.
  2012{\natexlab{a}},
  \href{http://dx.doi.org/10.1111/j.1365-2966.2012.21809.x}{\JournalTitle{\mnras},
  427, 311}

\bibitem[{{Wise} {et~al.}(2014){Wise}, {Demchenko}, {Halicek}, {Norman},
  {Turk}, {Abel}, \& {Smith}}]{2014MNRAS.442.2560W}
{Wise}, J.~H., {Demchenko}, V.~G., {Halicek}, M.~T., {et~al.} 2014,
  \href{http://dx.doi.org/10.1093/mnras/stu979}{\JournalTitle{\mnras}, 442,
  2560}

\bibitem[{{Wise} {et~al.}(2012{\natexlab{b}}){Wise}, {Turk}, {Norman}, \&
  {Abel}}]{2012ApJ...745...50W}
{Wise}, J.~H., {Turk}, M.~J., {Norman}, M.~L., \& {Abel}, T.
  2012{\natexlab{b}},
  \href{http://dx.doi.org/10.1088/0004-637X/745/1/50}{\JournalTitle{\apj}, 745,
  50}

\bibitem[{{Woodward} \& {Colella}(1984)}]{Woodward84}
{Woodward}, P.~R., \& {Colella}, P. 1984, \JournalTitle{J. Comp. Phys.}, 54,
  174

\bibitem[{{Xu} {et~al.}(2016){Xu}, {Wise}, {Norman}, {Ahn}, \&
  {O'Shea}}]{2016ApJ...833...84X}
{Xu}, H., {Wise}, J.~H., {Norman}, M.~L., {Ahn}, K., \& {O'Shea}, B.~W. 2016,
  \href{http://dx.doi.org/10.3847/1538-4357/833/1/84}{\JournalTitle{\apj}, 833,
  84}

\bibitem[{{Yates} {et~al.}(2013){Yates}, {Henriques}, {Thomas}, {Kauffmann},
  {Johansson}, \& {White}}]{2013MNRAS.435.3500Y}
{Yates}, R.~M., {Henriques}, B., {Thomas}, P.~A., {et~al.} 2013,
  \href{http://dx.doi.org/10.1093/mnras/stt1542}{\JournalTitle{\mnras}, 435,
  3500}

\bibitem[{{Young} \& {Fryer}(2007)}]{2007ApJ...664.1033Y}
{Young}, P.~A., \& {Fryer}, C.~L. 2007,
  \href{http://dx.doi.org/10.1086/518081}{\JournalTitle{\apj}, 664, 1033}

\end{thebibliography}
